%% file: main.tex
\documentclass[submitting,floatfix]{nst}
\usepackage[draft]{changes}
\definechangesauthor[name={Revision}, color=blue]{rev}

\usepackage{subfigure,dcolumn}
\usepackage{overpic}
\usepackage[T2A,T1]{fontenc}
\usepackage[russian,english]{babel}
\usepackage{amsmath}

\newcommand{\QCD}{QCD}
\newcommand{\HENP}{high-energy nuclear physics}

\newcommand{\EoS}{equation of state}

\newcommand{\LFT}{lattice field theory}

\newcommand{\SBI}{simulation-based inference}

\newcommand{\LLM}{large language model}

\newcommand{\QML}{quantum machine learning}

\newcommand{\beq}{\begin{equation}}
\newcommand{\eeq}{\end{equation}}
\newcommand{\ba}{\begin{array}}
\newcommand{\ea}{\end{array}}
\newcommand{\bea}{\begin{align}}
\newcommand{\eea}{\end{align}}
\newcommand{\bi}{\begin{itemize}}
\newcommand{\ei}{\end{itemize}}
\newcommand{\ben}{\begin{enumerate}}
\newcommand{\een}{\end{enumerate}}
\newcommand{\bc}{\begin{center}}
\newcommand{\ec}{\end{center}}
\newcommand{\bl}{\begin{flushleft}}
\newcommand{\el}{\end{flushleft}}
\newcommand{\br}{\begin{flushright}}
\newcommand{\er}{\end{flushright}}

\usepackage{listings}
\usepackage{tikz,xcolor}
\definecolor{lime}{HTML}{A6CE39}
\DeclareRobustCommand{\orcidicon}{
	\begin{tikzpicture}
	\draw[lime, fill=lime] (0,0) 
	circle [radius=0.16] 
	node[white] {{\fontfamily{qag}\selectfont \tiny ID}};
	\draw[white, fill=white] (-0.0625,0.095) 
	circle [radius=0.007];
	\end{tikzpicture}
	\hspace{-2mm}
}

\foreach \x in {A, ..., Z}{%
	\expandafter\xdef\csname orcid\x\endcsname{\noexpand\href{https://orcid.org/\csname orcidauthor\x\endcsname}{\noexpand\orcidicon}}
}

\usepackage{url}
\usepackage{hyperref}
\begin{document}
\input{meta}

\begin{abstract}
Machine learning (ML) in high-energy nuclear physics (HENP) is entering a new stage in which physical knowledge is incorporated more directly into data analysis, simulation, and physics inference. This mini-review focuses on developments that have matured in the past several years. Whereas earlier applications emphasized event classification, pattern recognition, and surrogate models for selected observables, recent work has moved toward physics-integrated workflows: calibrated Bayesian extraction of QCD matter properties, dense-matter equation-of-state inference from heavy-ion and neutron-star data, generative event modeling, neural unfolding of weak physical signals, differentiable inverse solvers, gauge-equivariant and diffusion-based lattice-field samplers, and neural reconstruction of model functions in holographic QCD. We survey recent applications of ML in heavy-ion collisions, neutron-star physics, lattice QFT, and holographic or continuum QCD. The emphasis is not on ML architectures alone, but on how they enter concrete physics workflows, how physical constraints such as symmetries, conservation laws, causality, thermodynamic stability, and topology are imposed, and how uncertainty quantification and validation determine whether an AI-assisted result can support a reliable physics conclusion.

\end{abstract}

\maketitle
\nolinenumbers
\tableofcontents

\input{introduction}

\input{HIC_Theory_with_AI}

\input{Bayesian_Inference_Related}
\input{HIC_Experiments_with_AI}
\input{Holographic_QCD_with_machine_learning}
\input{Neutron_Star}
\input{Lattice_QFT}
\input{LLM_and_agents}

\input{summary_outlook}

\begin{acknowledgments}
This work was partially supported by the National Natural Science Foundation of China under Contract  No.\ 12547102, 12688101, 12405154, 
92570117, 12435009, 12575140, the CUHK-Shenzhen University development fund under grant No. UDF01003041 and UDF03003041, Shenzhen Peacock fund under No. 2023TC0179, Ministry of Science and Technology of China under Grant No. 2024YFA1611004 and Shenzhen general fund under No.\ JCYJ20250604141001001, and the STCSM under Grant No. 23590780100 and 23JC1400200.
\end{acknowledgments}

\bibliography{main.bib}
\end{document}

%% file: meta.tex
\title{Machine Learning Meets High-Energy Nuclear Physics: From Pattern Recognition to Physics-Integrated Discovery}
\author{Xun Chen\orcidA{}}\thanks{Email:  chenxun@usc.edu.cn}
\affiliation{School of Nuclear Science and Technology, University of South China, Hengyang 421001, China}
\affiliation{INFN --- Istituto Nazionale di Fisica Nucleare --- Sezione di Bari, Via Orabona 4, 70125 Bari, Italy}
\author{Weiyao Ke\orcidB{}}\thanks{Email:  weiyaoke@ccnu.edu.cn}
\affiliation{Key Laboratory of Quark and Lepton Physics (MOE) \& Institute of Particle Physics, Central China Normal University, Wuhan, Hubei 430079, China}
\affiliation{Southern Center for Nuclear-Science Theory (SCNT), Institute of Modern Physics, Chinese Academy of Sciences, Huizhou, Guangdong 516000, China}
\author{Yu-Gang Ma\orcidC{}}\thanks{Email:  mayugang@ecnu.edu.cn}
\affiliation{School of Physics, East China Normal University, Shanghai 200241, China}
\affiliation{Key Laboratory of Nuclear Physics and Ion-beam Application (MOE), Institute of Modern Physics, Fudan University, Shanghai 200433, China}
\affiliation{Shanghai Research Center for Theoretical Nuclear Physics,
NSFC and Fudan University, Shanghai 200438, China}

\author{Long-Gang Pang\orcidD{}}\thanks{Email:  lgpang@ccnu.edu.cn}
\affiliation{Key Laboratory of Quark and Lepton Physics (MOE) \& Institute of Particle Physics, Central China Normal University, Wuhan, Hubei 430079, China}

\author{Kai Zhou\orcidE{}}\thanks{Email:  zhoukai@cuhk.edu.cn}
\affiliation{School of Science and Engineering, The Chinese University of Hong Kong (Shenzhen), \\
Shenzhen, Guangdong, 518172, China}
\affiliation{School of Artificial Intelligence, The Chinese University of Hong Kong (Shenzhen), \\
Shenzhen, Guangdong, 518172, China}

%% file: introduction.tex
\section{Introduction}
\label{sec:intro} 
A central goal of high-energy nuclear physics (HENP) is to understand the properties and dynamics of strongly interacting QCD matter under extreme conditions. Ultra-relativistic heavy-ion collisions create a short-lived droplets of quark--gluon plasma (QGP) \cite{MaYG_arxiv,He:2023zin,Shuryak2004QGP, GyulassyMcLerran2005, HeinzSnellings2013,ChenJH:2024,ShouQY:2024} that allows the study of QCD collective dynamics and phase diagrams in the high-temperature and vanishing baryon chemical potential region \cite{BraunMunzinger2016, Bzdak2019Fluctuations}, whereas collisions at lower beam energies produce baryon-rich QCD matter and probe the phase diagram at larger baryon chemical potential
\cite{Bzdak2019Fluctuations,An2022BEST,Du2024BES}.
Beyond collider experiments, neutron stars provide a complementary probe of cold dense matter. Measurements of neutron-star masses and radii from radio and X-ray observations, together with tidal-deformability constraints extracted from gravitational-wave signals, can be used to constrain the equation of state (EoS) of neutron-star matter~\cite{Abbott2017GW170817,Miller2019NICER,Annala2020NS}. 
From the theory side, lattice QCD supplies first-principles information at zero and small baryon chemical potential \cite{Aoki2006Lattice,Borsanyi2014Lattice,Bazavov2019HotQCD}, while continuum and holographic approaches extend the discussion toward real-time transport, spectral functions, and finite-density phenomenology \cite{Romatschke2019Hydro,CasalderreySolana2011AdS}. 
Each area constrains a different region of the QCD phase diagram or a different aspect of real-time dynamics; together, they provide a more complete understanding of strongly interacting QCD matter. Across these areas, many central scientific tasks in HENP tasks involve both forward and inverse problems. The Machine Learning (ML) methods can assist both by accelerating forward calculations and enabling inference from high-dimensional observables, particularly when the quantities of interest are connected to measurements through computationally expensive models.

This mini-review should be read as an update to the earlier review\cite{He:2023zin} that introduced the broad landscape of ML applications in HENP, ranging from event classification and learned observables to surrogate modeling and early generative approaches. The timing of the present update is motivated not merely by the growth in publication volume, but by a qualitative change in how ML enters HENP. Since Ref~\cite{He:2023zin}, several previously separate directions have matured simultaneously from proof-of-principle studies into components of quantitative physics workflows: full-event generative modeling, calibrated Bayesian and simulation-based inference, differentiable inverse solvers, symmetry-preserviing lattice samplers, and detector-level or online deployment. This convergence provides a natural point at which to reassess standards of physical validation, model dependence, uncertainty propagation, and reproducibility. Rather than repeating the general motivation for ML in HENP, we therefore focus on developments that have become sufficiently integrated with iphysics workflows to affect quantitative inference.

The experimental ML literature is already broad and highly collaboration- and detector-specific. Accordingly, the present mini-review is weighted toward theory, simulation, and physics inference. Section IV presents selected detector-facing examples rather than an exhaustive account of the many ML tools used in tracking, clustering, particle identification, triggering, calibration, data-quality monitoring, background rejection, and unfolding. This scope is stated explicitly so that the selection of examples is not mistaken for a comprehensive survey of experimental practice.

In heavy-ion collisions (HIC), the forward problem maps fluctuating nuclear initial conditions and medium properties to final-state particle spectra and correlations through pre-equilibrium dynamics, relativistic viscous hydrodynamics or transport, particlization, hadronic rescattering, and, where necessary, detector response. The related inverse problem reverses this logic: because the microscopic evolution cannot be observed directly, the properties of the produced matter must be inferred by confronting model predictions with final-state measurements while consistently accounting for experimental and theoretical uncertainties~\cite{Bernhard2019Bayesian,Pratt2015}. The inverse problem is therefore not a literal event-by-event reversal of a one-to-one map. The experimentally relevant map is many-to-one because of fluctuating initial conditions, quantum and thermal fluctuations, stochastic collision terms in effective transport descriptions, dissipative evolution, hadronization and decays, unobserved degrees of freedom, finite acceptance, and detector response. Many microscopic histories can consequently lead to statistically indistinguishable final states. Only in idealized limits with complete state information, such as nondissipative hydrodynamic evolution at the level of the equations, can formal reversibility be discussed. In realistic analyses, ML seeks a calibrated posterior over latent properties or an unfolding of ensemble-level information; it cannot uniquely reconstruct the microscopic history of an individual collision.

In neutron-star studies, the forward problem maps the nuclear EoS through the Tolman--Oppenheimer--Volkoff (TOV) equations to macroscopic mass--radius and tidal-deformability relations. X-ray pulse-profile and gravitational-wave models then connect these properties to observational data. The inverse problem is to constrain the unknown EoS function using astronomical measurements~\cite{Abbott2017GW170817,Miller2019NICER}, often supplemented by constraints from nuclear theory and experiments. Neutron-star inference is additionally data-limited in a way that differs qualitatively from collider analyses. Only a modest number of objects currently provide informative mass-radius, pulse-profile, tidal-deformability, or moment-of-inertia constraints, and the corresponding likelihoods are broad, heterogeneous, and model-dependent. Large simulated training sets do not create new independent observations; prior sensitivity, posterior calibration, and robustness to simulator mismatch therefore remain central.

Lattice field theory presents a related but distinct set of computational challenges. The forward calculation samples field configurations from the Euclidean path integral and uses them to evaluate correlation functions and other lattice observables. This sampling can be computationally expensive because of long autocorrelation times and topological freezing, while calculations at finite baryon chemical potential encounter the sign problem~\cite{Aarts2016SignProblem}. The corresponding inverse problem is to reconstruct real-time spectral information from finite and noisy Euclidean correlators, an intrinsically ill-posed task further complicated by statistical and systematic uncertainties.

Finally, in bottom-up holographic QCD, the forward problem maps a chosen bulk action, including its potentials and couplings, to thermodynamic and transport observables of the boundary theory. The inverse problem can be formulated either as the calibration of a parameterized bulk model towards experimental data or as the direct reconstruction of unknown bulk functions from available boundary-theory data~\cite{CasalderreySolana2011AdS}. Once constrained, the resulting model can be used to extrapolate into regions where direct calculations are unavailable.

Despite their different physical inputs and observables, these examples share a common computational structure: a forward map connects microscopic properties and model assumptions to observable quantities, while the inverse problem uses finite, incomplete, and often noisy or high-dimensional data to constrain both the physical properties and the viable theoretical descriptions. Statistically controlled inference generally requires repeated evaluations of the forward map across a large parameter or model space, which can make direct sampling prohibitively expensive and demands consistent propagation of all relevant uncertainties.

Within this common computational structure, the recent development of ML in HENP can be understood as a shift from isolated applications toward integration into complete physics workflows. Early studies often used ML as a direct map from observables to physical labels or parameters, or as a surrogate model that accelerated conventional Bayesian analyses. More recent work reflects a broader methodological scope, a closer incorporation of physical knowledge, and more stringent standards for validation and uncertainty quantification. These developments build on the progress summarized in several recent reviews~\cite{He:2023zin,2303.15136v2,Ma:2023zfj,Aarts:2025gyp,He:2026,AI:NP}. Here, we focus on methods that have matured in recent years and on the specific roles they play in physics analyses.

\begin{figure*}[htb]
{\includegraphics[width=1.0\textwidth]{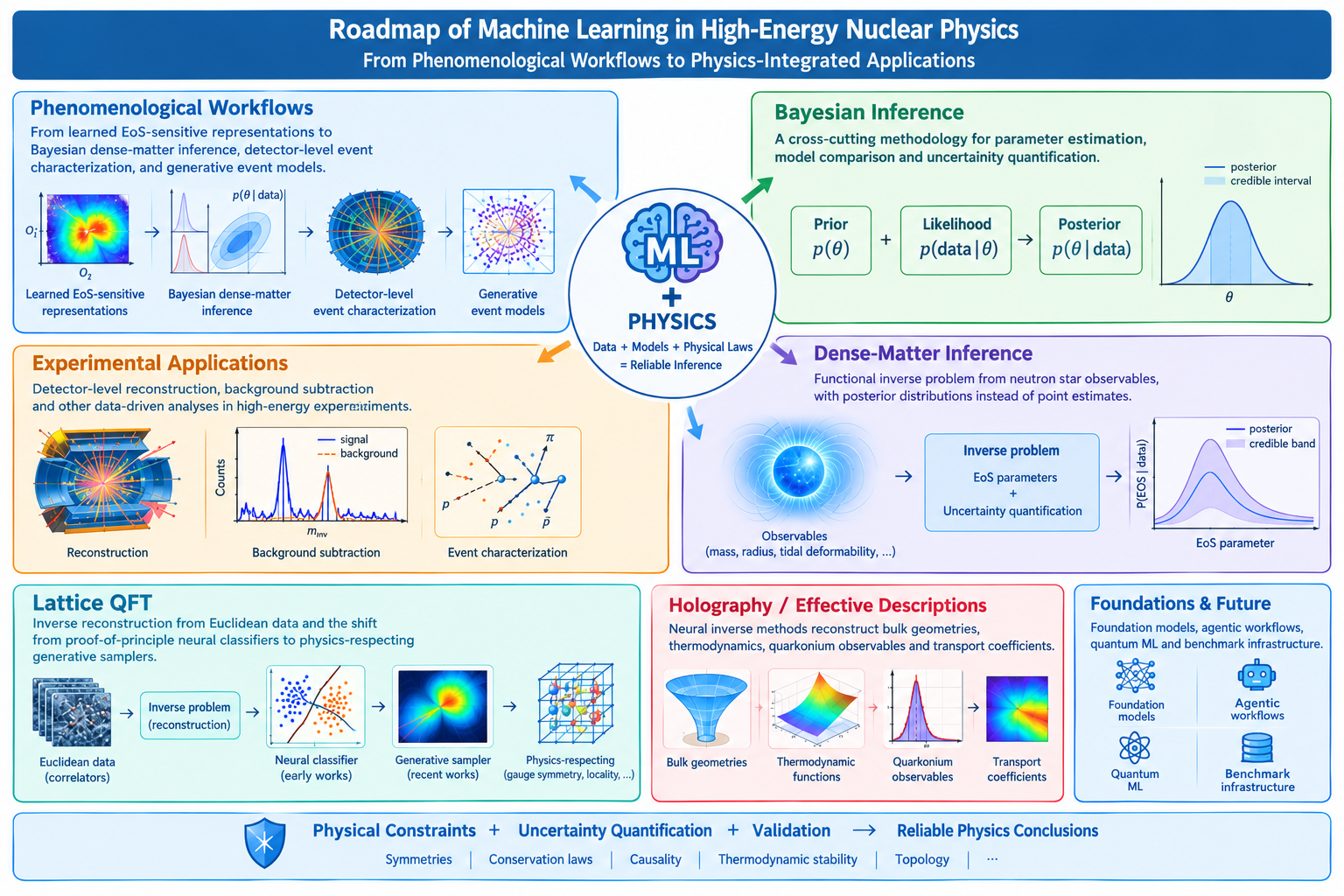}}
{\caption{Schematic organization of the review on machine learning in high-energy nuclear physics. The review is structured around seven interconnected themes: phenomenological workflows, Bayesian inference, experimental applications, dense-matter inference, lattice QFT, holographic and effective descriptions, and emerging foundations including foundation models, agentic scientific workflows, quantum machine learning, and benchmark infrastructure. The central hub emphasizes the integration of machine learning with physical models and data across these areas. The bottom layer highlights the common requirements for reliable physics inference—physical constraints such as symmetries, conservation laws, causality, thermodynamic stability, and topology, together with uncertainty quantification and rigorous validation.}\label{fig:myg_nst}}
\end{figure*}

At the methodological level, ML can now contribute at several distinct stages. Representation-learning methods extract informative features from high-dimensional inputs such as particle-level collision events and lattice field configurations. Surrogate models replace expensive components of forward calculations, reducing the cost of repeated model evaluations in Bayesian analyses. Neural likelihood and posterior estimators enable simulation-based inference when an explicit likelihood is unavailable or prohibitively expensive. Generative models, meanwhile, can accelerate the production of collision events or field configurations. These methods address different parts of the forward and inverse problems and should therefore be evaluated according to their specific roles.

At the physical level, known structures can increasingly be incorporated directly into the ML procedure. Equivariant architectures can enforce selected symmetries by construction, while physics-informed objectives and differentiable solvers can incorporate evolution equations or penalize violations of conservation laws, causality, and thermodynamic stability. 

In situations where ML do not by themselves guarantee physical consistency, the relevant constraints must still be tested over the full domain in which the model is applied. This leads to a broader standard of validation. For example, a classifier trained to distinguish simulated EoS labels is physically informative only if its predictions remain robust against changes in the dynamical model, experimental acceptance, centrality definition, and nuisance parameters. A generative model for heavy-ion events must reproduce not only single-particle spectra but also multi-particle correlations, event-plane correlations, rare fluctuations, and any relevant parameter dependence. An EoS inference framework must satisfy thermodynamic stability and causality while providing statistically calibrated uncertainties. A lattice-QFT sampler must reproduce the target distribution and controlled observables, with its autocorrelation times, sampling of topological sectors, and acceptance or reweighting corrections explicitly quantified. Reliable scientific applications therefore require known equations, symmetries, uncertainties, and validation tests to remain visible throughout the ML workflow.

These developments take different forms across the areas considered here. In heavy-ion physics, applications have expanded from event classification to calibrated parameter inference, generative event modeling, detector-level reconstruction, and the extraction of weak signals from complex backgrounds. In neutron-star physics, neural representations and differentiable inference pipelines permit more flexible reconstruction of the EoS, reducing reliance on a small number of predetermined parameters. In lattice QFT, generative methods have progressed from demonstrations in scalar field theories to gauge-equivariant samplers in gauge field theories with explicit tests of exactness, autocorrelation, and topology. In holographic QCD and related continuum approaches, ML is increasingly used to reconstruct bulk potentials and other model functions from boundary observables and to propagate uncertainties into subsequent predictions.

The content of this review is organized as follows. Section~\ref{sec:representation} discusses how ML is changing phenomenological workflows: from learned EoS-sensitive representations to Bayesian dense-matter inference, detector-level event characterization, and generative event models.
Section~\ref{sec:probabilistic} treats Bayesian inference as a cross-cutting methodology and discusses.. Section~\ref{sec:hic_exp} turns to experimental applications such as detector-level reconstruction, background subtraction.
Section~\ref{sec:neutron-star} treats dense-matter inference as a functional inverse problem from neutron star observables, emphasizing posterior distributions rather than point estimates.  Section~\ref{sec:lattice} focuses on inverse reconstruction from Euclidean data and the recent transition from proof-of-principle neural classifiers to physics-respecting generative samplers. Section~\ref{sec:holo} reviews ML-assisted holographic and effective descriptions, where neural inverse methods reconstruct bulk geometries, thermodynamic functions, quarkonium observables, and transport coefficients. Section~\ref{sec:foundation-agent} closes the technical survey with foundation models, agentic scientific workflow, quantum machine learning and benchmark infrastructre. In all cases, the goal is not to present ML as a black box, but to identify where physical constraints, uncertainty quantification, and validation must enter. The structure of this article can be summarized by Fig. ~\ref{fig:myg_nst}.

%% file: HIC_Theory_with_AI.tex
\section{AI for HIC Phenomenology Studies}
\label{sec:representation}
Relativistic heavy-ion collisions remain the most direct terrestrial probe of hot and dense baryon-rich QCD matter.  From the viewpoint of ML, they are challenging because the physics of interest is not an isolated physical parameters but a set of latent quantities---initial-state energy deposition and fluctuations, the QCD EoS, the shear and bulk viscosities, jet-quenching parameters, details in fluid-particle matching (particlization) and in hadronic afterburners, and possible critical or anomalous-transport signals---encoded in high-dimensional final-state distributions. ML methods are now fused at three levels: to learn representations of the final-state information for posterior-inference of QCD matter properties or desired physics, to accelerate or calibrate the event-level simulations, and to reconstruct latent physical information with quantified uncertainty. The present section will focus on theory-side phenomenological developments recently; detector-facing experimental uses are discussed separately in Sec.~\ref{sec:hic_exp}.

\subsection{Event representations and learned observables}

The EoS-meter in Ref.~\cite{Pang:2016vdc} provided an early and useful exploration of incorporating modern deep learning into HIC physics studies. A convolutional network trained on final-state pion's spectra was able to distinguish different QCD-transition scenarios in HIC simulations~\cite{Pang:2016vdc,Du:2019civ,Du:2020poe}.  The important lesson was not that a classifier alone can determine the QCD phase diagram, but that high-dimensional spectra may retain information about the thermodynamic response of the medium after dynamical evolution.  This observation motivated later studies of spinodal clumping, event-level transition recognition, and criticality-sensitive representations~\cite{Steinheimer:2019iso,Steinheimer:2021hoc,Huang:2021iux}.  The more recent question is how such information can be incorporated into calibrated inference with uncertainties.

Phase transition learning has also been tested with lower energy heavy-ion data e.g., in Ref.~\cite{Wang:2020tgb}, the unsupervised learning and a confusion scheme was used to classify liquid-like and gas-like patterns from final-state nuclear-reaction data, obtaining a limiting temperature consistent with the caloric-curve method. The density and beam-energy regime differ from relativistic QCD searches; yet the lesson is also that a learned classifier becomes useful after it is tied to a conventional physics observable and a closure test, as emphasized in the recent review~\cite{Ma:2023zfj}.

The representation of heavy-ion collision events has since become more faithful to the underlying data structure. In the CBM-energy regime, PointNet-based approaches treat an event as an unordered point cloud of particles and have been used for centrality (impact-parameter) and EoS-related tasks~\cite{OmanaKuttan:2020brq, OmanaKuttan:2020btb}. Point-cloud networks are natural for HIC because final-state particles form permutation-invariant sets rather than images (in terms of matrices) on a fixed grid. More recent work has used point-cloud networks to learn collective-flow patterns between small and large colliding systems~\cite{Guo:2023zfk} and transformer-based self-supervised learning to construct compact latent features for HIC events~\cite{Zhang:2025wqc}. The methodological shift is from supervised classification of a few labels to representation learning that can support multiple downstream tasks and can be interrogated with tools such as saliency, SHAP values or comparisons with physics-engineered observables.

Physics-informed neural networks have been used in quasi-parton description to construct a four-dimensional QCD EoS~\cite{Li:2026hnl} over thermodynamic variables from first principle lattice QCD evaluations. This line of work complements lattice QCD input, hydrodynamic calibration, and controlled effective models by building thermodynamic identities and physical constraints into neural parameterizations, before they are used in phenomenology.

\subsection{Generative event models and fast simulation}

Generative AI is rapidly becoming one of the most active areas of ML for heavy-ion physics studies, especially for simulation speed-up.
Earlier surrogate models often predicted selected observables or hydrodynamic fields on average.  Recent works aim to generate full event-level information. The goal is not merely to draw realistic event images, but to learn conditional probability distributions over colision output. DiffHIC introduced an end-to-end conditional diffusion model that maps initial entropy-density profiles and transport information to final charged-particle two dimensional spectra~\cite{Sun:2024lgo}, where the integrated and differential flow observables, higher-order correlations and fluctuations are reproduced well. Differently, point-cloud diffusion models such as HEIDi instead generate events represented as particle cloud consisting of complete lists of final-state hadrons, including momenta and particle species, and therefore preserve the natural unordered-particle structure of collision outputs~\cite{OmanaKuttan:2024mwr,OmanaKuttan:2025qjj}.  This is a substantial step toward AI-assisted event generation for next-generation heavy-ion experiments. 

\begin{figure}[t]
    \centering
    \includegraphics[width=0.98\linewidth]{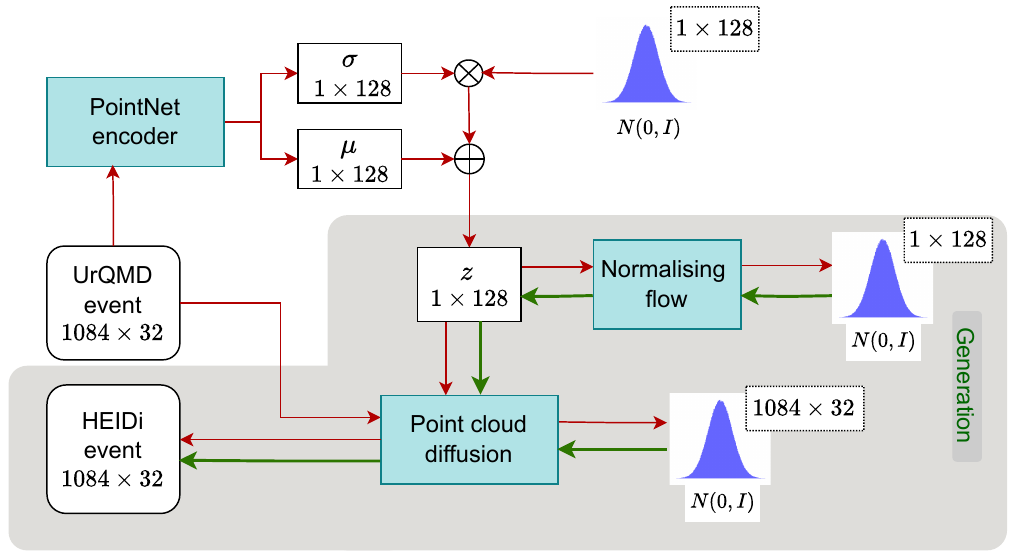}
    \caption{Network structure of the HEIDi point-cloud generative model.  A PointNet encoder, normalizing-flow latent decoder, and point-cloud diffusion module are combined to generate complete event outputs.  Adapted from Ref.~\cite{OmanaKuttan:2025qjj}; see also Ref.~\cite{OmanaKuttan:2024mwr}.}
    \label{fig:hic_heidi_architecture}
\end{figure}

Generative models are also being developed for more specialized heavy-ion tasks. Denoising diffusion has been tested for whole-event simulation in experimental environments~\cite{Torbunov:2024iki}; diffusion and image-reconstruction methods have been used for heavy-ion background subtraction~\cite{Qureshi:2025ylv}; Flow-matching models have also been introduced for jet-induced hydrodynamic response, generating final-state hadron spectra from initial $\gamma$-jet information with large speedup relative to full CoLBT-hydro calculations~\cite{Wu:2026pdi}. Such methods are attractive because hard probes and medium response require many repeated simulations over jet energy, position, direction, and medium evolution. A fast conditional generator could make Bayesian jet-tomography analysis substantially more scalable.

Generative event models should be judged by physics validation rather than visual plausibility alone.  Reproducing particle spectra is only a first test.  A useful generator should also reproduce multiplicity fluctuations, species-dependent correlations, anisotropic-flow distributions, event-plane correlations, centrality and system-size trends, and rare-event tails.  If the generator is used in Bayesian calibration, it must preserve parameter dependence and provide uncertainty estimates.  If it is used for detector studies, it must be connected to reconstruction and acceptance effects.  These requirements are more demanding than standard image-generation metrics, and they motivate domain-specific benchmarks for heavy-ion generative AI.

The foundation-model language has recently appeared in this context because a sufficiently general event generator could serve as a reusable representation of heavy-ion collision outputs.  Such a model would not be a universal simulator of QCD matter by itself.  Its value would come from being a fast, differentiable, and conditionable component that can be inserted into inference, detector optimization, and rare-signal searches.  In this sense, heavy-ion generative modeling is moving from single-observable emulation toward a more ambitious event-distribution learning program.

\subsection{Subtle signals identification from HICs}

Another recent trend is the use of ML to isolate weak or entangled physical signals in HICs. Hard probes and rare signals pose a different challenge essentially: the physics signal is often small compared with background fluctuations. Deep Learning has been used to classify quenched jets and to test how robust such classification remains under underlying-event contamination and model variation~\cite{Apolinario:2021olp,Liu:2022hzd,ArrudaGoncalves:2025wtb,Qureshi:2024ceh}. A related direction uses Bayesian and information-field methods to constrain the jet transport coefficient from multiple observables~\cite{Xie:2022ght}.
The target is a controlled inverse workflow that connects hard probes to medium properties with quantified uncertainty. These applications are important because jet-medium response contains differential information about the QGP transport properties, but the relevant structures are embedded in a evolving and fluctuating soft background. 

Nuclear deformation and initial nucleon distributions affect final-state flow and transverse-momentum correlations, but the inverse map is complicated by hydrodynamic response and event-by-event fluctuations.  Interpretable neural approaches have been used to connect deformation information to final-state observables~\cite{Pang:2019aqb}, and Bayesian imaging has been applied to infer nucleon distributions in isobar collisions~\cite{Cheng:2023ucp}.  Recently, self-attention mechanisms from transformers have been applied to detect nucleon–nucleon correlations using final-state hadrons produced in heavy-ion collisions \cite{Huang:2025uvc}. A transformer without positional embedding naturally respects the permutation symmetry of a particle cloud. Furthermore, the study of nucleon–nucleon short-range correlations demonstrates that averaging learned features over multiple events performs remarkably well for identifying subtle signals.
These studies are examples of a broader strategy: use ML to construct or discover observables that are sensitive to specific initial-state structures while suppressing nuisance variations.

\begin{figure}[t]
    \centering
    \includegraphics[width=1.0\linewidth]{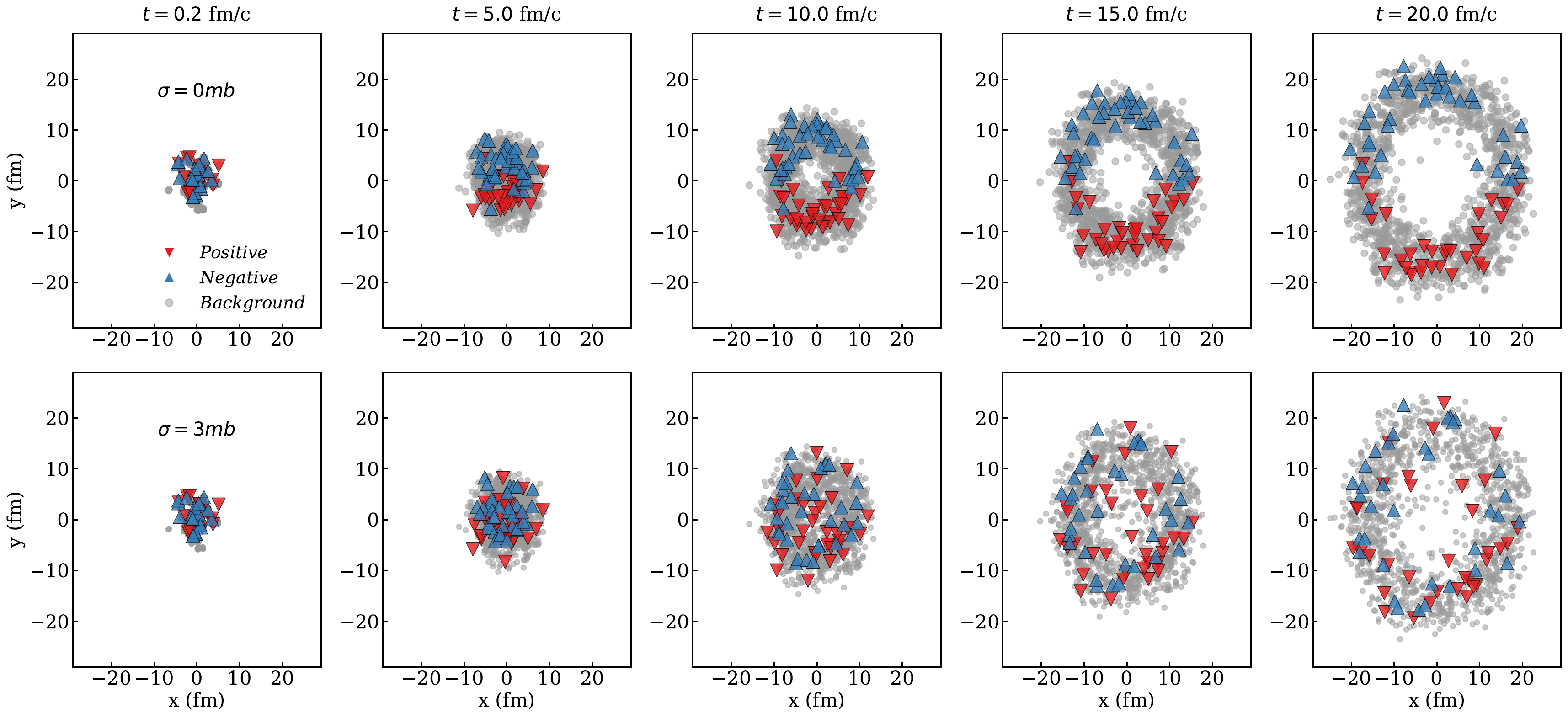}
    \caption{Space-time evolution of a simulated CME charge-separation pattern used in a neural-unfolding study.  The figure illustrates how interactions and late-stage evolution degrade the signal that the network attempts to reconstruct.  Adapted from Ref.~\cite{Guo:2025wry}.}
    \label{fig:hic_cme_unfolding}
\end{figure}

The chiral magnetic effect (CME) provides another instructive case.  Traditional CME searches are limited by large backgrounds from flow, resonance decays, local charge conservation, and final-state interactions.  A neural CME-meter was proposed earlier as a way to learn charge-separation information from simulated events~\cite{Zhao:2021yjo}.  More recent neural-unfolding work targets the inverse problem more directly: reconstruct the time-reversed or less-degraded CME signal from final-state information and quantify how interactions obscure the initial charge separation~\cite{Guo:2025wry}. Fig.~\ref{fig:hic_cme_unfolding} shows the space-time degradation that such a network learns to invert. The significance of this development is methodological as well as phenomenological.  ML is no longer used only to label events; it is used to define an unfolding map between an experimentally accessible distribution and a latent physical signal.

The common lesson from these examples is that heavy-ion ML should be tied to a physical hypothesis.  A high validation accuracy is not enough.  The trained model should identify which information is used, how it changes under alternative generators or detector conditions, and how the result is propagated into a physics uncertainty.  The most useful applications will combine learned representations with posterior inference and posterior-predictive checks.

\subsection{Symbolic regression and interpretable model discovery}
\label{sec:sr}

Most ML applications in HENP are optimized for prediction: a model is
trained to reproduce an observable, while the learned mapping remains
difficult to express in the analytical language of physics. Symbolic
regression (SR) provides a complementary approach by searching directly
over mathematical expressions, learning both their functional form and
numerical parameters from data. It therefore offers a route from
data-driven prediction to interpretable relations and, potentially, to
physics-informed model discovery
\cite{makke2024review,cranmer2023pysr}.

The challenge is that the space of candidate expressions is discrete and
combinatorially large. Modern SR methods address this problem using
evolutionary search, sparse regression, or neural generative models for
expression trees \cite{makke2024review}. For example, NeSymReS employs an
encoder--decoder transformer that maps numerical data to a
permutation-invariant latent representation and autoregressively generates
candidate equation skeletons, whose numerical coefficients are subsequently
optimized \cite{biggio2021nesymres}. The methodological objective,
however, is not merely to find a compact fit: a useful SR result should
identify a functional structure that remains interpretable, generalizes
beyond the data used for inference, and survives independent physical
consistency tests.

Applications to real experimental data in HENP remain limited, although
SR has already demonstrated the ability to recover known physical
relations from complex or noisy datasets
\cite{lemos2023orbital,reinbold2021robust,udrescu2020aifeynman}. Open-source
frameworks such as PySR have further lowered the practical barrier to
symbolic model discovery \cite{cranmer2023pysr}. A representative HENP
application is the study of Makke and Chawla
\cite{makke2024tsallis}, which applies a pretrained NeSymReS model directly
to measured transverse-momentum-dependent charged-hadron multiplicities
from COMPASS SIDIS data and ALICE $pp$ collisions
\cite{aghasyan2018compass,alice2013energy,alice2016pseudorapidity,alice2019multiplicity}.
Rather than assuming a specific functional form, the analysis searches for
analytical expressions directly from the measured distributions. A
frequently recovered structure is
\begin{equation}
  f(u) \propto \frac{1}{1+c\,u^n},
  \qquad u \equiv p_T^2,\qquad n=2,3,4,
\end{equation}
with the best-performing form given by
\begin{equation}
  f_3(u)=\frac{c_0}{1+c_1u^3}.
\end{equation}
Its structural similarity to the Tsallis distribution
\cite{tsallis1988possible} illustrates how SR can recover compact
relations without imposing the corresponding phenomenological ansatz
a priori.

More importantly, this example highlights what is required before an
SR result can be interpreted as physics rather than simply as an accurate
fit. First, candidate expressions must satisfy basic physical requirements,
such as dimensional consistency and the exclusion of mathematically
allowed but physically implausible functional structures. Second,
generalization provides a nontrivial test: an equation skeleton inferred
from one subset of kinematic intervals can be refitted and tested in
intervals that were not used during inference. In the above analysis, the
same functional structure describes multiple unseen $(x,Q^2)$ intervals.
As illustrated in Fig.~\ref{fig:sr_generalization}, the
SR-learned form $f_3(u)=c_0/(1+c_1u^3)$ reproduces the measured
$p_T^2$ dependence across several kinematic intervals, including
intervals that were not used to infer the equation skeleton. This
cross-interval behavior provides a nontrivial test that the learned
relation captures a common functional structure rather than a local
statistical fluctuation.

A further validation comes from the extracted parameters, which exhibit
consistency with an independent relation motivated by the Gaussian
description of the average transverse momentum. Such a comparison is
particularly important for SR: agreement with an independent physical
relation provides a physics-level consistency check that is stronger than
the numerical quality of the original fit alone. In this sense,
cross-domain generalization and independent physical consistency provide
complementary tests of whether a symbolic expression has captured useful
physics.

These requirements also expose a central limitation of current SR
approaches. Uncertainty quantification, failure-rate characterization,
and systematic reporting of unsuccessful inferences remain underdeveloped,
particularly for neural SR methods. A compact and interpretable expression
is therefore not, by itself, evidence for a new physical law. At its
present stage, SR is better viewed as a hypothesis-generation framework:
it proposes candidate analytical structures that must subsequently be
tested against symmetries, dimensions, known limits, independent
observables, and uncertainty estimates. The longer-term opportunity for
HENP is to combine symbolic search with such physics constraints, turning
\emph{data-driven functional discovery} into a systematic pipeline for
\emph{interpretable and testable model discovery}.

\begin{figure*}[t]
  \centering
  \includegraphics[width=0.8\textwidth]{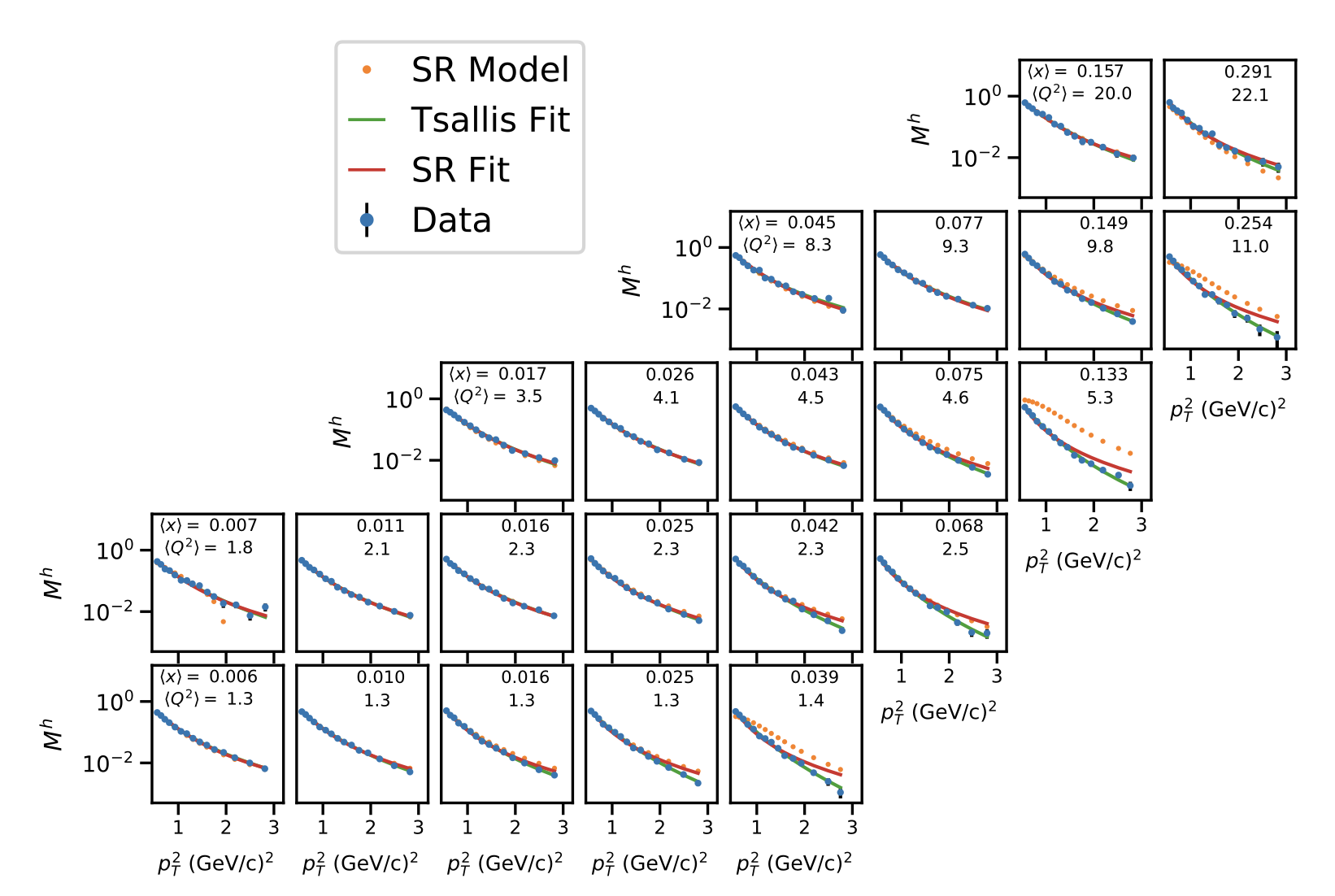}
  \caption{Cross-interval generalization of a symbolic-regression (SR)
  model for charged-hadron multiplicities. Differential multiplicities as
  a function of $p_T^2$, with $p_T^2>0.5~(\mathrm{GeV}/c)^2$, are shown for
  several $(x,Q^2)$ intervals in the second $z$ bin. The SR models are
  compared with a Tsallis fit and with the best-learned SR form
  $f_3(u)=c_0/(1+c_1u^3)$. The same equation skeleton, inferred from a
  subset of intervals, describes additional intervals not used during
  inference, illustrating the potential of SR for interpretable
  cross-domain generalization. Adapted from Ref.~\cite{makke2024tsallis}.}
  \label{fig:sr_generalization}
  \label{fig:sr_generalization}
\end{figure*}

%% file: Bayesian_Inference_Related.tex
\section{Bayesian Inference}
\label{sec:probabilistic}

Given a model $\mathcal{M}$ with parameters collectively denoted by $\boldsymbol{\theta}$, Bayesian inference determines their probability distribution conditioned on data $\boldsymbol{D}$:
\begin{align}
p(\boldsymbol{\theta}\mid\boldsymbol{D},\mathcal{M})
=
\frac{
p(\boldsymbol{D}\mid\boldsymbol{\theta},\mathcal{M})
p(\boldsymbol{\theta}\mid\mathcal{M})
}{
p(\boldsymbol{D}\mid\mathcal{M})
}.
\end{align}
Here, $p(\boldsymbol{\theta}\mid\mathcal{M})$ is the prior, which represents knowledge or assumptions about the parameters before comparison with the data. The likelihood
$p(\boldsymbol{D}\mid\boldsymbol{\theta},\mathcal{M})$ quantifies the probability of obtaining the observed data for a given parameter set, while the posterior
$p(\boldsymbol{\theta}\mid\boldsymbol{D},\mathcal{M})$ incorporates the constraints supplied by the data. The denominator,
\begin{align}
p(\boldsymbol{D}\mid\mathcal{M})
=
\int d\boldsymbol{\theta}\,
p(\boldsymbol{D}\mid\boldsymbol{\theta},\mathcal{M})
p(\boldsymbol{\theta}\mid\mathcal{M}),
\end{align}
is the Bayesian evidence. It normalizes the posterior and quantifies the compatibility of the model with the data after averaging over the prior parameter space. We use boldface symbols for vectors and ordinary symbols for individual parameters, data points, or observables.

Bayesian inference is a statistical framework rather than an ML method by itself. In HENP applications, ML commonly enters by accelerating the forward calculation or by learning otherwise intractable likelihoods or posterior distributions.

For sufficiently simple problems, the likelihood can be derived explicitly. In many heavy-ion applications, however, it is approximated by a multivariate Gaussian,
\begin{align}
\log p(\boldsymbol{D}\mid\boldsymbol{\theta})
&=
-\frac{1}{2}
\left[
\boldsymbol{y}(\boldsymbol{\theta})-\boldsymbol{D}
\right]^T
\boldsymbol{\Sigma}^{-1}
\left[
\boldsymbol{y}(\boldsymbol{\theta})-\boldsymbol{D}
\right]
\nonumber\\
&\quad
-\frac{1}{2}\log\det\boldsymbol{\Sigma}
-\frac{N}{2}\log(2\pi),
\end{align}
where $\boldsymbol{y}(\boldsymbol{\theta})$ is a vector of $N$ model observables and $\boldsymbol{\Sigma}$ is their total covariance matrix. Depending on the analysis, $\boldsymbol{\Sigma}$ may include experimental statistical and systematic uncertainties, finite-simulation statistics, emulator uncertainty, and estimates of model discrepancy. The Gaussian form is an assumption and should be tested whenever non-Gaussian uncertainties or low-statistics observables are important.

Heavy-ion observables are commonly estimated by averaging over many fluctuating collision events. Obtaining sufficient statistics at every point in a high-dimensional parameter space can therefore require a prohibitive number of model calculations. This difficulty is often addressed using a surrogate model, such as a Gaussian-process (GP) emulator.

The full physical model is first evaluated at $M$ representative design points, $\left\{
\boldsymbol{\theta}_i,
y_i=y(\boldsymbol{\theta}_i)
\right\}, i=1,2,\cdots M$, chosen to cover the parameter region of interest. For a scalar observable with its mean removed, a GP treats $y(\boldsymbol{\theta})$ as a random function characterized by a covariance kernel
$k(\boldsymbol{\theta},\boldsymbol{\theta}')$. 
The predictive mean and variance at a new point $\boldsymbol{\theta}$ are
\begin{align}
\mu(\boldsymbol{\theta})
&=
\boldsymbol{k}(\boldsymbol{\theta}, \boldsymbol{\theta}_i)
[\boldsymbol{k}(\boldsymbol{\theta}_i, \boldsymbol{\theta}_j)]^{-1}
\boldsymbol{y}(\boldsymbol{\theta}_j),
\\
\sigma^2(\boldsymbol{\theta})
&=
k(\boldsymbol{\theta},\boldsymbol{\theta})
-
\boldsymbol{k}(\boldsymbol{\theta}, \boldsymbol{\theta}_i)
[\boldsymbol{k}(\boldsymbol{\theta}_i, \boldsymbol{\theta}_j)]^{-1}
\boldsymbol{k}(\boldsymbol{\theta}_j, \boldsymbol{\theta})
\end{align}
A commonly used choice is the squared-exponential kernel,
\begin{align}
k(\boldsymbol{x},\boldsymbol{x}')
=
\sigma_f^2
\exp\left[
-\frac{1}{2}
\sum_{a}
\frac{(x_a-x'_a)^2}{\ell_a^2}
\right],
\end{align}
where $\sigma_f$ controls the overall variation of the function and $\ell_a$ is the correlation length along parameter direction $a$~\cite{RasmussenWilliams2006,Pratt2015,Bernhard2016}. Mat\'ern and composite kernels can be used when different assumptions about smoothness or noise are appropriate. The kernel hyperparameters are usually determined by maximizing the GP marginal likelihood or by sampling their posterior distribution. Regardless of the kernel choice, the emulator must be validated against model calculations that were not used for training. 

When many correlated observables are included, the emulation can be stabilized using principal-component analysis (PCA). Separate emulators are then trained for the retained few major principal components. The observables are approximately reconstructed from the inverse transformation.

The combination of PCA, GP emulation, and Markov-chain Monte Carlo sampling was introduced into heavy-ion phenomenology in Refs.~\cite{Pratt2015,Bernhard2016} and has enabled systematic analyses of initial-state parameters and QGP transport coefficients in high-dimensional parameter spaces~\cite{JETSCAPE:2017eso,Bernhard:2019bmu}.

More recent work has explored deep neural networks as scalable and accurate surrogates for expensive forward calculations~\cite{Cox:2024mkz}. Normalizing flows play a different role: as conditional density estimators, they can learn likelihoods or posterior distributions directly from simulated data and thereby enable simulation-based or likelihood-free inference when an explicit likelihood is unavailable~\cite{Cranmer2020,Brandes:2024vhw,Tao:2026yvi}. Posterior sampling can also be accelerated using Hamiltonian Monte Carlo (HMC) when gradients are available~\cite{Brooks2011,Cheng:2023ucp}, while nested sampling provides both posterior samples and an estimate of the Bayesian evidence~\cite{Skilling2006Nested,Dobler:2025eps}. These methods address different computational bottlenecks and should be compared through closure tests, posterior calibration, and computational cost rather than through sampling speed alone.

\subsection{Bayesian extraction of the dense nuclear-matter EoS}

\begin{figure}[t]
    \centering
    \includegraphics[width=0.9\linewidth]{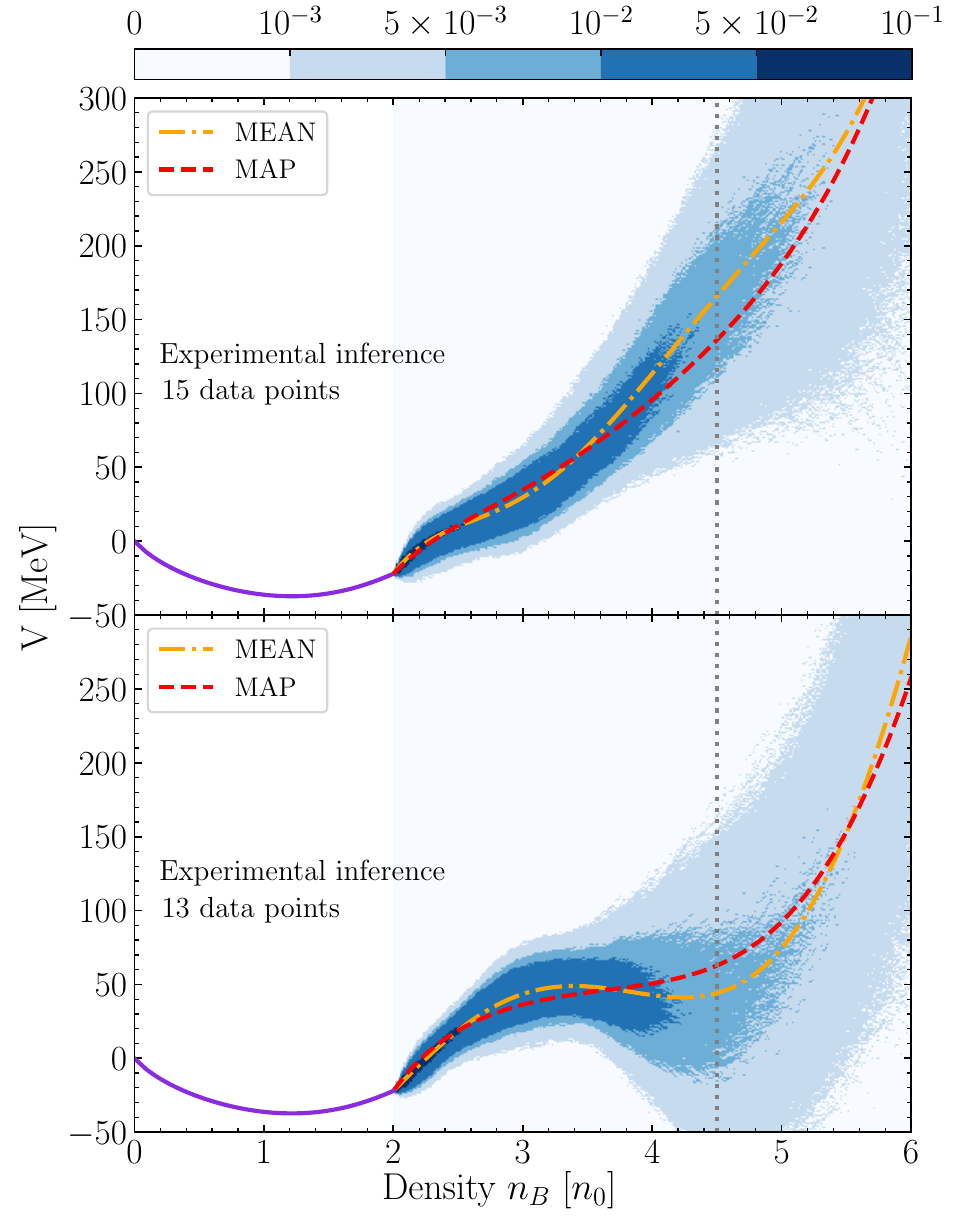}
    \caption{Bayesian inference of the high-density nuclear EoS from heavy-ion data. The posterior bands illustrate how transport-model calculations and experimental observables constrain a density-dependent mean-field potential.  Adapted from Ref.~\cite{OmanaKuttan:2022aml}.}
    \label{fig:hic_dense_eos_bayes}
\end{figure}

A representative recent development is the Bayesian extraction of the dense nuclear-matter EoS from heavy-ion data~\cite{OmanaKuttan:2022aml}. In that work, a flexible density-dependent mean field was embedded into a transport-model analysis and constrained by proton mean transverse kinetic energies and elliptic flows in the beam-energy range relevant for baryon-rich matter.  This marks a conceptual shift from pattern recognition to quantitative inference: the output is a posterior distribution over EoS functions rather than a class label. Fig.~\ref{fig:hic_dense_eos_bayes} gives the Bayesian inference of the high-density nuclear EoS from heavy-ion data \cite{OmanaKuttan:2022aml}. It also highlights why ML and Bayesian methods are inseparable in modern heavy-ion phenomenology.  Emulators and sampling strategies are needed because direct exploration of the model space is too costly, but the scientific result must still be a physics posterior conditioned on data and modeling assumptions.

Bayesian constraints on nuclear matter inputs are also being developed outside collider-energy observables. Recently, the BUU transport calculation of the iso-vector giant dipole resonance was used to infer the iso-vector effective mass and the neutron-proton effective mass splitting from the energy-weighted sum rule and width~\cite{Song:2025bvt}. This is a low energy nuclear structure example, but it belongs to the same statistical program: uncertain effective interactions are propagated through a forward model and constrained by measured response function.

This direction is particularly relevant for connecting heavy-ion collisions to neutron-star physics. Heavy-ion observables probe hot, dynamical, and often non-equilibrium matter, while neutron stars probe cold beta-equilibrated neutron-rich matter.  The two systems do not follow the same thermodynamic trajectory, but they constrain overlapping regions of the dense-QCD theory space.  A useful near-term goal is therefore not to force a single EoS reconstruction from all data, but to construct hierarchical analyses in which heavy-ion data, chiral effective theory, neutron-star observations, perturbative QCD calculations, and lattice constraints each enter with their own nuisance parameters and validity ranges.  ML contributes by accelerating the forward maps and by representing functional uncertainty in a way that can be propagated across sectors.

\subsection{Transport properties of baryon-rich QCD matter}
\label{sec:finite-density-transport}

Early extraction of the QGP transprot properties focused on LHC and top RHIC energeis, where EoS and transport coefficients were primarily parameterized as function of temperature. Lowering the collision energy increases nuclear stopping and allows heavy-ion experiments to probe QCD matter at appreciable net-baryon density. The relevant transport properties therefore are the specific shear and bulk viscosities to entropy ratio, $\eta(T,\mu_B)/s$ and $ \zeta(T,\mu_B)/s$, and the baryon diffusion coefficient $\kappa_B(T,\mu_B)$, all as functions of both temperature $T$ and baryon chemical potential $\mu_B$. Their extraction is more difficult than at top RHIC and LHC energies because they are correlated with nuclear stopping, the longitudinal distributions of energy and baryon density, the finite-density equation of state. Consequently, realistic analyses require $(3+1)$-dimensional models and observables over a broad rapidity range. And a challenge is to distinguish genuine finite-density transport effects from changes in nuclear stopping, longitudinal initial conditions, the equation of state, and the hadronic phase.

Earlier Bayesian study in this direct Auvinen \textit{et al.} analyzed Au--Au collisions at
$\sqrt{s_{\rm NN}}=19.6$, 39, and 62.4~GeV using a transport--hydrodynamics hybrid model~\cite{Auvinen:2017fjw} and an effective viscosity as a function of beam energy is extracted. Since then, a major extension was achieved~\cite{Shen:2023awv}, using an event-by-event $(3+1)$D dynamical initialization model coupled to MUSIC hydrodynamics and UrQMD hadronic transport. The calculation employed a finite-density lattice-QCD-based equation of state. The analysis included data of particle yields, mean transverse momenta, anisotropic flow, and pseudorapidity-dependent observables in Au--Au collisions at $\sqrt{s_{\rm NN}}=7.7$, 19.6, and 200~GeV. The parameter space included nuclear-stopping parameters, the values of the specific viscosity at $\mu_B=0$, $0.2$, and $0.4$~GeV, and parameters controlling the magnitude and temperature dependence of the bulk viscosity. The analysis constrained the average initial-state rapidity loss and favored an increase of the specific shear with $\mu_B$ up to approximately $0.4$~GeV, and a preference for a larger bulk-viscosity peak at $\sqrt{s_{\rm NN}}=19.6$~GeV than at 7.7 or 200~GeV. 

A full posterior analysis of the finite-baryon region was further examined in Ref.~\cite{Jahan:2024wpj}. In addition to standard GP emulation, the authors used a heteroscedastic principal-component stochastic-kriging emulator that accounts for the varying statistical uncertainty of different observables. Training points sampled from preliminary high-probability posterior regions were added to the original Latin-hypercube design, improving the emulator specifically where the final inference placed substantial probability.Posterior sampling employed a preconditioned Monte Carlo algorithm in which a normalizing flow decorrelated the parameters before sequential Monte Carlo sampling. The analysis again obtained strong constraints on nuclear stopping and favored an increasing shear viscosity over the range $\mu_B\lesssim0.2$~GeV. Full simulations at 100 posterior samples were then used to validate the emulator-based inference and propagate parameter uncertainty to $p_T$-differential predictions.

The emulator dependence of the analysis was further explored in dedicated closure tests in Ref.~\cite{Roch:2024xhh}. Different GP implementations produced substantially different posterior widths even when their conventional prediction errors appeared reasonable. Emulators designed to include the heterogeneous Monte Carlo uncertainty recovered the planted parameter values more accurately than a standard GP implementation. Adding training points in the high-posterior-probability region reduced the information loss by roughly two orders of magnitude in their closure metric. This study demonstrates that the transport-coefficient posterior can be biased or artificially broadened by the surrogate model itself, making emulator-level closure tests essential.

Nevertheless, the extracted viscosities remain sensitive to the assumed dynamical framework. G\"otz, Karpenko, and Elfner performed an independent Bayesian analysis using SMASH-generated initial conditions, vHLLE hydrodynamics, and a SMASH hadronic afterburner~\cite{Gotz:2025wnv}. Their posterior favored a small shear viscosity in the high-temperature QGP phase and a moderate-to-large value near particlization. However, the $\mu_B$ dependence of both the shear and bulk viscosities remained weakly constrained. These results emphasize that a narrow posterior obtained within a particular model does not by itself constitute a model-independent determination. A more systematic treatment of uncertainties associated with different dynamical frameworks will therefore be essential in future analyses.

Finally, a preliminary Bayesian determination of baryon diffusion coefficient was reported by Sarkar, Panday, and Chatterjee using STAR data at $\sqrt{s_{\rm NN}}=19.6$~GeV~\cite{Sarkar:2026hbb}. PCA and GP regression were used to emulate the hydrodynamic calculation, followed by MCMC sampling of the initial-condition and diffusion parameters. 
The application of Bayesian analysis to the extraction of baryon diffusion is promising but was presented in a short conference proceeding and was based on one collision energy and two centrality classes. A more definitive extraction will require simultaneous variation of the initial baryon deposition, stopping, viscosity, equation of state, and hadronic transport, together with closure tests and a broader set of rapidity-dependent observables.

\subsection{Bayesian jet tomography and ML-assisted jet analysis}

\label{sec:bayesian-jets}

Jets originate from hard scatterings during the early stage of a nuclear collision and subsequently evolve while traversing the QGP. Their energy loss, transverse-momentum broadening, modified shower structure, and induced medium response therefore carry information about the microscopic interactions between energetic partons and the medium. Bayesian inference has been used to constrain parametrized energy-loss distributions and transport coefficients, while ML tools have been introduced to accelerate the required simulations, reconstruct jet properties, and identify informative observables.

\begin{figure}[t]
    \centering
    \includegraphics[width=0.9\linewidth]{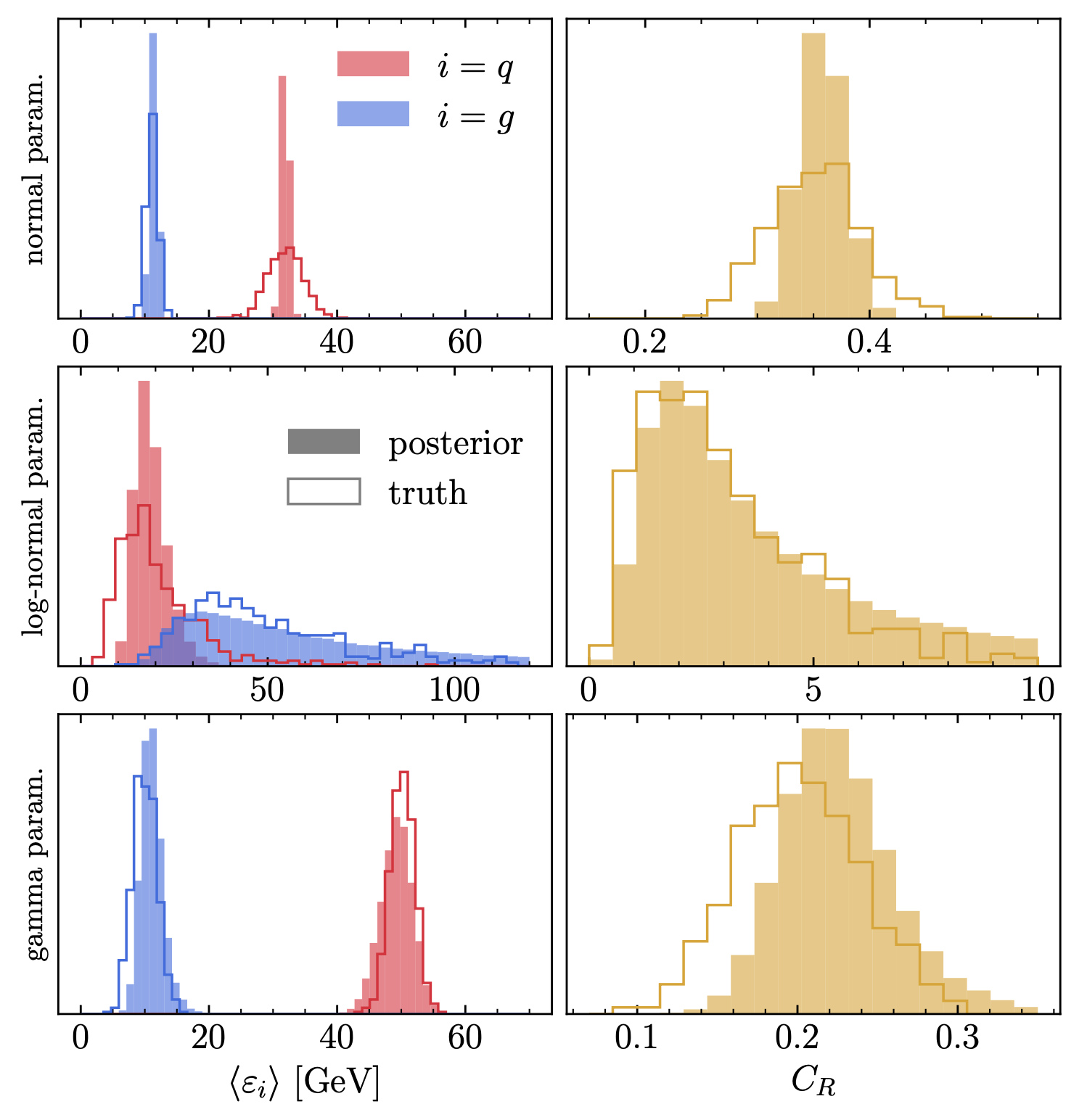}
    \caption{Prior and posterior distributions of the mean energy loss by the quark- and gluon-initiated jets obtained for the closure tests, for the three studied distributions, obtained from marginalizing the posteriors. The color ratio is also shown.  Adapted from Ref.~\cite{Falcao:2024zkw}.}
    \label{JHEP}
\end{figure}

An early data-driven extraction of jet energy loss was performed using inclusive and photon-triggered jet spectra in Pb--Pb collisions~\cite{He:2018gks}. The energy-loss distribution was characterized by its average and fluctuations, both inferred from experimental data. The extracted distributions were found to be consistent with Linear Boltzmann Transport calculations, demonstrating how Bayesian inference can connect effective energy-loss descriptions with microscopic transport simulations. Following this approach, the average energy loss of full jets was extracted from inclusive jet spectra for different collision systems and centralities~\cite{Wu:2023azi}. The suppression of $J/\psi$ production was also shown to provide sensitivity to the energy loss of gluon-initiated jets in the QGP~\cite{Zhang:2022rby}. More recently, Ref.~\cite{Falcao:2024zkw} employed flexible quenching-weight parametrizations of the parton energy-loss distribution to analyze inclusive and photon-tagged jet measurements. Their GP-assisted Bayesian analysis examined whether a common energy-loss distribution could describe different hard processes and attempted to separate the quenching of quark- and gluon-initiated jets. 
The results supported an approximately universal description and indicated a gluon-to-quark energy-loss ratio somewhat larger than that expected from naive Casimir scaling. 
To further validate the closure tests, they examined whether the posterior distributions for the mean quark/gluon energy loss $\langle \epsilon_i \rangle$ and the color ratio $C_R$ reproduce the true values used to generate the mock data. These quantities were derived from the posterior samples for each of the three parametrizations. As shown in Figure ~\ref{JHEP}, the posterior distributions generally agree well with the true distributions across all parametrizations, confirming the reliability of their framework.
However, closure tests showed that survival bias arising from the steeply falling jet spectrum strongly limits the reconstruction of the full energy-loss distribution and renders parts of the result sensitive to the assumed prior.

A more microscopic determination of the jet-medium interaction is via the extraction of the jet transport parameter $\hat{q} = d\langle k_\perp^2\rangle/dL$. Such studies were later carried out using microscopic transport models~\cite{Ke:2020clc,JETSCAPE:2021ehl}. A subsequent calibration included charged hadrons, $D$ mesons, and inclusive jets~\cite{JETSCAPE:2023ikg}. The authors introduced a closure-test metric related to the Kullback--Leibler divergence to quantify the information lost when a GP emulator replaces the full model. This metric was used to select the GP kernel and the number of retained principal components, as well as to identify parameters that were poorly reconstructed by the emulator. The work emphasized that emulator accuracy should be assessed through its effect on the inferred posterior, rather than only through pointwise prediction errors on a test set.

A subsequent analysis by the JETSCAPE Collaboration performed a joint calibration using inclusive-hadron and inclusive-jet suppression~\cite{Ehlers:2024miy}. Because the required simulation campaign remained extremely expensive, the authors combined GP emulation with active learning. Starting from an initial space-filling design, the active-learning algorithm selected additional parameter points in regions of large emulator uncertainty, thereby allocating the available simulation budget more efficiently. The combined analysis constrained the virtuality- and temperature-dependent $\hat q$, but also revealed tensions between calibrations based on different subsets of the data. High-$p_T$ hadrons and reconstructed jets yielded broadly compatible constraints, whereas lower-$p_T$ hadrons preferred a different parameter region. The analysis traced this inconsistency primarily to the modeled dependence of $\hat q$ on parton energy rather than to collision centrality. This is an important physical result: disagreements among posterior distributions inferred from different classes of observables can expose limitations of the model in different kinematic regions.

A complementary development addressed the prior dependence introduced by assuming a specific functional form for the temperature dependence of $\hat q$. Xie \textit{et al.} treated $\hat q(T)$ as an information field and constructed a nonparametric prior with controlled correlations over temperature~\cite{Xie:2022fak,Xie:2023qhat}. Unlike low-dimensional parametrizations, which can introduce artificial correlations between widely separated temperature regions, this formulation allows the data to constrain $\hat q(T)$ primarily where they have sensitivity while leaving poorly probed regions prior dominated. Applied to a global analysis of single-hadron, dihadron, and photon--hadron measurements at RHIC and the LHC, the method yielded a pronounced temperature dependence of $\hat q/T^3$ and showed how measurements from different collision energies and centralities progressively constrain different temperature regions. This information-field approach therefore reduces functional-parametrization bias and makes the prior dependence of the extracted transport coefficient more transparent.

ML has also been used to determine which aspects of jet substructure carry the most information about medium modification. Lai \textit{et al.} compared Particle Flow Networks, Energy Flow Networks, complete substructure bases, and sparse regression in distinguishing simulated jets in $pp$ and heavy-ion collisions~\cite{Lai:2021ckt}. They found that substantial discrimination power resides in soft and infrared- and collinear-unsafe information, but that much of this information is obscured by the fluctuating underlying event and can be further reduced by background subtraction. Sparse regression was then used to construct simpler observables that approximate the neural classifier while remaining more accessible to theoretical calculations.

\subsection{Open heavy flavor and quarkonium}

Open heavy flavor complements light hadrons and reconstructed jets by probing both partonic energy loss and the diffusion of heavy quarks through the QGP. The relevant quantities include the heavy-quark spatial diffusion coefficient $D_s(T,p)$, the momentum-broadening coefficient $\hat q$, and parameters controlling fragmentation and recombination.

An early Bayesian analysis calibrated an improved Langevin transport model to $D$-meson $R_{\rm AA}$ and $v_2$ data from RHIC and the LHC~\cite{Xu:2017obm}. The analysis constrained the temperature and momentum dependence of $2\pi T D_s$ and demonstrated that a common transport description could reproduce several collision systems. More recent analysis have increased both the physical scope and the statistical validation of this program. 

A more recent analysis simultaneously inferred the temperature dependence of $2\pi T D_s$ and $\hat q/T^3$ from $D$-meson observables~\cite{Xue:2025hrw}. The study further constrained the ratio $\hat q/\kappa$, where $\kappa$ is the heavy-quark momentum-diffusion coefficient, and found a non-monotonic temperature dependence that deviates substantially from the naive expectation $\hat q/\kappa=2$.

Heavy-flavor observables have also been used to constrain a broader description of the medium. Within a quasiparticle linear Boltzmann framework, Bayesian calibration to $D$-meson $R_{\rm AA}$ and $v_2$ was used to constrain both the effective coupling governing the QCD equation of state and the heavy-quark transport coefficient~\cite{Liu:2023rfi}. 

Quarkonium extends the hard-probe program from the transport of individual heavy quarks to the in-medium evolution of a correlated heavy-quark pair. A recent study parameterized the real and imaginary parts of the heavy-quark potential and propagated them through a time-dependent Schr\"odinger equation coupled to a hydrodynamic background~\cite{Liu:2026uav}. Convolutional neural networks were trained as surrogates for the nonlinear map from the potential to bottomonium $R_{\rm AA}$ across several RHIC and LHC collision systems. Principal-component analysis and Gaussian-process regression were used to augment the training data, and stochastic-gradient Langevin dynamics was then employed to sample the posterior distribution of the potential parameters. Within the adopted potential and medium model, the imaginary part was more strongly constrained and provided the dominant contribution to bottomonium suppression, whereas the inferred real part remained comparatively close to the vacuum Cornell form. This result is a strong example of ML entering several stages of a Bayesian workflow, although its physical interpretation must still be tested against variations of the hydrodynamic background, feed-down contributions, cold-nuclear-matter effects, and the functional form assumed for the potential.

\subsection{Bayesian constraints on initial-state and pre-equilibrium dynamics}
\label{sec:early-time-bayes}

The early stage of a heavy-ion collision converts the structure of the incoming nuclei into the energy--momentum tensor and conserved-charge distributions that initialize the subsequent hydrodynamic evolution. Because this short-lived stage is not directly observable, its properties must be inferred from final-state measurements and are often correlated with the extracted transport properties of the QGP. Recent Bayesian studies have therefore moved beyond parameter estimation within a fixed dynamical model. They compare alternative descriptions of the initial state and pre-equilibrium evolution, examine theoretical consistency and causality, and propagate model and emulator uncertainties into the inferred posteriors. These developments help distinguish features of the early-time dynamics that are constrained by data from those that remain dependent on the assumed theoretical description.

Ref.~\cite{Bay1} challenges the conventional practice of using parametric initial-condition models (e.g., TRENTo) by replacing them with SMASH (Simulating Many Accelerated Strongly-interacting Hadrons) transport to generate initial states for a (3+1)D viscous hydrodynamic + hadronic afterburner hybrid. The Bayesian analysis, calibrated to Pb–Pb collisions at \(\sqrt{s_{NN}}=2.76\) TeV, reveals that transport-generated initial conditions produce significantly different posterior constraints on the specific shear viscosity (\(\eta/s\)), which is an important transport coefficient for nuclear matter \cite{Bernhard,DengXG,DengXG2}, compared to parametric models. Importantly, this work demonstrates that the initial-state modeling is not a mere nuisance, it directly couples to the extracted QGP properties. The physical, non-equilibrium-driven initial profiles lead to tighter constraints on early-time dynamics, implying that a more realistic description of the pre-hydrodynamic stage is essential for unbiased tomography of the QGP.

\begin{figure}[t]
    \centering
    \includegraphics[width=0.9\linewidth]{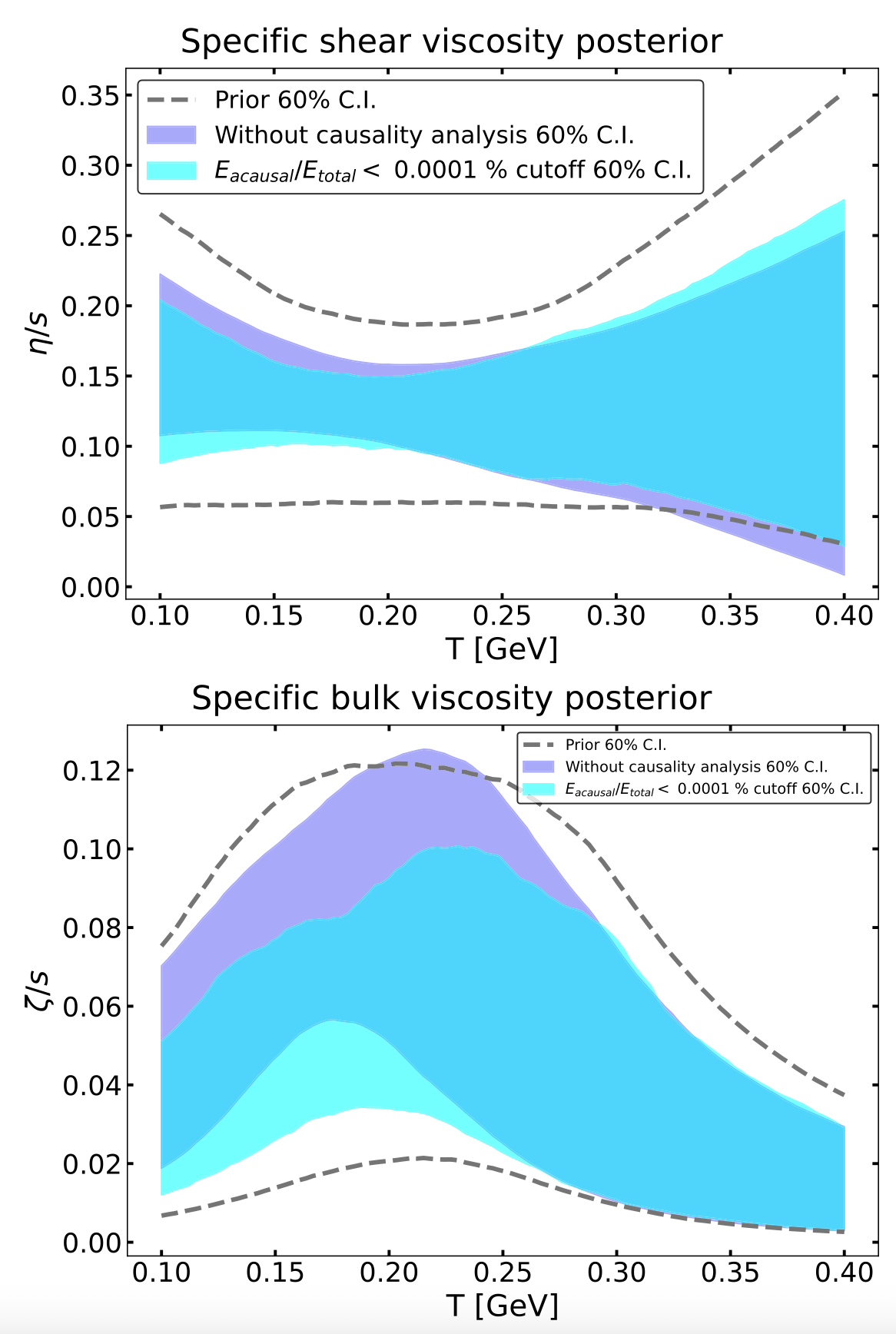}
    \caption{Specific shear and bulk viscosities as a function of temperature. Here we show the 60\% credible interval for the full prior, the full posterior with no causality restriction, and when the initial acausal energy fraction is not allowed to exceed $10^{-4}$. We can see in particular that bulk viscosity is significantly biased by acausal regimes \cite{Bay2}.}
    \label{fig:Bay2}
\end{figure}

Ref.~\cite{Bay2} addresses a fundamental theoretical issue: the validity of relativistic hydrodynamics in the early, far-from-equilibrium stage. The authors introduce causality constraints by penalizing or excluding regions in the evolution where the dissipative corrections violate the physically required conditions (e.g., positivity of the effective pressure or causality). Using a Bayesian analysis of flow observables in Au–Au collisions at RHIC, they find that imposing causality significantly shifts the posterior distributions of both the initial-condition parameters and the temperature-dependent specific bulk viscosity. The most striking result is that causality-violating parameter sets, which are statistically acceptable in standard analyses, become strongly disfavored once constraints are applied. This suggests that the early-time dynamics, when hydrodynamics is least justified, must be described by non-fluid approaches, otherwise, the inferred transport coefficients are biased by unphysical solutions. This paper directly quantifies how the frontier of the hydrodynamic description affects the entire Bayesian inference. Fig.~\ref{fig:Bay2} illustrates that specific shear and bulk viscosities as a function of temperature is  biased by acausal regimes \cite{Bay2}.

Ref.~\cite{Bay3} introduces a rigorous framework for incorporating model discrepancy, the systematic error arising from the theoretical model itself—into the Bayesian analysis. Unlike standard approaches that assume the model can exactly describe reality with tuned parameters, this work explicitly models the discrepancy between the simulation output and actual experimental data using a Gaussian process. Applied to hadron spectra and flow in Pb–Pb collisions, the analysis shows that a significant portion of the data-model mismatch is absorbed by the discrepancy term rather than by parameter variation. Crucially, when model discrepancy is included, the inferred values of \(\eta/s\) and the initial-state entropy deposition broaden substantially, and the early-time shear-stress tensor initialization parameters become less constrained. This implies that the theoretical uncertainty from the pre-equilibrium modeling is comparable to or larger than experimental and statistical uncertainties, and ignoring it leads to overconfident, potentially inaccurate parameter estimates.

In addition, there are also more sophisticated methods have been developed to model the initial stage and early-time evolution.
Recent analyses have connected CGC initial conditions more directly to the soft observables measured in heavy-ion collisions. Heffernan et al.\ calibrated an IP-Glasma--hydrodynamics--hadronic-transport framework to Pb+Pb data~\cite{Heffernan:2023utr,Heffernan:2023gye}. Transfer learning reduced the cost of constructing emulators for related model variants, while Bayesian model averaging propagated uncertainty associated with the hydrodynamic-to-hadronic transition. The analysis constrained both initial-state and QGP parameters and found that the available data did not statistically favor a temperature-dependent shear viscosity over a constant parametrization once model complexity was taken into account. The methodological advance here is therefore not only faster emulation, but also the use of ML-assisted transfer learning and Bayesian model averaging to treat differences among physical model choices.

\subsection{Bayesian and machine-learning constraints on the nuclear initial state: from partonic structure to nuclear shape}
\label{sec:initial-state}

The initial stages of high-energy nuclear collisions are controlled by the spatial and momentum distributions of quarks and gluons in the incoming nuclei. At small Bjorken $x$, nonlinear gluon dynamics and event-by-event fluctuations determine the initial energy-density profile supplied to the subsequent pre-equilibrium and hydrodynamic evolution. Nuclear parton distribution functions, in contrast, provide the collinear partonic input for hard processes and cold-nuclear-matter baselines. Bayesian inference and ML enter these two problems differently: Bayesian calibration of Color Glass Condensate (CGC) calculations constrains model parameters and fluctuating geometry, while neural-network PDF analyses provide flexible functional parametrizations and data-driven uncertainty ensembles.

Bayesian methods have also been applied directly to the initial condition of small-$x$ evolution. Casuga, Karhunen, and M\"antysaari inferred parameters of the nonperturbative initial condition for the Balitsky--Kovchegov equation from HERA structure-function data~\cite{Casuga:2023dcf}. A GP emulator and MCMC sampler made it possible to propagate the resulting posterior through the nonlinear evolution. The data favored an anomalous dimension close to unity, avoiding positivity problems encountered by some earlier parameter choices, and the posterior was propagated to proton--proton and future electron--ion observables. This is an important advance over a best-fit determination because it quantifies how uncertainty in the small-$x$ initial condition affects subsequent predictions.

A subsequent global analysis combined coherent and incoherent $J/\psi$ photoproduction on proton and lead targets~\cite{Mantysaari:2025ltq}. Principal-component Gaussian-process emulators were trained on full CGC calculations, with additional training points selected from an intermediate posterior to improve accuracy in the high-likelihood region. This posterior-focused refinement is an elementary form of active learning. The study found that simultaneously describing proton and nuclear data remained difficult unless an additional normalization factor was introduced, thereby using Bayesian analysis to reveal model tension rather than only to determine parameter values. Such failures of posterior predictive agreement are physically informative because they identify missing ingredients in the assumed nuclear small-$x$ description.

Besides the parton-level wave function of the nuclei, relativistic nuclear collisions are increasingly used to study the nucleon-level structure of the colliding nuclei. Nuclear $\alpha$-clustering structure \cite{ZhangS_2017,HeJJ_2021}, nuclear radii, surface diffuseness, neutron skins, and intrinsic deformations influence the fluctuating initial geometry and are subsequently reflected in particle multiplicities, anisotropic flow, mean transverse momentum, and correlations among these observables \cite{Ma_2023,STAR_2024,JiaNST,GiacaloneNST,SchenkeeNST}. Extracting nuclear structure from such measurements is nevertheless an inverse problem. In these applications, ML commonly enters through Gaussian-process (GP) emulators of the collision model or, more recently, through neural density estimators that directly approximate the posterior distribution.

A prominent application is the extraction of the neutron skin of ${}^{208}\mathrm{Pb}$ from LHC data by Giacalone, Nijs, and van der Schee~\cite{Giacalone:2023cet}. In this analysis, the neutron diffuseness was included among the parameters of a multistage heavy-ion collision model and constrained using observables in Pb--Pb collisions, together with the total p--Pb cross section. The observables included particle yields, mean transverse momenta, anisotropic-flow coefficients, and correlations between elliptic flow and mean transverse momentum. The resulting posterior yielded
\begin{equation}
    \Delta r_{np}({}^{208}\mathrm{Pb})
    = 0.217 \pm 0.058~\mathrm{fm},
\end{equation}
which was further translated into a constraint on the slope of the nuclear symmetry energy. An important methodological result was the identification of correlations between the neutron skin and parameters controlling the initial entropy deposition. This demonstrates why the nuclear and collision-model parameters should be inferred jointly rather than varied one at a time.

The sensitivity to nuclear deformation parameters was investigated by Cheng \textit{et al.} using Bayesian imaging of ${}^{96}\mathrm{Ru}+{}^{96}\mathrm{Ru}$ and ${}^{96}\mathrm{Zr}+{}^{96}\mathrm{Zr}$ collisions~\cite{Cheng:2023ucp}. The nuclear densities were parametrized by the Woods--Saxon radius $R$, diffuseness $a$, quadrupole deformation $\beta_2$, and octuple deformation $\beta_3$. A Monte Carlo Glauber model was evaluated at 1001 Latin-hypercube design points, after which principal-component analysis and GP emulation were used to interpolate its predictions. Posterior sampling was performed with the No-U-Turn Sampler, an adaptive Hamiltonian Monte Carlo algorithm. Since the ``data'' in this study were generated by the same model, the analysis should be interpreted as a closure test rather than as an extraction from experimental measurements. It nevertheless established that multiplicity, elliptic and triangular eccentricities, and radial-size information provide complementary sensitivity to $R$, $a$, $\beta_2$, and $\beta_3$. It also showed that isobar ratios alone mainly constrain differences between the two nuclei and leave degeneracies in their absolute structures. Combining the ratios with selected single-system observables substantially improves the simultaneous reconstruction.

A more explicitly ML-based approach was introduced by Tao \textit{et al.} for the inference of the quadrupole and hexadecapole deformations $(\beta_2,\beta_4)$ from simulated heavy-ion initial states~\cite{Tao:2026yvi}. Permutation-invariant neural networks were used to encode ensembles of nucleon configurations or entropy-density profiles, while conditional normalizing flows learned the posterior directly from simulations. This constitutes simulation-based inference: the network is trained on parameter--simulation pairs without requiring an explicit analytic likelihood. The study found that aggregating information from multiple collision events is essential for suppressing event-by-event fluctuations. Quadrupole deformation was reconstructed more accurately than the weaker $\beta_4$ signal, and central collisions carried more deformation information than peripheral collisions. Posterior coverage and probability-integral-transform tests were used to assess statistical calibration. This work demonstrates the potential of neural posterior estimation, although the analysis was restricted to simulated initial-state information and has not yet produced an extraction from final-state experimental data.

These examples illustrate a progression from model comparison, through GP-assisted Bayesian parameter estimation, to neural simulation-based inference of nuclear structure. At present, Bayesian extraction of more detailed structures, such as two-nucleon correlations or clustering, remains considerably less developed.

%% file: HIC_Experiments_with_AI.tex
\section{HIC Experiments with AI}
\label{sec:hic_exp}

Experimental uses of machine learning in heavy-ion physics differ from theory-side applications because the model can become part of the measurement chain. It must work with detector acceptance, pileup, backgrounds, high occupancy, finite resolution, calibration drift and online latency. The relevant quantities are therefore efficiency, resolution, stability over detector conditions, and the effect on final physics observables. 

The QGP is extremely short-lived (approximately $10^{-23}$ seconds) and cannot be observed directly, making its study one of the most challenging endeavors in modern nuclear physics. Traditional approaches rely on a combination of sophisticated theoretical models, computationally expensive numerical simulations, and statistical comparisons with experimental observables.
This section presents selected examples of AI applications in heavy-ion collision experiments, organized into three thematic subsections.

\subsection{Jet Quenching Classification, Background Subtraction, and Unfolding}

Jet quenching, i.e. the energy loss and modification of high-energy jets as they traverse the QGP, is one of the most powerful probes of the medium's properties \cite{Soloveva}. However, jet reconstruction in heavy-ion collisions suffers from a notoriously large and fluctuating thermal background. Reconstructed jets always include underlying event contamination, making it difficult to isolate genuine quenching effects from background-induced modifications \cite{Chen_NIM}.

Deep-learning methods have explored how more information can be recovered from individual jets. Haake and Loizides trained regressors using jet and constituent properties to estimate the hard jet momentum in the presence of a fluctuating heavy-ion background~\cite{Haake:2018hqn}. In simulation, the ML estimator achieved better momentum resolution than conventional area-based subtraction, particularly for low-$p_T$ jets. This improvement could extend the kinematic reach of jet measurements, but its application to data requires validation against mismodeling of the background and the fragmentation pattern.

A major advance in this area is the use of machine learning to classify quenched jets on a jet-by-jet basis. Apolin\'ario {\it et al.} established a rigorous baseline by embedding proton-proton jets into realistic heavy-ion backgrounds, ensuring that classification models learn to distinguish true QGP-induced modifications rather than background artifacts1 \cite{Apolinario:2021olp}. Their study demonstrated that a simple Boosted Decision Tree remains robust in scenarios where both medium response and underlying event are present, supporting portability to experimental contexts \cite{Apolinario:2021olp}.

More sophisticated architectures have also been explored. Wu {\it  et al.} employed Long Short-Term Memory (LSTM) neural networks trained on jet substructure incorporating parton shower history to predict jet-by-jet quenching levels \cite{Wu_JHEP}. Using photon-jet samples from the Jewel event generator, they showed that LSTM predictions strongly correlate with true jet energy loss. The method was validated against Delphes detector simulations, confirming its ability to identify quenching effects in realistic environments \cite{Wu_JHEP}.

Background subtraction itself has been revolutionized by deep learning. Qureshi and Kunnawalkam Elayavalli introduced DeepSub, the first machine learning-based approach for full-event background subtraction \cite{Qure}. DeepSub utilizes a Swin Transformer-based model to denoise jet images and disentangle hard jets from the heavy-ion background. The method significantly outperforms existing subtraction techniques, reproducing key jet observables such as jet $p_T$, mass, and substructure observables at the sub-percent to percent level. Meanwhile, a Dense Neural Network (DNN) approach was developed to reconstruct jet momentum within a QGP background \cite{Li}. Training on quenched jets from the Linear Boltzmann Transport (LBT) model significantly reduced prediction bias compared to conventional Area-based and Constituent Subtraction methods.

Du, Pablos, and Tywoniuk used deep neural networks to estimate the initial, unquenched energy of a reconstructed jet from its final-state constituents~\cite{Du:2021pqa}. Conditioning jet samples on the reconstructed initial energy reduces the survival bias associated with selecting jets by their final energy. In model studies, this procedure provided improved access to the spatial distribution and orientation of jet production and even allowed the production region of dijet pairs to be localized. The method therefore suggests a route toward event-by-event jet tomography, although the inferred initial energy and production geometry remain dependent on the jet-quenching model used to generate the training data.

Finally, measurements of jet observables are affected by detector-response smearing, with effects that are often correlated across multiple observables. Recent high-dimensional extensions of OmniFold have been developed to unfold jet measurements in the simultaneous presence of underlying-event background and detector effects~\cite{Falcao:2025jom}. Simulation studies showed that incorporating auxiliary information from up to 18 correlated observables can improve the unfolding of jet-substructure distributions.

\subsection{Real-Time Online Triggering and Data Processing with AI}

Modern high-energy heavy-ion collision experiments face an extraordinary data challenge. At the ALICE experiment at the LHC, collisions occur at rates of up to 750,000 proton–proton collisions or 50,000 heavy-ion collisions per second \cite{ALICE_26}.  In head-on heavy-ion collisions, thousands of particles can be produced simultaneously, creating exceptionally high track densities that must be reconstructed in real time. The ALICE detector produces data at rates of up to 1 terabyte per second, handled by a large online computing system equipped with around 2800 GPUs \cite{ALICE_26}.  Under such extreme conditions, traditional data processing algorithms often become bottlenecks, and AI-based solutions have emerged as indispensable tools for real-time event selection, tracking, and reconstruction.

\subsubsection{Machine Learning-Based Online Triggers for Heavy-Flavor Event Selection}

The sPHENIX experiment at the Relativistic Heavy Ion Collider (RHIC) is designed to investigate the physics of the strongly coupled quark–gluon plasma (QGP) \cite{Chen_NIM}. Heavy-flavor hadrons—containing charm and beauty quarks—serve as key probes of the QGP produced during heavy-ion collisions. However, the production rates of heavy-flavor events are notably low: in p+p collisions at $\sqrt{s}$ = 200 GeV, charm and beauty events account for only 0.3\% and 0.003\%, respectively. Compounding this challenge, the sPHENIX detector's readout event rate is limited to 15 kHz due to constraints in the outer calorimeter readout system, meaning the experiment can collect only approximately 0.5\% of full detector data  \cite{Chen_NIM}.

\begin{figure}[t]
    \centering
    \includegraphics[width=0.98\linewidth]{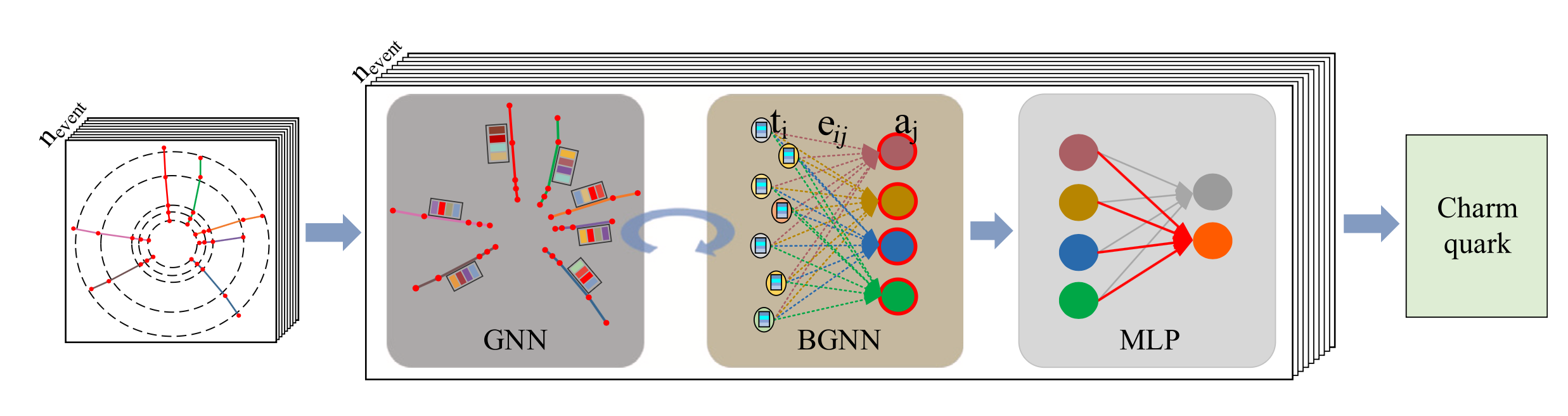}
    \caption{The GNN model extracts and encodes the topological structure of reconstructed tracks. Subsequently, the BGNN model captures the correlations among these reconstructed
tracks. Finally, HF quark events are identified by an MLP model~\cite{Chen_NIM}.}
    \label{fig:Chen_NIM}
\end{figure}

To overcome this limitation, a dedicated online trigger system based on machine learning has been developed. The algorithms, primarily built on Multi-Layer Perceptrons (MLPs), Graph Neural Networks (GNNs), and Bipartite Graph Neural Networks (BGNN), use only spatial track hit information from fast silicon detectors as input. This design choice is critical for online deployment—the algorithm must make decisions within extremely tight latency constraints, and using minimal, low-level detector information reduces computational overhead. Evaluated on Monte Carlo simulations of charm-quark events in p+p collisions at $\sqrt{s}$  = 200 GeV, the system achieves a track reconstruction efficiency of 0.945, with precision and recall for identifying heavy-flavor events both around 0.79. This work demonstrates that graph-based neural architectures can effectively identify rare heavy-flavor signatures in real time, significantly enhancing the physics reach of the sPHENIX experiment despite its readout rate limitations. Fig.~\ref{fig:Chen_NIM} illustrates that the flow-chart to identify HF quark events \cite{Chen_NIM}.

A related end-to-end AI-enabled triggering solution has been developed for relativistic heavy-ion colliders and future Electron-Ion Colliders \cite{Xuan}. This system, based on Graph Neural Networks, classifies trigger events versus background events, makes online decisions to retain signal data, and enables efficient data acquisition. The integration of GNNs into the trigger chain represents a paradigm shift: rather than relying on hand-crafted thresholds and cuts, the system learns directly from data to distinguish signal from background, adapting to the complex correlations present in high-multiplicity heavy-ion events. 

\subsubsection{CNN-Based Online Trigger for QGP Event Selection}

A complementary approach to online triggering focuses on directly selecting events associated with QGP formation. Soloveva et al. developed a convolutional neural network (CNN)-based trigger concept that encodes events as compact multidimensional histograms of reconstructed particle content, including particle species, momentum magnitude, and angular information. The method was first evaluated within the Parton-Hadron-String Dynamics (PHSD) framework, where microscopic QGP-related labels are available, and independently validated against Ultra-relativistic Quantum Molecular Dynamics (UrQMD) simulations \cite{Soloveva}.

For realistic deployment, a lightweight C++ inference package, ANN4FLES, is employed at the physics-analysis stage after tracking and topology reconstruction. For Au+Au collisions at 30 AGeV, the classification accuracy decreases from 95.1\% on generator-level PHSD events to 83.7\% after full reconstruction, while retaining practical separation power for online event selection. Importantly, SHAP-based interpretability analysis was used to identify the dominant particle-species contributions to the network decision, providing physics insight into what the network has learned. This work illustrates a key advantage of AI-based triggers: not only do they perform the selection task, but they can also be interrogated to reveal which physical features drive their decisions \cite{Soloveva}.

\subsubsection{ Neural Networks for Real-Time Cluster Finding and Track Reconstruction}

Beyond event-level triggering, AI is being deployed at the lowest levels of data processing—the reconstruction of individual detector signals. In October 2025, the ALICE Collaboration deployed, for the first time, a neural-network-based algorithm for real-time processing in the experiment \cite{2026_ALICE_Machine_Learning}. Its purpose is to reconstruct clusters from the raw readout of the Time Projection Chamber (TPC)—small groups of signals generated as charged particles traverse the detector. These clusters form the fundamental input for particle tracking.

The challenge is substantial: in head-on heavy-ion collisions, thousands of charged particles are produced simultaneously, and their signals in the TPC can overlap significantly. The neural network not only reconstructs clusters but also assigns to each a probability that it originates either from background noise or from a real particle traversing the detector. This approach effectively acts as a smart filter that removes noise and separates overlapping features, producing a cleaner, more interpretable representation of the event. The deployment of this algorithm marks a significant milestone—it demonstrates that neural networks can operate within the extreme computational constraints of real-time data processing at the LHC, using smaller models and lower numerical precision optimized for specific hardware.

\begin{figure}[t]
    \centering
    \includegraphics[width=0.98\linewidth]{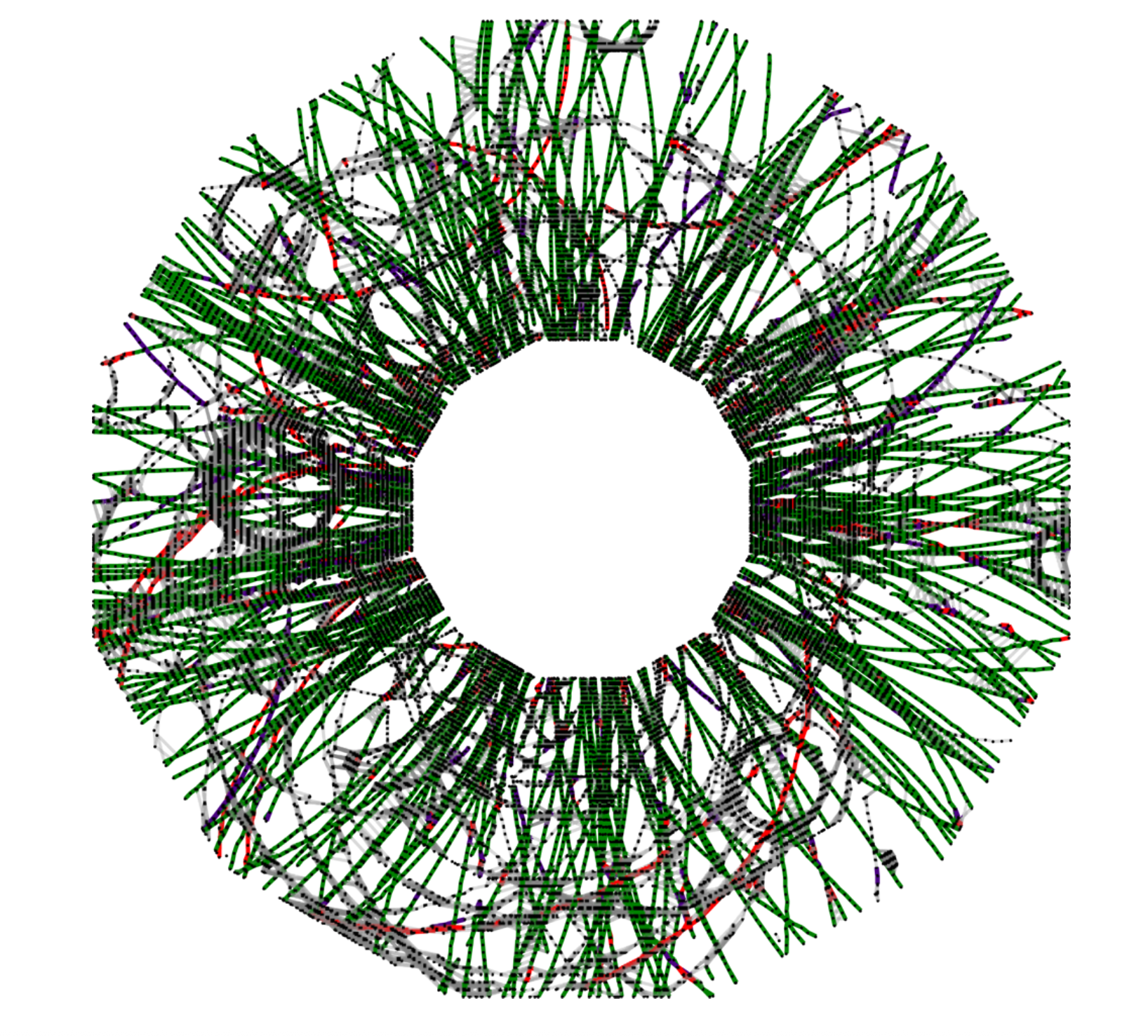}
    \caption{Example of track reconstruction results obtained using the
GNN model and projected onto the xy-plane. The green and purple
lines represent true and false positive track segments, respectively,
whereas the gray and red lines correspond to true and false negative
track segments, respectively ~\cite{NICA}.}
    \label{fig:NICA}
\end{figure}

A Graph Neural Network with attention mechanism and two-stage aggregation has also been developed for particle track reconstruction in the Time-Projection Chamber of the MPD detector at the NICA accelerator complex \cite{NICA}. These developments highlight a broader trend: tracking and clustering tasks, which are both highly complex and naturally parallelizable, are particularly well-suited for machine learning approaches. Previously developed algorithms have now been matched or surpassed in their performance by trainable approaches such as neural networks. For instance, an example of track reconstruction results obtained using the GNN model and projected onto the xy-plane was shown in Fig.~\ref{fig:NICA}.

\subsubsection{Contrastive Metric Learning for Stable Shower Reconstruction}
Recently, a new AI-powered event-processing algorithm is being readied to cope with the torrent of data that will stream from the Large Hadron Collider
upgrade \cite{Jar_2026,Mar_2026}. The work presents a clustering framework for point cloud segmentation in high-granularity calorimeters based on supervised contrastive metric learning (CML). Unlike object-centric approaches that predict cluster assignments or condensation points, CML learns a latent embedding space where hits from the same particle shower are pulled together while hits from different showers are pushed apart. Clustering is performed as a separate post-training readout using a density-based procedure, decoupling representation learning from cluster formation.

\begin{figure}[t]
    \centering
    \includegraphics[width=1.1\linewidth]{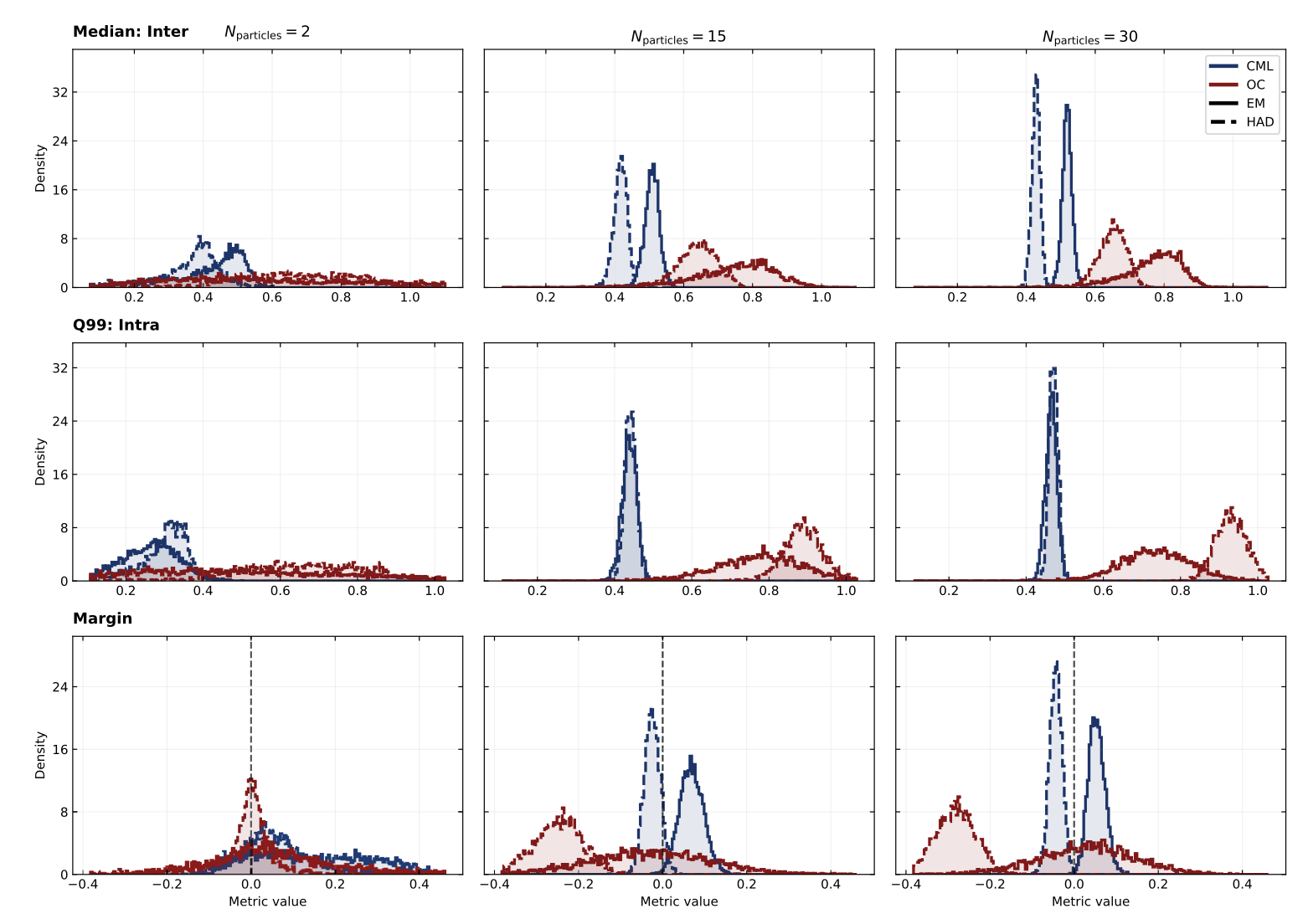}
    \caption{Distributions of embedding-geometry quantities evaluated at the event level for EM and HAD showers across increasing particle multiplicity $N_{particles}$. The top panel shows the median distance between different showers ($d^{median}_{inter}$), the middle panel shows the tail of within-shower distances ($d^{Q99}_{intra}$), and the bottom panel presents the separation margin ($\Delta = d^{median}_{inter} - d^{Q99}_{intra}$). A narrow margin centered at positive or near-zero values signals a stable clustering scale, whereas broad or negative margins indicate growing ambiguity between showers. Adapted from Ref.~\cite{Jar_2026}.}
    \label{fig:Jam}
\end{figure}

The method is rigorously compared against object condensation (OC) using identical graph neural network backbones and matched embedding dimensionality on simulated electromagnetic and hadronic shower datasets. Geometric analysis reveals that CML produces narrow and stable separation margin distributions, whereas OC yields broad and often negative margins, indicating intrinsic ambiguity in clustering decisions. Fig.~\ref{fig:Jam} demonstrates the event-level embedding-geometry distributions for simulated electro-magnetic (EM) and hadronic (HAD) showers at increasing generated particle multiplicity ($N_{particles}$). This geometric advantage translates directly into superior reconstruction performance: CML achieves higher efficiency, higher purity, more stable multiplicity ratios, and improved energy resolution, with the largest gains observed at high multiplicities where shower overlap is most severe.

In mixed-particle training scenarios, CML maintains consistent performance across particle types, while OC degrades substantially, particularly for electromagnetic showers. The results demonstrate that learning a similarity-based representation with a stable geometric structure is more effective than optimizing object-centric clustering variables for dense point cloud segmentation. The authors conclude that contrastive metric learning combined with density-based readout offers a robust and promising alternative to object-centric methods for reconstruction in highly granular detectors, with potential applicability beyond high-energy physics to general point cloud segmentation problems where object boundaries are ambiguous and overlap is common.

%% file: Holographic_QCD_with_machine_learning.tex
\section{Holographic QCD with machine learning}
\label{sec:holo}

The Anti-de Sitter/Conformal Field Theory (AdS/CFT) correspondence \cite{Maldacena:1997re,Gubser:1998bc,Witten:1998qj}, more broadly referred to as holographic duality, has become one of the central frameworks for studying strongly coupled quantum many-body systems. Its core idea is that certain quantum field theories can be equivalently described by a gravitational theory in one higher dimension. In practice, this duality provides an alternative language for non-perturbative dynamics: difficult field-theory questions may be translated into geometric problems in the bulk spacetime. Over the past two decades, this perspective has had a major impact not only in high-energy theory, but also in condensed matter physics, hydrodynamics, and quantum information.

For QCD and QCD-like theories, holography is especially attractive because many of the most important phenomena of interest---confinement, chiral symmetry breaking, real-time transport, and finite-density dynamics---occur precisely in regimes where perturbation theory is unreliable. Since a fully controlled top-down holographic dual of real QCD is not available, much of the phenomenological progress has relied on effective bottom-up constructions. These include hard-wall and soft-wall AdS/QCD
\cite{Erlich:2005qh,DaRold:2005mxj,Karch:2006pv},
improved holographic QCD
\cite{Gursoy:2007cb,Gursoy:2010fj},
Einstein-dilaton and Einstein-Maxwell-dilaton models
\cite{Gubser:2008yx,DeWolfe:2010he},
and V-QCD/Einstein-Dilaton-Flavor-type models
\cite{Jarvinen:2011qe}. Related top-down or string-inspired constructions, such as the Sakai--Sugimoto model
\cite{Sakai:2004cn,Sakai:2005yt}, have also provided important guidance. These frameworks have been remarkably successful at reproducing qualitative and even semi-quantitative features of hadron physics and the quark-gluon plasma. At the same time, traditional model building often depends on phenomenological ans\"atze for the warp factor, dilaton profile, or coupling functions, which introduces an element of arbitrariness.

\begin{figure}[htbp]
    \centering
    \includegraphics[width=0.95\linewidth]
    {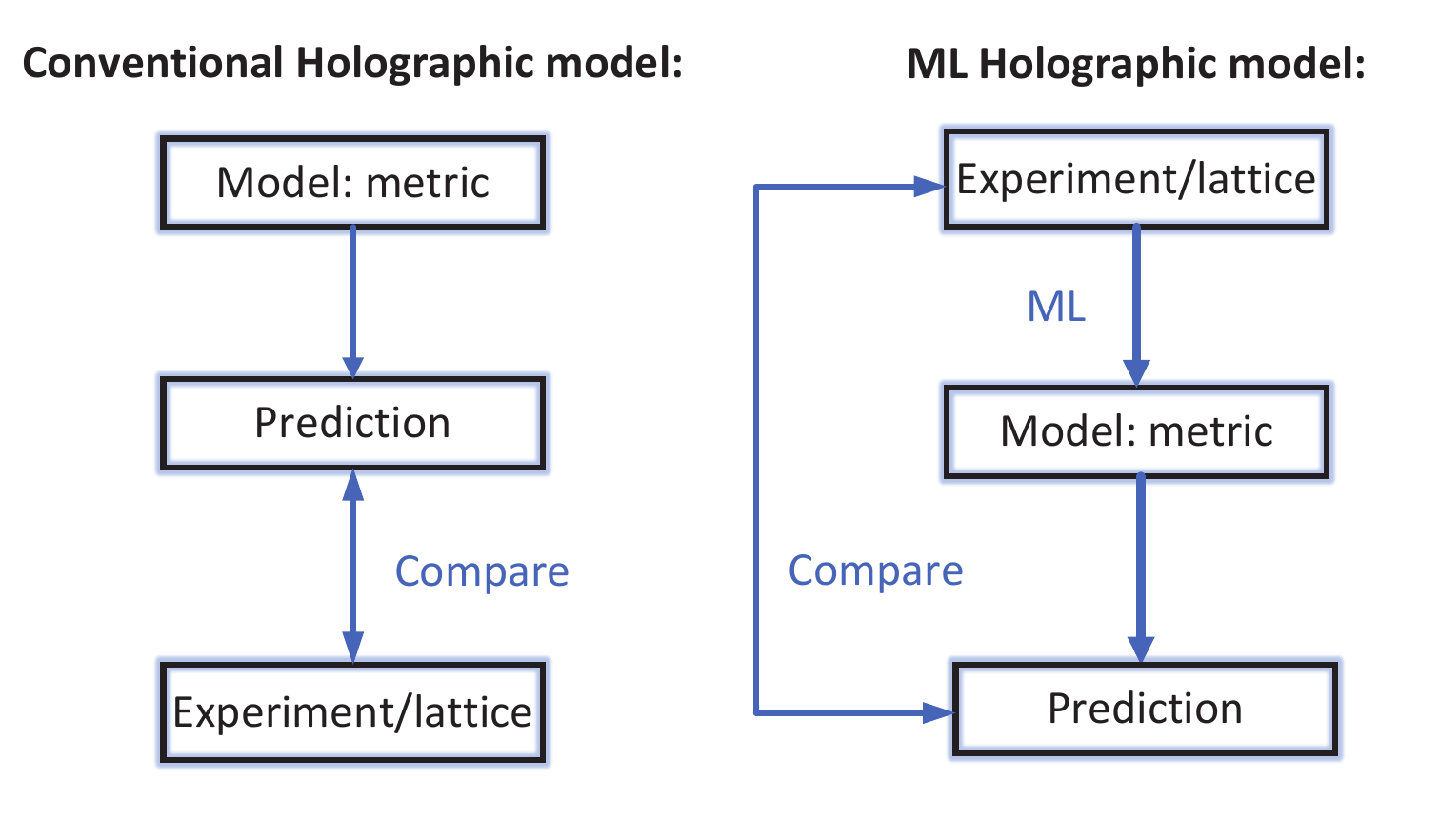}
    \caption{
    Schematic comparison of conventional holographic modeling
    (left) and ML-assisted holographic inference (right).
    In the conventional forward workflow, a specified bulk model
    yields observables that are compared with experimental or
    lattice data. In the ML-assisted workflow, selected data
    constrain the holographic model through a learning procedure,
    after which further observables are calculated. This figure is reproduced from Fig.~2 of Ref.~\cite{Chen:2024mmd}. }
    \label{fig:holographic_workflow}
\end{figure}

Machine learning provides a systematic way to constrain this model-building freedom with data. As illustrated in Fig.~\ref{fig:holographic_workflow}, conventional forward modeling starts from a specified bulk model and compares its predictions with experimental or lattice results, whereas ML-assisted inference
uses boundary observables to constrain model parameters or reconstruct bulk functions, which then serve as the basis for further predictions \cite{Chen:2024mmd}. This shifts the emphasis toward data-driven inference without eliminating model dependence; physical constraints and uncertainty quantification therefore
remain important. The resulting research program may be organized into several interconnected directions, ranging from conceptual
links between renormalization and deep learning to concrete applications in spectroscopy, thermodynamics, quarkonium physics, proton structure, and transport.

\subsection{From Renormalization Group to Deep Generative Models}

An important intellectual precursor to ML-assisted holography is the long-recognized relation between the renormalization group (RG) and the extra radial dimension in AdS. In holography, the radial coordinate is naturally interpreted as an energy scale, with the ultraviolet (UV) boundary corresponding to short-distance physics and the infrared (IR) interior encoding long-distance dynamics \cite{McGreevy:2009xe}. This picture suggests that the emergent bulk geometry may be understood as a geometric representation of coarse graining.

This RG/geometric intuition is also closely related to tensor-network approaches to holography, where entanglement renormalization and MERA provide discrete realizations of an emergent radial direction \cite{Swingle:2009bg}. In this broader context, deep learning offers another hierarchical language for representing coarse graining and latent bulk degrees of freedom. Early influential works~\cite{Gan:2017nyt,Hashimoto:2019bih} argued that neural-network architectures can be viewed as concrete realizations of holographic or RG-like structures. In Ref.~\cite{Gan:2017nyt}, the hierarchical organization of a deep neural network (DNN) was interpreted as an RG flow, and it was suggested that repeated coarse graining naturally generates a hyperbolic geometry reminiscent of AdS. In this RG/geometric intuition, the depth of the network plays the role of the holographic direction, while the compression of information across layers mirrors the flow from UV degrees of freedom to effective IR descriptions.

A closely related probabilistic formulation was proposed in Ref.~\cite{Hashimoto:2019bih}, where the AdS/CFT correspondence was mapped to a deep Boltzmann machine (DBM). In that construction, visible units represent the boundary field-theory variables and hidden units represent bulk degrees of freedom distributed along the radial direction. The network energy function is associated with the bulk action, and the boundary generating functional is obtained by integrating over hidden variables. This is conceptually important because it recasts holography as a generative model: the bulk acts as a latent representation from which boundary configurations are produced.

These ideas were further developed in more explicitly constructive directions. Ref.~\cite{Hu:2019nea} proposed a neural-network renormalization-group algorithm based on flow-type generative models. The goal was to learn an exact holographic mapping between uncorrelated bulk variables and the interacting boundary field ensemble. When applied to the critical regime of a two-dimensional complex $\phi^4$ theory, the learned geometry of the latent manifold was found to be consistent with a three-dimensional hyperbolic space, providing numerical evidence that an AdS-like structure can emerge from data-driven learning.

\begin{figure}[htbp]
    \centering
    \includegraphics[width=0.8\linewidth]{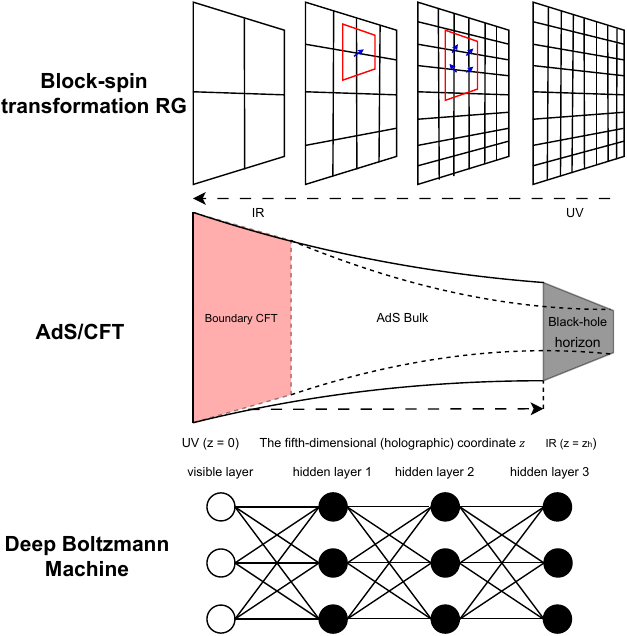}
    \caption{Schematic relation between holographic duality, renormalization-group flow, and deep generative architectures. The upper part illustrates the flow from the UV boundary theory to the IR through RG coarse graining, which is geometrized as motion along the AdS radial direction. The lower part shows the corresponding deep Boltzmann machine interpretation, where visible units encode boundary degrees of freedom and hidden layers represent emergent bulk scales. The figure summarizes the common intuition underlying holography, RG, tensor-network constructions, and machine-learning architectures inspired by Refs.~\cite{Swingle:2009bg,Gan:2017nyt,Hu:2019nea,Hashimoto:2019bih}.}
    \label{fig:holographic_dbm}
\end{figure}

Taken together, these studies established a conceptual bridge between RG, information compression, and generative learning. They also motivated a broader perspective that has become central in recent years: holography may be viewed not only as a formal duality, but also as an inverse-learning problem in which geometry is inferred from observables. This conceptual relation is summarized schematically in Fig.~\ref{fig:holographic_dbm}, which juxtaposes the RG interpretation of the AdS radial direction with the DBM picture of hidden bulk layers.

\subsection{Methodological Core: Geometric Reconstruction and Holographic Inverse Problems}

The most direct and influential line of work concerns the inverse problem itself: how to reconstruct a bulk spacetime, an effective action, or a set of coupling functions from boundary data. This direction forms the methodological backbone of the field, because essentially all later phenomenological applications rely on the ability to infer holographic models from observables rather than specifying them a priori. A central difficulty, however, is that holographic reconstruction is generally non-unique: different bulk potentials, coordinate choices, or matter couplings may reproduce the same limited set of boundary observables. The role of modern ML architectures is therefore not only to fit data efficiently, but also to incorporate physical priors, constraints, and uncertainty estimates that make the inverse problem better posed.

\subsubsection{Foundational frameworks and early works}

A first concrete realization of this program appeared in the so-called AdS/deep learning correspondence \cite{Hashimoto:2018ftp,Hashimoto:2018bnb}. In these studies, the radial evolution of bulk fields was represented by a deep neural network whose layers correspond to discretized positions in the bulk. Network weights were identified with geometric quantities such as metric components or interaction terms. By training on boundary observables, the network was able to reconstruct a bulk background that reproduces the data. A notable feature of these early models was that geometrically meaningful structures, such as a black-hole horizon, could emerge from the training procedure rather than being imposed from the beginning. This marked a qualitative shift from forward modeling to inverse reconstruction.

\subsubsection{Algorithmic refinements and physically informed architectures}

Subsequent studies improved both the numerical stability and the physical transparency of the reconstruction. One important step was the introduction of Neural Ordinary Differential Equations (Neural ODEs) \cite{Chen:2018wjc,Hashimoto:2020jug}. In contrast to discrete-layer networks, Neural ODEs treat the network depth as a continuous variable. This is especially natural in holography, where the radial direction is itself continuous. In practice, the method allows one to model the radial evolution of bulk fields or metric functions by solving trainable differential equations, reducing discretization artifacts and making the learned geometry smoother. These approaches have been used, for example, to reconstruct temperature-dependent holographic backgrounds more efficiently and to improve the extraction of thermodynamic quantities \cite{Hashimoto:2020jug}. Related pedagogical introductions to the AdS/deep-learning correspondence are given in \cite{Song:2020agw}.

Another major development is the use of physics-informed neural networks (PINNs) \cite{Raissi:2017zsi,Bea:2024xgv,Jeong:2025omu}. In PINN-based frameworks, the loss function is not based solely on matching data. It also contains residuals of the bulk equations of motion, such as Einstein equations, dilaton equations, or constraint equations. This makes the learned solution more interpretable and better aligned with the underlying theory. Instead of fitting observables in a purely black-box fashion, PINNs enforce that the reconstructed geometry is dynamically consistent. This is particularly useful when the available boundary data are limited, because the field equations provide a strong inductive bias. PINNs have also been applied to forward holographic calculations, for example to compute holographic entanglement entropy and entanglement-wedge cross sections for prescribed asymptotically AdS geometries \cite{Deb:2025kwx}.

A complementary variational strategy has also emerged in holographic flavor physics. Instead of enforcing the Euler--Lagrange equations through residual losses, one may parameterize a probe-brane embedding by a neural network and directly minimize the regularized DBI action, interpreted as a thermodynamic free energy. This action-level formulation is particularly natural in flavor-brane systems, where the physical embedding is selected through an extremization principle. It provides a useful counterpart to PINN-based reconstruction: while PINNs enforce the equations of motion through differential residuals, the variational approach imposes the underlying extremization principle directly at the level of the action \cite{Filev:2025mbt}.

More recently, new architectures have been explored for specific tasks. Transformer-based models \cite{Park:2023slm} have been used in settings where one aims to predict geometric information from nonlocal quantum data such as entanglement entropy. Since transformers are well suited to learning long-range correlations, they are natural candidates for reconstructing bulk features encoded in multipartite boundary information. Kolmogorov-Arnold Networks (KANs) \cite{Liu:2024swq} and related KAN-based holographic parametrizations \cite{Luo:2024iwf} are also attracting interest because they can yield more interpretable functional relations than standard multilayer perceptrons. In holographic reconstruction, this raises the possibility of obtaining approximate analytic forms for bulk potentials or coupling functions, rather than only numerical outputs.

\subsubsection{Boundary observables used as training data}

An important sign of maturity in this area is the diversity of observables now used as inputs. Early studies often focused on relatively simple thermodynamic or condensate data, but later work showed that the inverse problem can be posed using a broad range of boundary information. These inputs may be roughly organized into local equilibrium observables, real-time response functions, nonlocal probes, and hadronic structure observables. Each class constrains a different aspect of the holographic model. Thermodynamic data mainly fix the equilibrium black-hole background; transport and conductivity data are sensitive to near-horizon dynamics and matter couplings; entanglement entropy, Wilson loops, and static potentials probe extended bulk geometry and IR confinement physics; hadronic structure functions instead constrain bulk hadron wave functions and exchange kernels relevant for deep-inelastic scattering.

Real-time response data provide one class of input. For example, shear viscosity and related dissipative quantities were considered in \cite{Yan:2020wcd}, while optical conductivity was used in \cite{Li:2022zjc,Ahn:2024gjf}. In such cases, the reconstructed geometry is constrained by the requirement that it reproduces frequency-dependent response functions, which are sensitive to both the near-horizon region and the bulk matter couplings.

Another important input comes from quantum information observables. Entanglement entropy \cite{Ryu:2006bv,Hubeny:2007xt} has been employed in \cite{Park:2022fqy,Park:2023slm,Ahn:2024jkk} as a probe of bulk geometry, motivated by the fact that holographic entanglement directly relates boundary entanglement patterns to extremal surfaces in the bulk. These studies are conceptually significant because they connect data-driven reconstruction to the broader idea that spacetime geometry is encoded in quantum information.

Nonlocal observables such as Wilson loops and static quark-antiquark potentials have also been used \cite{Hashimoto:2020mrx,Luo:2024iwf}. Since these probes are particularly sensitive to confinement physics and the IR structure of the geometry, they complement thermodynamic inputs that primarily constrain the equilibrium black-hole background.

A related extension concerns hadronic structure functions. Ref.~\cite{Kou:2026xsx} introduced a physics-guided neural network for the proton structure function $F_2$ in holographic QCD, embedding the AdS$_5$ Dirac equation and the holographic string-diffusion kernel directly into the learning architecture. This illustrates how physics-integrated learning can extend holographic QCD beyond thermodynamics and spectroscopy toward deep-inelastic scattering and the transition between resonance-dominated and diffractive regimes.

Finally, equation-of-state data and other thermodynamic observables remain central ingredients in many reconstruction programs \cite{Bea:2024xgv}. Overall, this methodological direction has evolved from proof-of-principle demonstrations into a flexible toolbox of inverse solvers. It now combines neural architectures, differential-equation solvers, and explicit physical constraints, thereby providing the technical foundation for data-driven holographic phenomenology.

\subsection{Vacuum Structure and Hadron Spectroscopy}

A second major application area concerns the vacuum structure of QCD and the spectroscopy of hadronic states. Here the goal is to reproduce hadronic observables associated with confinement, chiral symmetry breaking, and bound-state spectra, and then to infer the corresponding bulk background or action.

In soft-wall and related AdS/QCD models, hadron masses depend sensitively on the warp factor and dilaton profile. This makes spectroscopy a natural target for inverse reconstruction. Refs.~\cite{Akutagawa:2020yeo,Hashimoto:2021ihd}, for example, used meson spectral data such as the $\rho$ and $a_2$ trajectories to reconstruct effective holographic backgrounds or dilaton potentials. More recent extensions have considered broader unflavoured meson spectra, including scalar, vector, axial-vector, and tensor channels \cite{ThomasArun:2025uyi}. The basic strategy is to determine bulk functions in such a way that the eigenvalue problem for normalizable bulk modes reproduces the observed Regge-like structure of meson masses. This line of work demonstrates that ML can extract effective backgrounds directly from hadronic data, reducing the need for hand-tuned soft-wall profiles.

The reconstruction can be further pushed in Einstein-dilaton, IHQCD-type, and soft-wall-inspired settings. In such constructions, one often seeks to determine not only the background solution but also the dilaton profile or potential, because these functions control the running coupling, confinement scale, and IR physics of the theory. Studies such as \cite{Hashimoto:2022eij,ThomasArun:2025uyi,Zhang:2026zoz} used chiral condensates, unflavoured meson spectra, or heavy-quarkonium spectra as inputs to infer the shape of the dilaton sector. In this sense, the problem shifts from fitting a background to learning part of the holographic action. This is a more ambitious inverse problem, since multiple sectors of observables may have to be reproduced simultaneously.

An especially interesting development is the attempt to unify vacuum spectroscopy with finite-temperature physics in a single model. Ref.~\cite{Chen:2025kqb} is an example of this trend.
Fig.~\ref{fig:Chen} illustrates the sequential stages of the holographic reconstruction workflow. The process begins with the transformation of glueball spectra data into the reconstructed metric potential $V_g(z)$  via Schr\"odinger inversion. From this, the composite derivative function $B_p(Z) =3A'(z)-\Psi'(z)$ is extracted through potential decomposition. The workflow concludes by integrating equation-of-state (EOS) constraints with the derived $B_p(z)$ to reconstruct the warp factor $A(z)$. With the neural-network parametrization of  $A(z)$ constrained by both the EOS and the glueball spectrum, the entropy density is computed and compared against lattice results, showing good agreement.

There, an ML-assisted Einstein-dilaton model was constrained at the same time by lattice thermodynamic data and by the scalar glueball spectrum of pure Yang-Mills theory. The significance of this result is not only technical but conceptual: it suggests that finite-temperature thermodynamics and zero-temperature bound-state physics can be described consistently within one reconstructed background, rather than by separate models specialized to different observables.

{\color{red}
\begin{figure}[htbp]
    \centering
    \includegraphics[width=0.95\linewidth]{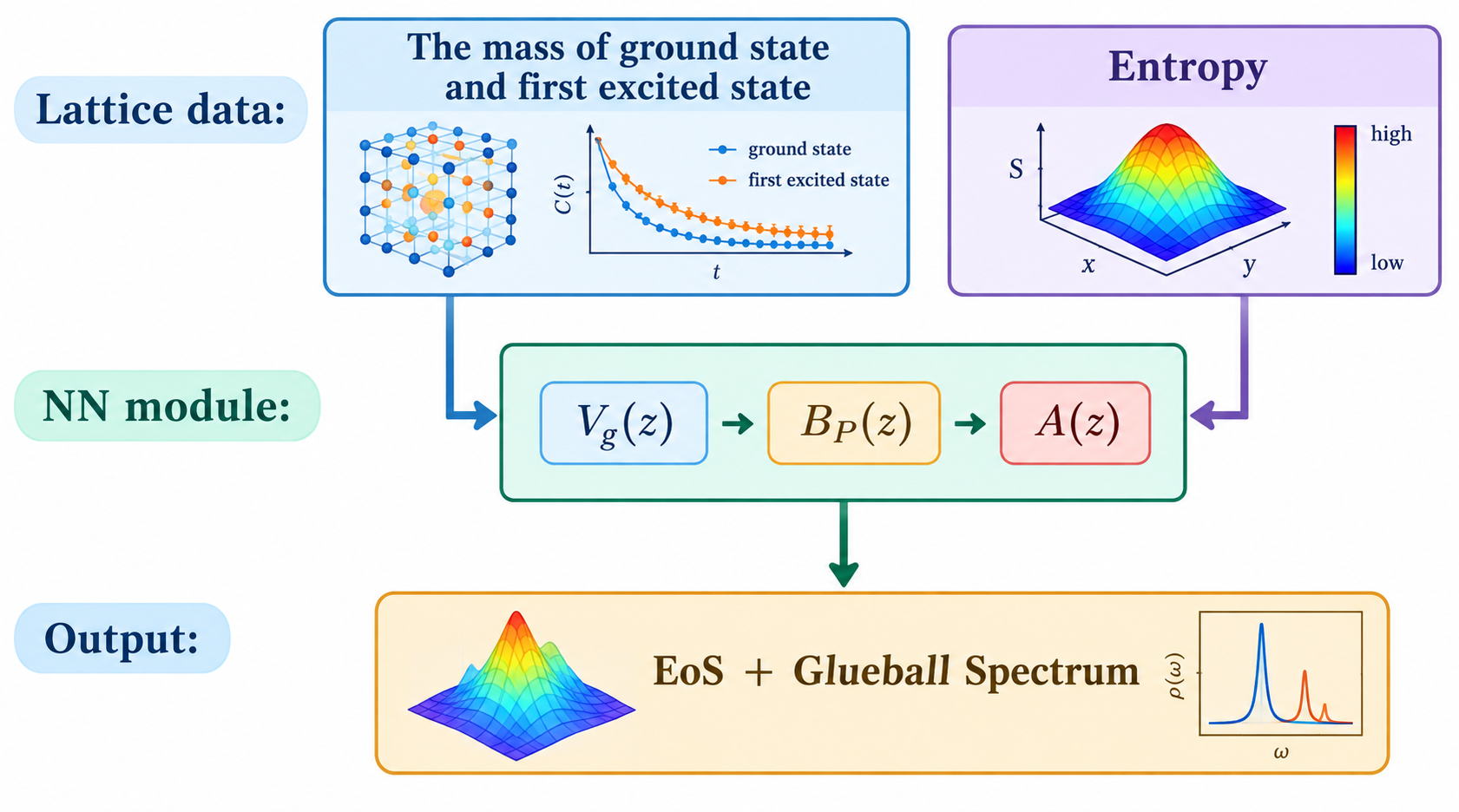}
    \caption{A sketch of holographic reconstruction workflow.  }
    \label{fig:Chen}
\end{figure}
}

More broadly, this direction shows that data-driven holography can reverse-engineer the non-perturbative QCD vacuum from phenomenology. Rather than treating hadron spectroscopy and confinement diagnostics as qualitative benchmarks, the newer literature uses them as quantitative constraints on the effective five-dimensional theory.

\subsection{QCD Thermodynamics and the Phase Diagram}

Finite-temperature and finite-density QCD is arguably the most active arena for ML-assisted holographic model building. The central goal is to calibrate a holographic effective theory against reliable lattice-QCD data at vanishing baryon chemical potential and then use the model to extrapolate into the finite-density region, where lattice methods face severe difficulties.

A common starting point is to fit the QCD equation of state, including quantities such as pressure, entropy density, trace anomaly, and baryon-number susceptibilities. In this context, Einstein-Dilaton-Flavor models \cite{Shen:2025yrn} incorporate the effects of $2+1$ dynamical flavors and allow one to address both thermodynamic and chiral observables. By optimizing model parameters with ML tools, such constructions can reproduce the lattice crossover behavior and study how the phase structure varies with quark masses, including applications to the Columbia plot. Related EMD-plus-flavor constructions \cite{Deng:2026aht} have used neural-ODE reconstruction and back-reaction tests to constrain the QCD equation of state from lattice data.

Einstein-Maxwell-Dilaton models provide another widely used framework \cite{Chen:2024ckb,Zeng:2025tcz}. In these models, the metric and dilaton encode the gluonic sector, while a bulk gauge field captures the effects of baryon chemical potential. ML-based reconstruction is used to infer metric functions and gauge-dilaton couplings from lattice input. Fully data-driven EMD realizations such as Refs.~\cite{Zeng:2025tcz,Zhu:2026dmc} represent the warp factor and gauge-dilaton coupling by neural networks and learn these functions directly from lattice or hadron-resonance-gas data. Complementary approaches retain analytical holographic ans\"atze and use machine learning or gradient-based optimization to calibrate their parameters against lattice data \cite{Chen:2024mmd,Chen:2025goz}.

Once calibrated at $\mu_B=0$, these models can be continued to finite baryon chemical potential, where holography becomes especially valuable. Several studies have used ML-assisted holographic models to explore the possible emergence and location of the QCD critical endpoint in the $(T,\mu_B)$ plane \cite{Zeng:2025tcz,Zhu:2026dmc,Chen:2025goz}, while its dependence on the number of flavors has been investigated explicitly in Ref.~\cite{Chen:2024mmd}. Although different model assumptions lead to different numerical values, the broader significance is that the CEP search can be formulated as a constrained inference problem rather than a purely phenomenological scan.

External magnetic fields add another layer of complexity. Ref.~\cite{Cai:2024eqa} used Neural ODE techniques to investigate the phase diagram in the presence of strong magnetic fields.
Figure \ref{fig:Cai} illustrates the neural ODE workflow. Boundary conditions are imposed at the ultraviolet (UV) boundary and the black hole event horizon. The neural ODE numerically solves the equations of motion to produce a trial equation of state, which is iteratively optimized via backpropagation to minimize deviations from lattice QCD data. This procedure efficiently converges to an optimal magnetic coupling, yielding good agreement with lattice data and enabling model predictions across unexplored magnetic fields and chemical potentials. The resulting phase structure can become considerably richer, with indications of nontrivial critical behavior and even the possibility of multiple critical endpoints in some parameter regimes. Such results illustrate how flexible neural reconstruction methods are in extending holographic models beyond standard thermodynamic settings.

{\color{red}
\begin{figure}[htbp]
    \centering
    \includegraphics[width=0.8\linewidth]{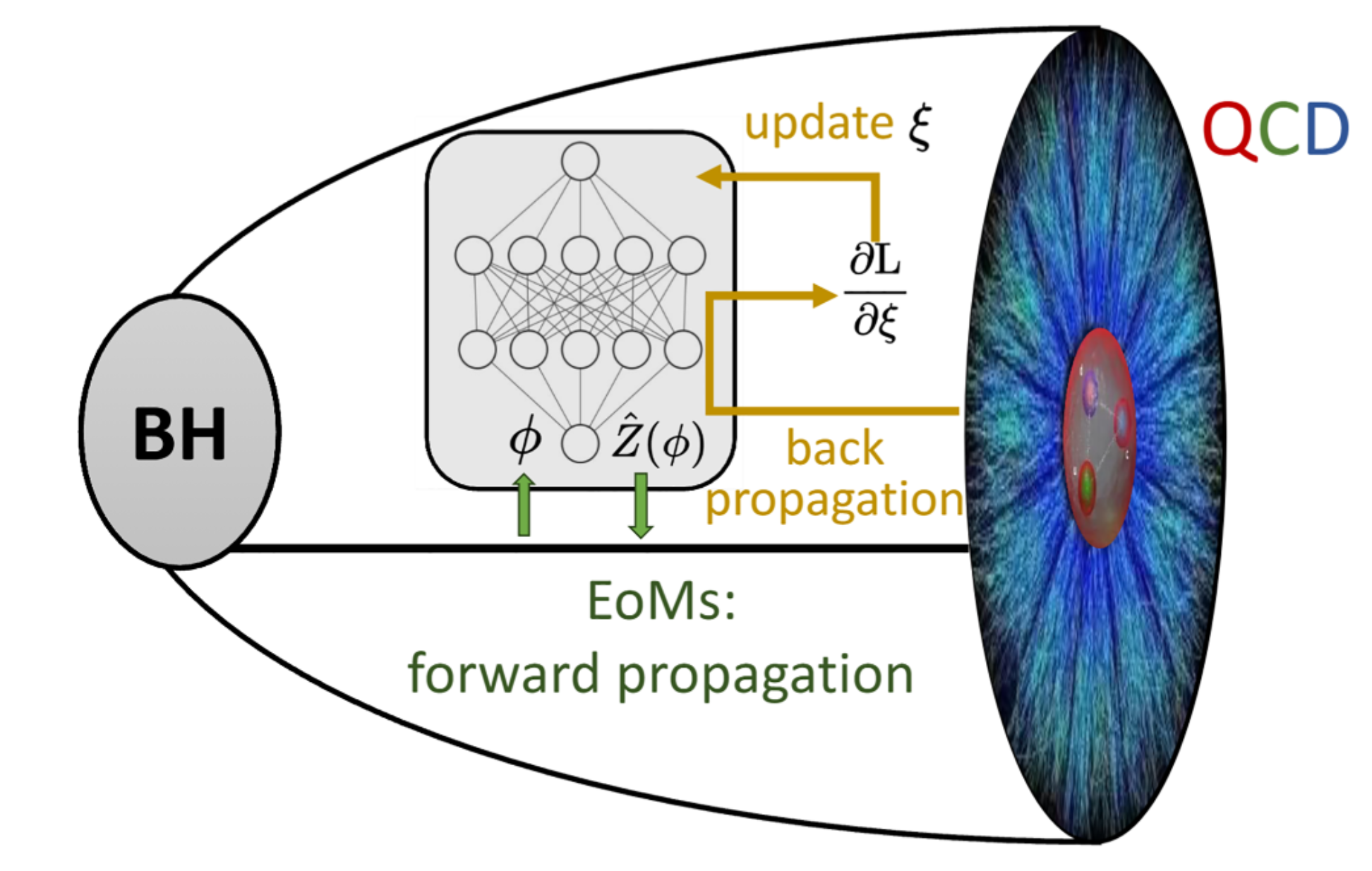}
    \caption{Schematic of the neural ODE framework for solving the inverse problem in QCD phase diagram analysis. Initial conditions at the black hole (BH) horizon are propagated to QCD boundary data through the equations of motion (EoMs). The neural network, initialized as a trial function for the coupling  $Z(\phi)$, is optimized via backpropagation to minimize the loss function $L$, which quantifies the deviation from lattice QCD data. The parameters  $\xi$ are updated using gradient descent. Further technical details are given in Appendix in Ref.~\cite{Cai:2024eqa}. Adapted from Ref.~\cite{Cai:2024eqa}. }
    \label{fig:Cai}
\end{figure}
}

A particularly important advance is the incorporation of uncertainty quantification. Bayesian approaches \cite{Hippert:2023bel,Zhu:2025gxo} allow one to infer posterior distributions for model parameters and derived observables rather than quoting only a single best-fit result. In practice, this means that one can attach credible intervals to the predicted equation of state, the phase boundary, or the CEP location. For a field that historically relied heavily on model intuition, this is a major step toward statistical rigor.

In summary, ML has substantially strengthened the role of holographic QCD in finite-temperature and finite-density phenomenology. It enables efficient calibration to lattice data, facilitates extrapolation into otherwise inaccessible regions of the phase diagram, and increasingly supports uncertainty-controlled predictions.

\subsection{Heavy Quarkonium in the Quark-Gluon Plasma}

Heavy quarkonia, especially charmonium and bottomonium, are among the most informative probes~\cite{Zhou:2014kka} of the quark-gluon plasma (QGP). Because the heavy quark mass introduces a relatively hard scale, quarkonium states can survive above the deconfinement crossover for some temperature range, and their eventual melting encodes valuable information about in-medium screening and dissipation. Holographically, these systems are often studied either through spectral functions computed in thermal black-hole backgrounds or through Wilson-loop and static-potential observables. This builds on earlier holographic studies of meson melting and thermal spectral functions in flavor-brane constructions \cite{Mateos:2006nu,Myers:2007we}.

Beyond spectral functions, Ref.~\cite{Zhu:2025xiz} used a Bayesian holographic QCD model calibrated to lattice-QCD thermodynamic data to study heavy-quarkonium thermodynamic observables, including the interquark distance, potential energy, entropy, binding energy, and internal energy at finite temperature and baryon chemical potential. This provides an uncertainty-aware, potential-based description of how the hot and dense medium affects quarkonium dissociation.

Related potential-based work includes Ref.~\cite{Mansouri:2024uwc}, which used a machine-learned holographic background to extract the complex static quarkonium potential and the temperature dependence of the dissociation length. A complementary spectral-function analysis was carried out in Ref.~\cite{Zhu:2026pqm}, where a data-driven EMD background was combined with a heavy-quark probe sector to study charmonium and bottomonium dissociation at finite temperature and baryon chemical potential. Together with related anisotropic spectral-function studies \cite{Chang:2024ksq}, this work supports a common qualitative pattern: excited states are more fragile than ground states, spectral peaks broaden and decrease in height as the medium becomes hotter or denser, and anisotropy can enhance dissociation. This sequential pattern, in which excited states such as 2S modes melt before the corresponding 1S ground states, is qualitatively consistent with the standard sequential-suppression picture of quarkonium in a hot medium.

This direction is important because it moves holographic QCD beyond equilibrium bulk thermodynamics and toward in-medium hard-probe dynamics. Spectral functions, complex potentials, screening lengths, and dissociation temperatures provide complementary diagnostics that are closer to heavy-ion phenomenology than the equation of state alone. As a result, quarkonium studies provide a testing ground for whether data-driven holographic models can describe not only the equation of state of the plasma but also the fate of hard probes embedded in it.

\subsection{Transport Properties of the Quark-Gluon Plasma}

Depending on the workflow, transport observables can either be used as inputs for reconstruction or treated as predictions after the equilibrium background has been calibrated. Once a holographic background has been calibrated to reproduce thermodynamic data, it becomes possible to compute a wide class of transport coefficients and hard-probe observables \cite{Kovtun:2004de,Gubser:2006bz,Herzog:2006gh,Liu:2006ug}. This is one of the most practical advantages of holography, because real-time transport is difficult to obtain from first principles in QCD, especially at finite density.

Ref.~\cite{Dai:2025dir} used a KAN-based holographic background to compute the heavy-quark drag force, the jet-quenching parameter, and the heavy-quark diffusion coefficient. Ref.~\cite{Chen:2025fpd} extended this program to a Bayesian holographic setting, calculating these hard-probe observables together with shear and bulk viscosities while propagating posterior uncertainty. These observables are obtained from probe-string configurations and bulk fluctuation equations in the reconstructed geometry. Near-horizon data play a central role for several transport coefficients, while the full radial background can also enter hard-probe observables. What is new in the ML-assisted setting is that these transport predictions are made on top of backgrounds that have already been quantitatively constrained by lattice or phenomenological data.

An especially promising step is the use of Bayesian methods in transport calculations \cite{Chen:2025fpd}. Rather than providing a single curve for each coefficient, the analysis propagates model uncertainty into the transport sector and reports credible intervals, for example at the 68\% and 95\% levels. This makes it possible to compare holographic predictions more meaningfully with lattice estimates and phenomenological constraints extracted from heavy-ion measurements, including ALICE data. Even when the uncertainties are still substantial, the framework represents a clear improvement over purely deterministic model fitting.

This line of work highlights the predictive potential of data-driven holography. Once a model has been trained on equilibrium observables, one can use it as a ``holographic simulator'' to infer real-time properties that are difficult to measure directly and difficult to calculate in traditional approaches. In that sense, transport studies are where the practical value of the entire reconstruction program becomes visible.

\subsection{Short Summary and Outlook}

The combination of machine learning and holographic QCD has developed into a coherent research program with a recognizable methodological core and a rapidly expanding range of applications. Conceptually, it builds on the common logic of RG flow, emergent geometry, and hierarchical latent representations. Methodologically, it formulates holographic model building as an inverse problem in which bulk geometries, matter couplings, or effective actions are inferred from boundary observables. Phenomenologically, it now spans vacuum spectroscopy, finite-temperature and finite-density thermodynamics, proton structure, heavy-quarkonium dynamics, and plasma transport.

The main challenge is no longer simply whether neural networks can fit holographic data, but whether the reconstructed bulk structures are unique, physically interpretable, and predictive outside the training regime. Future progress will likely require richer datasets, stronger theoretical priors, transparent architectures, and systematic uncertainty quantification. In this sense, ML-assisted holographic QCD is moving the field from qualitative model building toward a more constrained and uncertainty-aware framework for strong-coupling phenomenology.

%% file: Neutron_Star.tex
\section{Neutron Star}
\label{sec:neutron-star}
Neutron stars probe QCD matter in a regime complementary to the QGP produced in heavy-ion collisions: the matter is cold, beta-equilibrated and in macroscopic gravitational equilibrium, strongly interacting, and compressed to several times nuclear saturation density.  The main physics unknown is the dense-matter EoS, usually represented as a relation among pressure, energy density, baryon density, and composition \cite{CaiBJ:MPL,CaiBJ:EPJA,CaiBJ:PRD}. Observations do not measure this relation directly.  Instead, they constrain neutron stars' masses, radii, tidal deformabilities, moments of inertia, cooling behavior, and gravitational-wave signals, which are indirect functions of the EoS.  The theoretical task is therefore an inverse problem: infer an EoS from observables after passing through the TOV equations and, for mergers, dynamical simulations, while ensuring that the physical constraints to be satisfied, such as stability, causality, and high-density asymptotic behavior, are respected.

\subsection{Neural EoS Reconstruction}

Traditional neutron-star inference uses parameterized EoS families or non-parametric Bayesian priors and confronts them with astrophysical likelihoods.  ML adds complementary strategies for representing functions, generating admissible EoS ensembles, and accelerating inverse maps.  Early deep-learning studies learned direct maps from neutron-star observables to the underlying EoS~\cite{Fujimoto:2019hxv,Fujimoto:2021dce}. This direct inversion view is simple and fast, but it is sensitive to the assumed training ensemble. Subsequent work has therefore paid more attention to uncertainty quantification, data augmentation, and the use of probability distributions instead of isolated measurements
~\cite{Fujimoto:2024crn}. In this line of work, the main question is no longer whether a network can reproduce a test set. The question is which part of the inferred EoS is actually constrained by the data. 

\begin{figure}[t]
    \centering
    \includegraphics[width=0.94\linewidth]{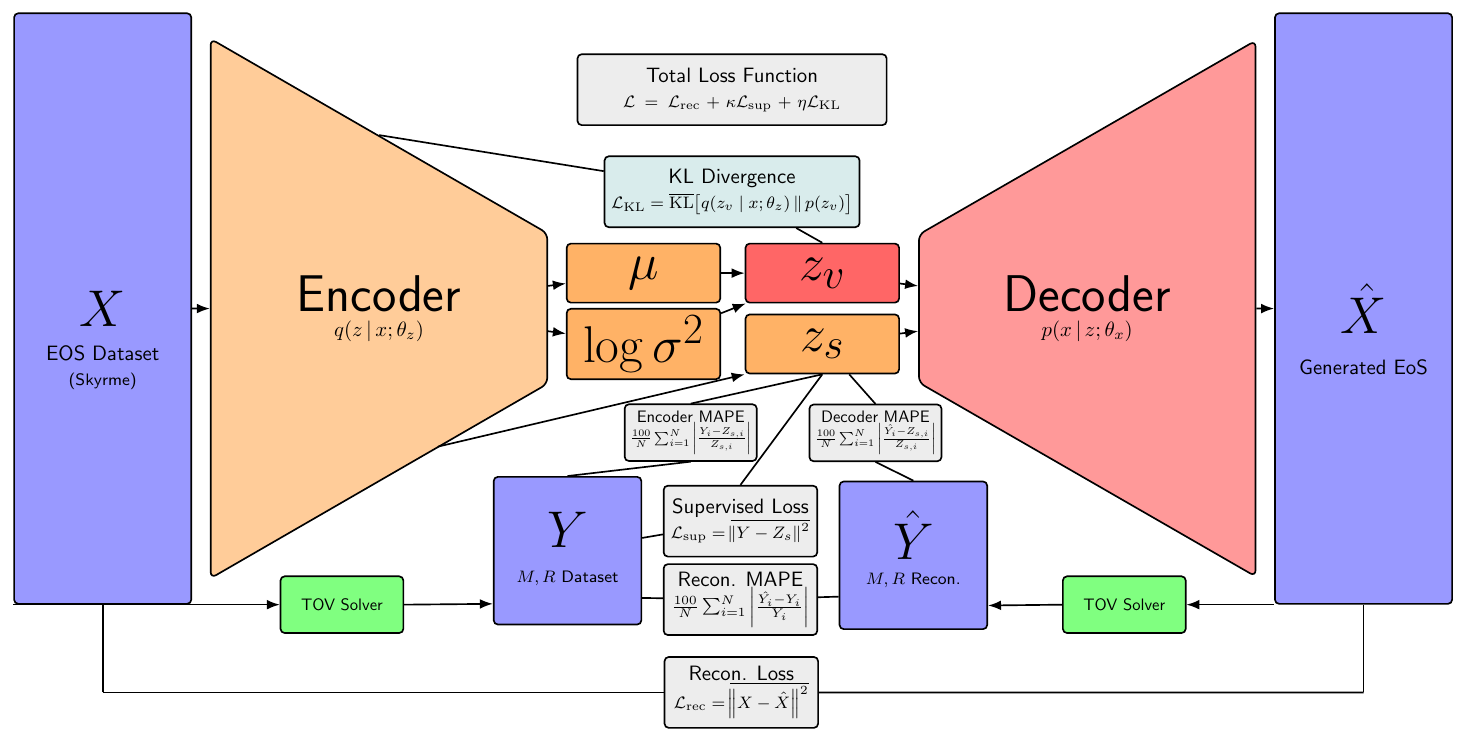}
    \caption{Structured variational-autoencoder representation of neutron-star EoS ensembles. The encoder compresses an EoS into supervised stellar observables and additional latent variables. The decoder reconstructs or generates EoS curves that are then checked against physical and astrophysical constraints. Adapted from Ref.~\cite{Ross:2026ozd}.}
    \label{fig:ns_vae_framework}
\end{figure}

Generative models give another way to organize the EoS space. Conditional variational-autoencoders can reconstruct pressure curves from a set of mass-radius observations after training on physically filtered EoS samples~\cite{Ferreira:2024rnf}. Structured variational autoencoders can also separate supervised stellar quantities, such as $M_{max}$ and $R_{1.4}$, from additional latent variables that describe residual EoS freedom.
~\cite{Ross:2026ozd}. Figure~\ref{fig:ns_vae_framework} shows this idea. The latent variables are not physical observables by themselves, but they provide a compact coordinate system for generating and testing EoS candidates. These studies clarify that ML inference is meaningful only when trained on physically admissible EoS ensembles and tested against distribution shifts. 

\begin{figure}[t]
    \centering
    \includegraphics[width=0.96\linewidth]{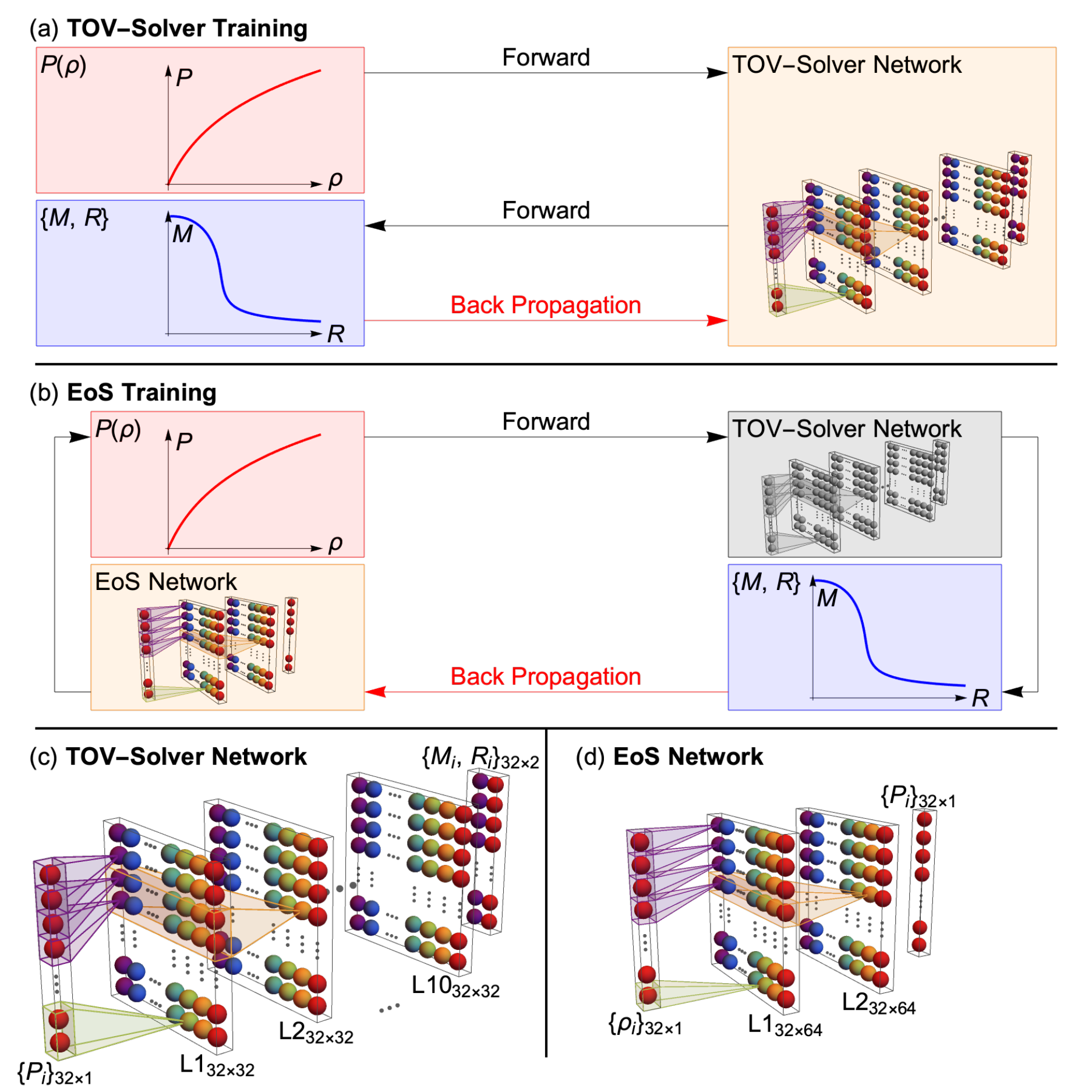}
    \caption{Automatic-differentiation strategy for neutron-star EoS reconstruction.  A neural representation of the EoS is coupled to a differentiable TOV solver so that observational residuals can be back-propagated to the EoS model.  Adapted from Ref.~\cite{Soma:2022vbb}.}
    \label{fig:ns_tov_ad}
\end{figure}

A more physics-driven direction represents the EoS itself by a neural network and couples it to the differentiable stellar-structure equations' solver. 
Neural reconstruction from mass-radius or tidal data has been developed using automatic differentiation (AD) and trained TOV-solver networks~\cite{Soma:2022qnv, Soma:2022vbb}. Ref.~\cite{Soma:2022qnv} demonstrated the devised AD-based neural reconstruction on mock data, and Ref.~\cite{Soma:2022vbb} applied this method to the accumulated mass-radius real observations. Figure.~\ref{fig:ns_tov_ad} illustrates the workflow where known physical equations define the map from a candidate EoS to the observable space. Ref.~\cite{Li:2025obt} used automatic differentiation to invert the TOV solving more explicitly and discussed the possibility to identify possible first-order phase transition for dense QCD matter.  In this setting, a neural network represents the EoS, the TOV equations map it to mass-radius data, and gradients are back-propagated through the solver.  The advantage is conceptual as well as computational: the learning problem is tied to the known physical equations, rather than replacing them by a black-box map.

Recent work has moved from point reconstruction toward posterior inference via deep learning. Neural simulation-based inference can infer the EoS directly from telescope spectra~\cite{Brandes:2024vhw}, while normalizing flows provide rapid posterior sampling for neutron-star EoS constraints~\cite{McGinn:2024nkd}. Deep-learning EoS inference has also been studied in settings designed to include realistic observational uncertainties~\cite{Ventagli:2024xsh}. Simulation-based inference can be combined with universal relations to improve accuracy and reduce the dimensional burden of the inference problem~\cite{Kruger:2026bnz}. These methods are attractive because they can handle non-Gaussian and correlated posteriors, which are common when masses, radii, and tidal deformabilities constrain overlapping but not identical density ranges.

A persistent difficulty is that the high-density EoS is not uniquely determined by present data.  Many EoS curves can satisfy the observed two-solar-mass constraint, produce compatible radii, and yield similar tidal deformabilities.  ML does not remove this degeneracy; instead, it can expose and quantify it.  Physical constraints must therefore be built into the inference.  At low density the EoS should be consistent with nuclear theory, at high density it should not violate causality or thermodynamic stability, and in between it may allow rapid changes associated with new degrees of freedom or phase transitions.  Recent studies have used Bayesian and neural approaches to search for possible QCD phase transitions in neutron-star interiors~\cite{Li:2025obt}, and anomaly-detection ideas have been proposed to identify unusual EoS behavior in neural inference~\cite{Fukushima:2025ywz}.

\subsection{Physical constraints}

Present observations do not determine a unique high-density EoS. Many curves satisfy the two-solar-mass constraint, produce compatible radii, and yield similar tidal deformabilities. Machine Learning can expose and quantify this degeneracy, but it cannot eliminate it by itself. Its useful role is to show which density ranges remain uncertain and how the answer changes with priors, data choices, and physical filters. Physical constraints must therefore be built into the inference. At low density, the EoS should be consistent with nuclear theory. At high density, it must respect causality, thermodynamic stability and perturbative QCD results. At intermediate densities, rapid changes in the speed of sound may indicate new degrees of freedom or phase transitions.

This makes neutron-star inference a natural testing ground for uncertainty-aware Machine Learning. Bayesian and automatic differentiation neural network approaches have been used to search for possible QCD phase transitions in neutron-star interiors~\cite{Li:2025obt}, and anomaly-detection ideas have been proposed to identify unusual EoS behavior in neural inference~\cite{Fukushima:2025ywz}. Symbolic regression provides a complementary, interpretable route by learning compact relations between neutron-star observables and nuclear-matter properties~\cite{Patra:2025xtd, Patra:2026yii}. The defensible output is a calibrated statement about which classes of EoS behavior remain compatible with the data and priors, rather than a unique claim about core composition.

\subsection{Connections to HICs}

The connection to heavy-ion physics is important but subtle.  Heavy-ion collisions and neutron stars probe different thermodynamic trajectories of the QCD phase diagram: hot and dynamical matter in one case, cold and beta-equilibrated matter in the other. A direct merger of information from the two sets is therefore not automatic. Nevertheless, both are sensitive to the same underlying strong-interaction physics at finite baryon density. Huth et al. performed a joint analysis of heavy-ion constraints, multimessenger observations, and microscopic theory to constrain neutron-star matter~\cite{Huth:2021bsp}. The relevant complementarity
is physical: heavy-ion dynamics and stellar structure respond to overlapping aspects of the
strong interaction, but in different temperature and isospin regimes. A relation inferred for
approximately symmetric matter in a collision cannot simply be identified with the cold betaequilibrated
stellar EoS.

A promising long-term direction is a joint inference framework in which heavy-ion data, neutron-star observations, chiral effective theory, perturbative QCD constraints, and lattice information at small chemical potential are treated as complementary constraints on dense QCD matter.  For such a program, ML is useful not primarily as a fitting device, but as a way to represent functional uncertainty and propagate it through multiple forward maps.

\subsection{Gravitational-wave inference and multimessenger workflows}

ML also operates upstream of EoS reconstruction, in extracting source properties from gravitational-wave strain. Dingo-BNS combines physics-informed data compression, neural inference, and importance weighting for rapid binary-neutron-star parameter estimation ~\cite{Dax:2024mcn}. Its scientific role is to accelerate the characterization and localization of a merger, thereby supporting timely electromagnetic follow-up and subsequent matter inference. This task is distinct from learning a common EoS across several systems: waveform parameters, individual stellar tidal responses, and a population-level EoS posterior belong to different stages of the analysis.

Importance weighting is valuable because the original likelihood can check and correct an approximate neural posterior, subject to proposal support and adequate effective sample size. It does not make a finite weighted sample an exact independent posterior ensemble. More generally, gains in inference speed should be reported alongside waveform assumptions, detector-noise robustness, and retained tidal information ~\cite{Dax:2024mcn}. The same distinction matters when extending EoS learning to merger dynamics. A cold beta-equilibrated relation used for isolated stars does not alone specify the temperature, composition, and transport dependence needed for a merger calculation; this extension requires additional physical inputs rather than merely a larger network~\cite{Huth:2021bsp}. 

%% file: Lattice_QFT.tex

\section{AI for Lattice QFT: From Reconstruction to Discovery}
\label{sec:lattice}

Lattice QFT provides the first-principles non-perturbative framework that anchors much of QCD phenomenology. It determines thermodynamic properties at vanishing or small baryon chemical potential, supplies hadronic and transport inputs, and provides benchmarks for continuum and effective descriptions. Its computational challenges are equally well known: large configuration ensembles are expensive to generate, autocorrelation times can grow near criticality and in topological sectors, real-time observables require analytic continuation from Euclidean data, and finite-density QCD suffers from the sign problem. Machine learning has therefore entered lattice QFT through several complementary routes, ranging from inverse reconstruction of physical information and recognition of phases and correlations to generative sampling and sign-problem mitigation. More recently, the scope is expanding further: rather than using AI only to accelerate a calculation within a prescribed theory, emerging approaches seek to infer the underlying operators, Hamiltonians, and effective field theories directly from data. This progression—from reconstruction and learning, through generation, toward theory discovery—provides a useful organizing principle for AI applications in lattice QFT.

\subsection{Inverse reconstruction from Euclidean data}

Many lattice observables are computed in Euclidean time, whereas heavy-ion and hadron-structure applications often require real-time or spectral information.  Spectral-function reconstruction from Euclidean correlators is therefore a central inverse problem.  Automatic-differentiation approaches represent the spectral function by trainable parameters and optimize it through the integral relation to the Euclidean correlator~\cite{Wang:2021jou}.  Related analyses clarified how neural priors, regularization, and correlator uncertainty affect the reconstruction~\cite{Shi:2022yqw}.  Similar inverse-learning methods have been applied to the heavy-quark potential in the QGP, where lattice bottomonium information is used to infer in-medium potential components~\cite{Shi:2021qri}, also applied to extract the valence parton distribution function (PDF) of pion employing next-to-next-leading-order (NNLO) perturbative QCD matching to lattice calculations for leading-twist expansion for renormalization group invariant (RGI) ratios of bilocal pion matrix elements with NNLO Wilson coefficients~\cite{Gao:2022iex}.

The significance of these studies is that ML provides flexible parameterizations and optimization tools for an ill-posed inverse problem rather than removing the ill-posedness of analytic continuation. Positivity, smoothness, asymptotic behavior, sum rules, and correlator uncertainty in the correlator must still be treated explicitly and specified.  When these ingredients are included, neural reconstruction can complement maximum-entropy, Bayesian, and kernel-based methods; when they are absent, a network can easily return a plausible but non-unique answer.

\subsection{Learning phases, fields, and correlations}
Before the recent rise of generative AI applications in lattice QFT, ML was already used to identify phases and order parameters in lattice and statistical field theories.  Regressive and generative neural networks were applied to scalar field theory in Ref.~\cite{Zhou:2018ill}, and deep-learning stochastic processes were explored in connection with QCD phase-transition dynamics~\cite{Jiang:2021gsw}. These controlled systems remain valuable benchmarks because one can test whether a model learns genuine field correlations rather than artifacts of a small dataset.

For gauge theories, symmetry is the central constraint.  A generic neural network does not automatically respect gauge covariance or gauge invariance.  Gauge-equivariant convolutional networks~\cite{Favoni:2020reg} and equivariant flow-based samplers~\cite{Kanwar:2020xzo} address this by building the symmetry into the architecture.  This is more reliable than relying only on data augmentation.  The same lesson applies to topological sectors, chiral symmetry, and continuum extrapolation: a lattice ML method is useful only if it respects or controls the structures that make lattice QCD predictive.

\subsection{Generative models for lattice QFT, and sign-problems}

Generative modeling is now one of the most active ML directions in lattice QFT studies, which is one important branch of HENP.  Flow-based models were proposed as independent samplers for lattice field theory~\cite{Albergo:2019eim} and then extended to gauge theories~\cite{Kanwar:2020xzo}.  Recent work has studied correlated lattice-QCD ensembles generated by flow models~\cite{Abbott:2024kfc}, stochastic normalizing flows combine non-equilibrium Monte Carlo evolution with gauge-equivariant layers and have been implemented in 4-dimensional SU(3) gauge theory~\cite{Bulgarelli:2024brv}. Related work on SU(3) topology sampling uses open boundary conditions to reduce topological-charge autocorrelations and remove the boundary effects through a non-equilibrium switch back to periodic boundaries; a customized stochastic normalizing flow improves this strategy~\cite{Bonanno:2025pdp}.
Super-resolving normalizing flows pursue a related practical goal by mapping between lattice resolutions
~\cite{Bauer:2024byr}.  These efforts aim to reduce sampling cost and autocorrelation while preserving the exact target distribution or allowing controlled reweighting.

\begin{figure}[t]
    \centering
    \includegraphics[width=0.9\linewidth]{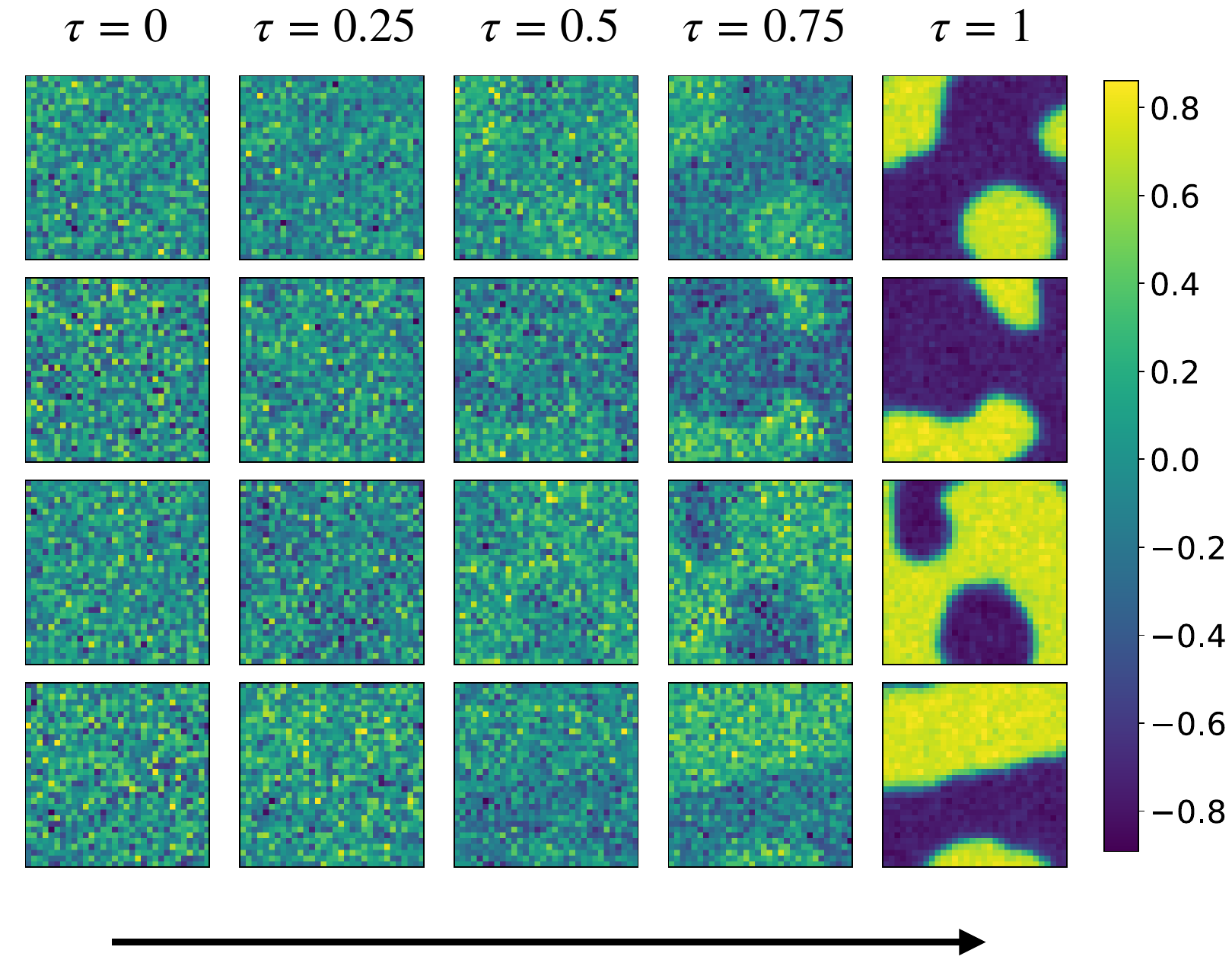}
    \caption{Example of field configurations generated during diffusion-model sampling of scalar lattice field theory.  The panels show a denoising trajectory toward broken-phase configurations.  Adapted from Ref.~\cite{Wang:2023neurips}; see also Ref.~\cite{Wang:2023fby}.}
    \label{fig:lattice_diffusion_configs}
\end{figure}

Diffusion models provide a complementary route.  Ref.~\cite{Wang:2023fby} established a direct connection between diffusion models and stochastic quantization and demonstrated diffusion-based global sampling in two-dimensional scalar lattice field theory.  The reverse diffusion process can be interpreted as a learned stochastic evolution from a simple prior toward the interacting field distribution.  Follow-up work analyzed how diffusion models learn higher-order cumulants~\cite{Aarts:2024cum}, and related studies combined complex Langevin dynamics with score-based and energy-based diffusion models in complex-weight systems~\cite{Aarts:2025cld}.  These developments are particularly relevant for finite-density and real-time problems, where the sampled distribution may be complex or difficult to represent by conventional probability densities. 

Gauge theory adds further constraints. Physics-conditioned diffusion models for lattice gauge theory then try to make the coupling or action parameters explicitly in the generator~\cite{Zhu:2024pcd} though not completely work well as shown in 2D U(1) case on topological sector yet. This would be important for parameter scans if the generalization or condition is achieved for different physical parameters: a single model can interpolate across ensembles only if observables, autocorrelations, and topological quantities remain correct.

\begin{figure}[t]
    \centering
    \includegraphics[width=0.9\linewidth]{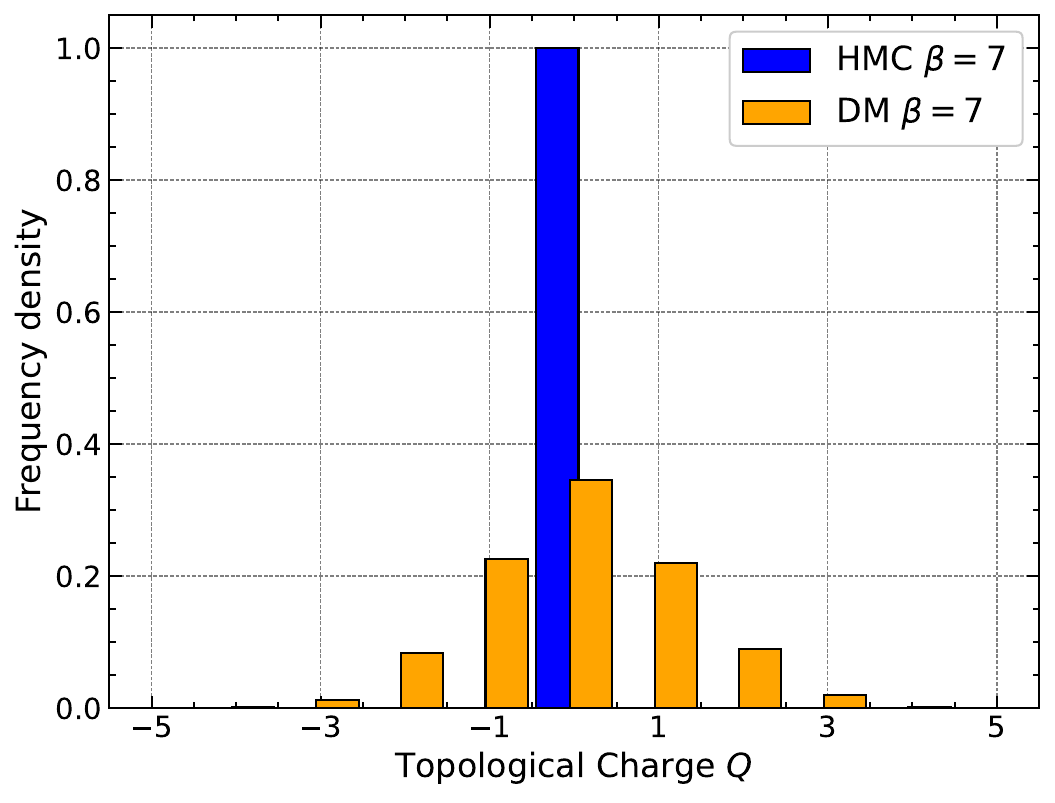}
    \caption{Physics-conditioned diffusion model for lattice gauge theory.  Conditioning on the inverse coupling allows a single generative model to represent multiple target ensembles and to test interpolation or extrapolation across parameter values.  Adapted from Ref.~\cite{Zhu:2024pcd}.}
    \label{fig:lattice_physics_conditioned}
\end{figure}

Recent stochastic-path approaches push the same idea of stochastic generation toward learned Markov-chain dynamics. The Stochastic Path Sampler (SPS) constructs trainable forward and backward stochastic kernels and uses a trajectory-balance-type objective to learn nonlocal paths through configuration space~\cite{Chen:2026bqw}. From the lattice perspective, the attraction is clear: a path model can propose updates that are less local than standard molecular-dynamics or heat-bath steps. The caution is equally clear. Unless the learned sampler is embedded in an exact accept/reject correction, unbiased reweighting, or a demonstrably controlled estimator, it is a proposal mechanism rather than a replacement for lattice sampling. SPS and related trajectory-balance ideas are therefore best read as part of a broader search for trainable proposal dynamics that reduce autocorrelation while preserving the target path-integral measure. For gauge theories, SPS-like methods must still prove gauge compatibility, topology sampling, acceptance behavior, autocorrelation reduction, and continuum scaling.

Physics-informed kernels (PIKs) introduce a particularly direct connection between a specified action and generative transport in the context of lattice QFT sampling. Ref.\cite{Kapust:2025ida} formulates the transformation through prescribed intermediate actions and kernels that can be solved and refined independently at different layers. The construction builds on physics-informed renormalization group
flows, which treat the effective description and the evolving field coordinates jointly. “kernel” denotes the field-transformation generator, not a Gaussian-process covariance
kernel. PIKs belong to a broader history of action-based transport, including trivializing maps and continuity-equation learning between Boltzmann distributions.

For sign problems, generative and variational methods can alter the integration manifold, learn phase-quenched corrections, or construct better proposal distributions. Neural path optimization and learned sign-fluctuation subtraction are promising, but they must be coupled to rigorous estimators and controlled benchmarks \cite{Lawrence:2022dba,Lawrence:2023cft,Hisayoshi:2025adi}. The complex extension of PIKs constructs transformations that preserve the integration weight while deforming the contour~\cite{Ihssen:2026njd}, where the reported demonstrations include zero-dimensional complex-weight theories and the real-time quantum harmonic oscillator. They provide a concrete bridge between transport equations and sign-problem mitigation A useful principle is to separate three notions: faster sampling of a positive distribution, variance reduction for a complex estimator, and unbiased recovery of the target observable. Only the last one is a complete solution.

Despite these advances, ML samplers must pass stringent lattice-QFT tests before they can be used in precision QCD.  Exactness requires an accept/reject correction, unbiased reweighting, or a controlled demonstration that residual bias is negligible.  Autocorrelation reduction must be measured for physical observables, not only for visual similarity of configurations.  In gauge theory, topological tunneling, continuum scaling, finite-volume behavior, and chiral properties are decisive benchmarks.  The field is therefore moving from proof-of-principle demonstrations toward validation standards that resemble conventional lattice methodology.

ML for lattice QFT is best understood as a set of physics-constrained numerical tools.  It can help reconstruct real-time information, learn field distributions, accelerate sampling, and explore sign-problem strategies.  Its long-term impact will depend on whether these tools can be combined with the rigor that makes lattice QCD first-principles: controlled systematics, continuum extrapolation, symmetry preservation, and reproducible error estimates.

\subsection{Learning Quantum Field Theories Across Scales}

The developments discussed above mainly use AI to reconstruct physical
observables, identify phases and correlations, or improve the sampling of a
theory whose microscopic formulation is already known. A more ambitious
direction is to reverse this logic: can machine learning infer the effective
theory itself from physical data? This question connects AI-based inverse
problems with Hamiltonian learning, effective field theory, and the
renormalization group. In this setting, the objective is not merely to
predict an observable, but to identify the relevant operators, their
couplings, and their scale dependence,
$
D \longrightarrow
\{O_i,g_i\}
\longrightarrow
H_{\rm eff},
$
where $D$ denotes measured or simulated data, $O_i$ are candidate operators,
and $g_i$ are the corresponding effective couplings.

\begin{figure}[t]
    \centering
    \includegraphics[width=0.95\linewidth]{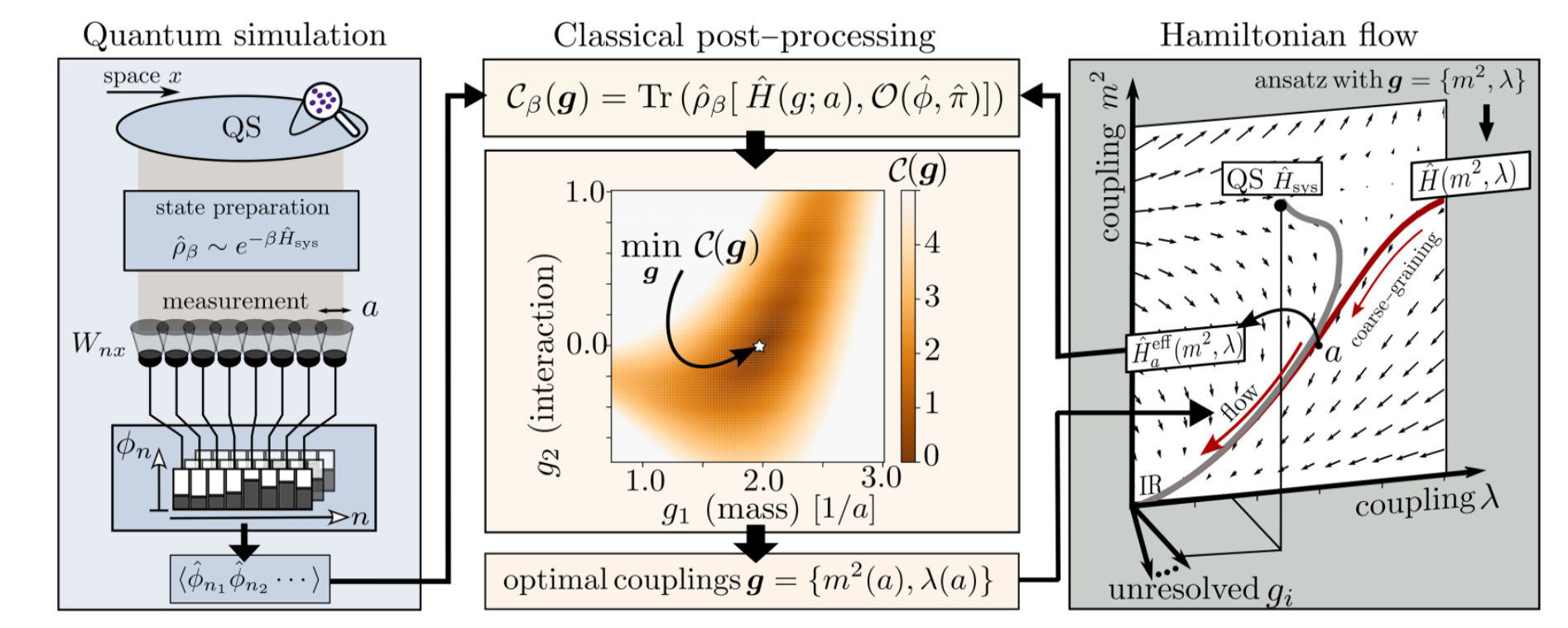}
    \caption{Schematic workflow of Hamiltonian learning for quantum field
    theories. Experimental or quantum-simulation data are measured at finite
    spatial resolution, from which field correlation functions are extracted.
    An effective QFT Hamiltonian is then inferred by learning its relevant
    operator content and coupling constants. Repeating the procedure at
    different resolution scales provides access to scale-dependent effective
    Hamiltonians, connecting Hamiltonian learning with the
    renormalization-group picture. Adapted from Ref.~\cite{Ott2024HamiltonianQFT}.}
    \label{fig:hamiltonian_learning_qft}
\end{figure}

A particularly relevant demonstration was provided by Hamiltonian learning
in quantum field theories using tunnel-coupled quantum gases
~\cite{Ott2024HamiltonianQFT}. As illustrated in
Fig.~\ref{fig:hamiltonian_learning_qft}, experimental measurements at finite
spatial resolution are first converted into field correlation functions.
These correlations are then used to determine the operator content and
coupling constants of an effective QFT Hamiltonian. In the demonstrated
system, the learning procedure distinguishes an interacting sine-Gordon
description from a free quadratic theory and extracts the corresponding
effective parameters. This example is conceptually important because the
machine-learning task is no longer restricted to reconstructing a particular
observable: the data are used to infer the dynamical theory that best
describes the observed correlations.

The role of the resolution scale provides a natural connection to the
renormalization group. In the Wilsonian picture, changing the resolution
changes the effective degrees of freedom and the values of the corresponding
couplings~\cite{Wilson1983RG}. Hamiltonian learning offers a data-driven
realization of this idea: rather than assuming the renormalization-group
trajectory in advance, one can infer effective Hamiltonians at different
resolution scales,
\[
H_{\rm eff}(\Lambda_1)
\;\rightarrow\;
H_{\rm eff}(\Lambda_2)
\;\rightarrow\;
H_{\rm eff}(\Lambda_3)
\;\rightarrow\;\cdots .
\]
The resulting flow,
$
\{g_i(\Lambda)\},
$
can in principle reveal which operators become relevant, irrelevant, or
approximately scale invariant over a given range. This establishes a
conceptual bridge between machine learning, Hamiltonian identification, and
the emergence of effective descriptions across scales.

For lattice QFT, this perspective suggests a broader research program.
Instead of using AI only to accelerate simulations within a prescribed
lattice action, one could seek to learn the effective operator basis and its
scale dependence directly from ensembles of lattice configurations or from
combined lattice and experimental data. A physics-informed implementation
would need to incorporate gauge invariance, locality, global symmetries,
conservation laws, and the appropriate continuum limit. Schematically,
\[
D_{\rm lattice}
\rightarrow 
\{O_i,g_i(a)\}
\rightarrow
H_{\rm eff}(a)
\rightarrow
{\rm continuum\ EFT}.
\]
Such an approach could complement conventional lattice calculations by
providing a systematic route from microscopic lattice dynamics to
coarse-grained effective descriptions.

The main challenge is that theory discovery is substantially more demanding
than parameter estimation. Different operator bases can describe the same
finite-resolution data, and learned Hamiltonians can suffer from
non-identifiability, finite-volume effects, discretization artifacts, and
statistical uncertainties. Consequently, a learned effective theory must be
tested not only by its fit to the training observables, but also by its
symmetry properties, predictive power for independent observables,
stability under changes of resolution, and consistency with known continuum
limits. These requirements are particularly important if AI-generated
Hamiltonians are to be interpreted as physical theories rather than merely
as flexible data representations.

Hamiltonian learning therefore represents a possible transition from
AI-assisted computation to AI-assisted theory discovery. In the longer term,
combining Hamiltonian learning with lattice QFT, effective field theory,
renormalization-group methods, and quantum simulation could provide a
data-driven framework for discovering how effective quantum field theories
emerge across scales. In this sense, the role of AI in lattice QFT may
ultimately extend beyond reconstructing observables or accelerating sampling
to learning the structure and flow of quantum field theories themselves.

%% file: LLM_and_agents.tex
\section{Foundation models, AI agents and Quantum machine learning}
\label{sec:foundation-agent}

\subsection{Foundation models for HENP}

The term foundation model refers to a large model pretrained on broad data and adapted to many downstream tasks. In \HENP, a foundation model need not be a \LLM. It might be a transformer or diffusion model pretrained on particle clouds, detector-level events, lattice configurations, hydrodynamic fields, correlators, or \EoS\ ensembles. The motivation is transfer: many tasks share low-level structure even when labels differ. For collider physics, foundation-model ideas are already being explored for jets and event-level representations \cite{Barman:2025lpm,Hallin:2025fmhep}. In heavy-ion physics, point-cloud diffusion models are a first step toward a foundation-style event generator because they aim to learn a broad representation of hadronic final states and support downstream generation and classification tasks \cite{OmanaKuttan:2024mwr, OmanaKuttan:2025fqc}.

For \HENP, the most plausible foundation models are domain-specific rather than universal. A heavy-ion event model should know about particle species, momentum coordinates, centrality, collision system, and detector acceptance. A lattice model should know about gauge groups, boundary conditions, lattice geometry, and actions. A neutron-star model should know about thermodynamic constraints and stellar-structure equations. Pretraining can be useful, but physics constraints and provenance are essential. Without them, a large model may become a powerful interpolator of simulation artifacts.

\subsection{LLM-based agents and autonomous scientific workflows}

An LLM agent is an AI system that leverages a large language model as its "brain" to autonomously accomplish tasks—not merely by generating responses, but by actively taking actions in its environment. Its key components include planning, task decomposition, reflection, chain-of-thought reasoning, execution, memory management, and the ability to utilize external tools. Collectively, these functionalities form a closed-loop cycle of reasoning, acting, and observation.

Large language model (LLM) agents have moved rapidly from proof of concept to deployed pipelines in nuclear and particle physics (2025--2026). In particle physics they organize around three complementary pillars. \emph{Simulation-native experimental design} is exemplified by GRACE~\cite{Hill:2026naa}, which treats detector design as a constrained search under physical law and has autonomously optimized electromagnetic calorimeters, muon spectrometers, the DarkSide-50 dark-matter detector, and the ProtoDUNE liquid-argon TPC. \emph{Autonomous data analysis} is demonstrated by LLM agents that, given only a dataset and a corpus of prior literature, reproduce electroweak, QCD, and Higgs-boson measurements on open LEP and CMS data end-to-end~\cite{Moreno:2026mqk} and automate HEP analysis~\cite{GendreauDistler:2025llmhep,Moreno:2026agentshep}. \emph{End-to-end collider phenomenology} is realized by ColliderAgent~\cite{Qiu:2026iby}, which carries a model from its Lagrangian through Feynman-diagram generation, matrix-element and cross-section evaluation, and detector-level simulation to exclusion limits, without package-specific glue code. The same paradigm extends to nuclear science: the Advanced Light Source has deployed the first language-model-driven accelerator control system~\cite{Hellert:2025jqi}, and agent--Monte-Carlo integrations now automate input-file generation, execution, and post-processing for reactor, radiation-shielding, and medical-physics simulations.

At the Institute of High Energy Physics (IHEP), the Dr. Sai multi-agent system has been deployed at the BESIII experiment, automating full-chain physics analysis from natural language queries to end-to-end workflow execution with over 90\% success rate~\cite{Li:2026krn}. At its core lies Xiwu, the first LLM specialized for high energy physics, which outperforms GPT-4 in domain-specific code generation and knowledge reasoning~\cite{Zhang:2024kws}. These systems represent a strategic roadmap toward "AI+HEP" and are being generalized across multiple HEP facilities in China.

What unites these efforts is a shift toward trust established by audit rather than assertion. ColliderAgent validates itself by reproducing published results across leptoquark, axion-like-particle, and higher-dimensional-operator scenarios, while companion benchmarks and prompting-strategy studies are standardizing how such systems are evaluated~\cite{Diefenbacher:2025zzn}. Persistent limitations temper this progress: anomalous outputs are flagged but not yet investigated in closed diagnostic loops, high-fidelity Geant4 or transport-code validation remains computationally expensive, and any direct interaction with hardware---accelerators or reactors---demands rigorous uncertainty quantification, hallucination mitigation, and mandatory human-in-the-loop approval for safety-critical actions. The trajectory therefore points not toward autonomous replacement but toward \emph{human--AI collaborative laboratories}, in which agents act as exploratory partners that rapidly survey vast design spaces, surface non-obvious optimizations, and preserve full reproducibility, amplifying human creativity as data volumes grow through the HL-LHC era and beyond.



A concrete agentic workflow for a heavy-ion Bayesian campaign might proceed as follows. The agent proposes a design of simulation points under human-specified priors; launches containerized simulations; checks energy conservation, multiplicity ranges, and hydrodynamic stability; trains an emulator; runs calibration diagnostics; compares posterior predictive observables to data; and produces a report listing failures and suggested next simulations. At every stage, the agent's actions are logged and the final scientific decision remains with the researcher. A lattice version would monitor acceptance, autocorrelation, and topology; a neutron-star version would check causality, stability, and posterior coverage; a literature agent would verify arXiv identifiers and journal metadata before adding references.

Ref.~\cite{Wang:2026cof} introduced CLVisc Agent, the first LLM agent applied to hydrodynamic studies in high-energy nuclear physics. Its meta-skill operates via a scan--compose--validate workflow: it automatically parses the CLVisc source tree, generates version-specific SKILLs, encapsulates the physical meanings of parameters such as $\tau_0$, $\eta/s$, $\zeta/s$, and $T_{\mathrm{frz}}$, and provides observable extractors. Running on kimi-cli, it handles end-to-end parameter-scan design, GPU-accelerated evolution, ensemble comparisons, and publication-quality plots. In Pb+Pb simulations, the high-temperature branch of $\eta/s$ is found to dominate viscous suppression; in O+O collisions, comparing four $^{16}\mathrm{O}$ structure models, the agent autonomously introduces $d_{\mathrm{area}}$ and $\beta_2$ as geometric diagnostics, yielding three distinct groups. This work proves that LLMs can autonomously close the research loop from code comprehension to physical interpretation, and the framework can be extended to SMASH, URQMD, and others, offering a paradigm for automated, reproducible discovery.

Recently, LQCDMaster was proposed, it  is a tool-augmented, skill-guided scientific computing agent that automatically converts natural-language lattice QCD research tasks into executable PyQUDA workflows, complete with measurement scripts, job-submission artifacts, and numerical outputs \cite{WangWei}. It operates through a plan-then-execute architecture, where a planner formulates a physics computation plan (subject to human review) and an executor generates the corresponding code. To ensure reliability, the system integrates expert-annotated LQCD skill libraries and a deterministic Wick-contraction tool that handles the algebraically fragile components of code generation. In benchmarks across 70 state-of-the-art LQCD tasks, LQCDMaster reproduced expert-written implementations exactly to machine precision in 63 cases, reducing implementation time from hours to minutes. It has also enabled novel computations—such as light-cone distribution amplitudes and multi-baryon spectra—that were previously unattempted, demonstrating its potential to lower barriers and accelerate discovery in fundamental physics.

\begin{figure}[t]
    \centering
    \includegraphics[width=0.94\linewidth]{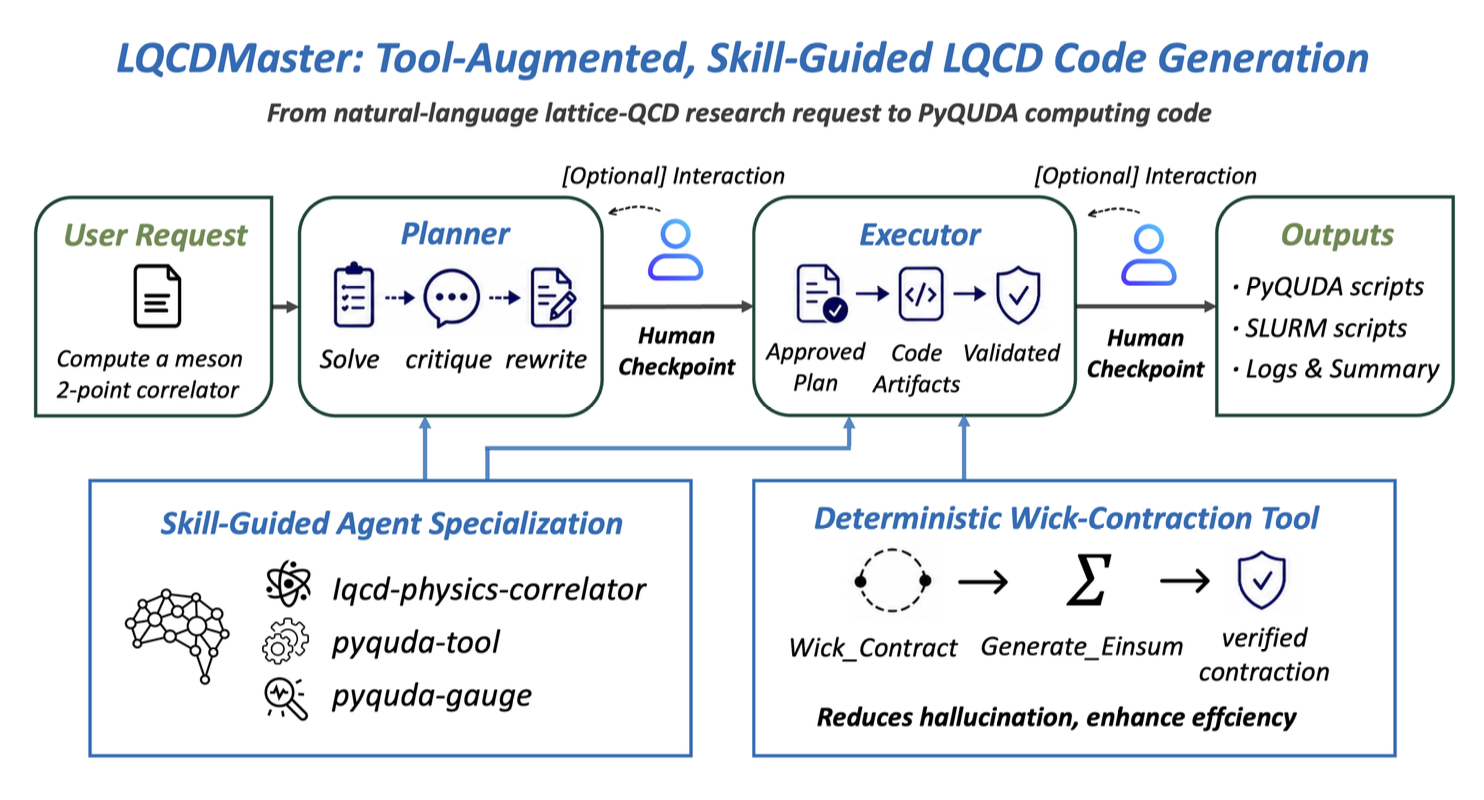}
    \caption{Agentic workflow of LQCDMaster. The planner converts natural-language LQCD tasks
into a structured scientific plan; the executor turns the approved plan into PyQUDA and SLURM
artifacts; expert-annotated domain skills and a deterministic Wick-contraction tool enhance reliability and efficiency. Adapted from Ref.~\cite{WangWei}.}
    \label{fig:Bal}
\end{figure}



\subsection{AI for High-Energy Nuclear Physics: From Symbolic Effective Laws to Scalable Many-Body Sampling}

Artificial intelligence is emerging as a transformative tool in high-energy nuclear physics, where the central challenge is to connect the fundamental theory of quantum chromodynamics (QCD) to the observable structure and dynamics of nucleons and nuclei. This subsection highlights two complementary frontiers of this effort. At one end lies AI-Newton \cite{Ai-newton}, a concept-driven discovery system that autonomously induces interpretable symbolic laws from noisy data—offering a pathway toward automated construction of effective interactions, scaling laws, and phenomenological models in nuclear and QCD physics. At the other end lies the Schr\"odinger Generator \cite{Sun}, a framework for high-dimensional integration and sampling that optimizes coordinate transformations to overcome the efficiency bottlenecks of conventional Monte Carlo methods, enabling stable simulations of nuclear quantum many-body states in dimensions exceeding 600 and faithfully reproducing short-range nucleon correlations in finite nuclei. Together, these works span the arc from symbolic effective laws to scalable many-body computation, reflecting two converging roles for AI in high-energy nuclear physics: one that extracts the effective laws governing strongly interacting matter, and one that makes the simulation of complex nuclear systems tractable.

AI-Newton is a concept-driven physical law discovery system developed \cite{Ai-newton}, capable of autonomously deriving fundamental physics laws from raw experimental data without supervision or prior physical knowledge. The system's core innovation lies in its three-layer knowledge base architecture—symbols, concepts, and laws—which mimics how human physicists organize knowledge, starting from simple concepts like position and time, progressively building more complex concepts such as velocity and acceleration, and ultimately forming complete physical laws. AI-Newton integrates symbolic regression with plausible reasoning techniques, using a recommendation engine to balance exploration and exploitation, and when a discovered general law fails in new scenarios, it can add new terms through plausible reasoning to revise the law. They validated the system on 46 Newtonian mechanics experiments covering scenarios such as free fall, collisions, and oscillations, with all data generated by solving differential equations and adding Gaussian noise to simulate real experimental conditions. Experimental results show that AI-Newton successfully rediscovered core physical laws including Newton's second law, conservation of energy, and universal gravitation, with the system averaging approximately 90 physical concepts and 50 general laws discovered. The research is considered an important step toward AI-driven autonomous scientific discovery, with the team now planning to apply the system to more challenging domains such as quantum theory.

\begin{figure*}[!t]
\centering 
\includegraphics[width=\linewidth]{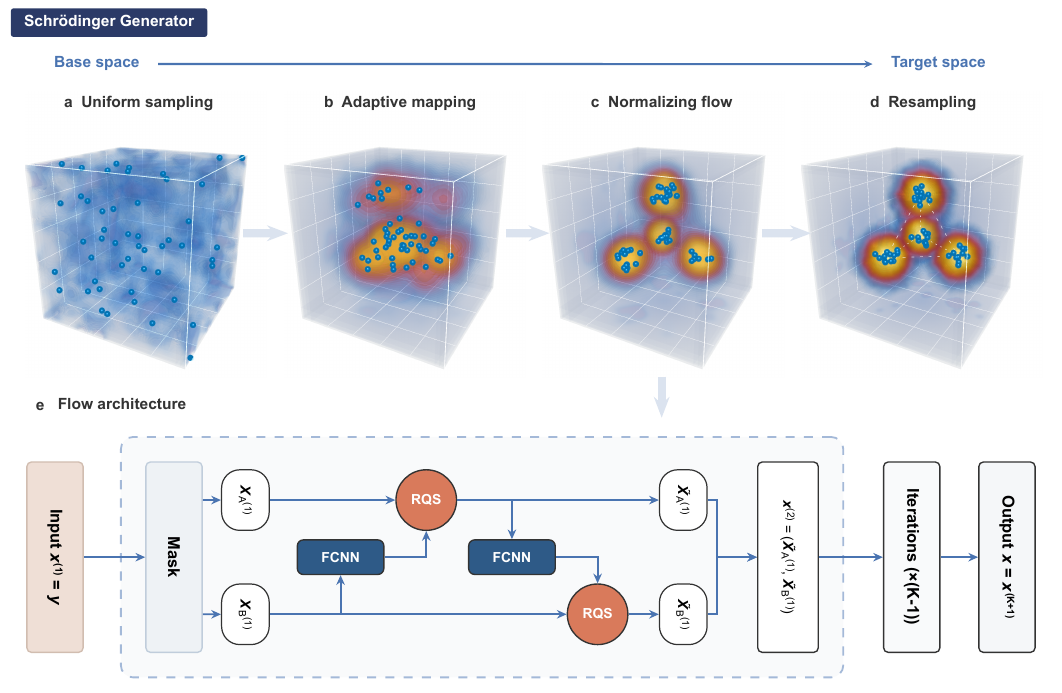} 
\caption{Schematic of the Schr\"odinger Generator for high-dimensional integration and sampling.
The framework combines an adaptive coordinate transformation that reduces per-dimension variance with a normalizing-flow transformation that learns high-dimensional correlations, enabling efficient integration and low-variance sampling from complex target distributions. Upper panels illustrate the evolution of the density distribution through successive stages: uniform initial sampling (a), adaptive importance mapping (b), flow-based transformation (c), and final resampling (d). Lower panel shows the architecture of the normalizing-flow transformation based on rational quadratic splines (RQS) with a fully connected neural network (FCNN) (e). Adapted from Ref.~\cite{Sun}.}
\label{pic:SG} 
\end{figure*}

On the other hand, integration and sampling in high dimensions are among the central challenges in modern science and technology, underpinning applications that range from quantum many-body physics to Bayesian inference and artificial intelligence. While conventional Monte Carlo methods are formally scalable, their efficiency degrades rapidly when strong correlations or sharp features are present in high-dimensional configuration space. Recently, a new framework, termed the "Schr\"odinger Generator", for integration and sampling based on the explicit optimization of coordinate transformations was proposed \cite{Sun}. The method decomposes the total Jacobian into two complementary components: an adaptive map that minimizes estimator variance by learning the marginal structure in each dimension, and a normalizing-flow-based transformation that captures non-factorizable correlations in the target distribution. A final resampling step guarantees unbiased sampling even when the learned transformation is imperfect. The overall workflow of "Schr\"odinger Generator" and the architecture of normalizing flow are illustrated in Fig.~\ref{pic:SG}.
They  demonstrate stable and scalable performance for nuclear quantum many-body states in dimensions exceeding 600, faithfully reproducing short-range correlations among nucleons in finite nuclei. The framework offers a physically transparent approach to high-dimensional stochastic integration and sampling, opening new possibilities for simulations of complex quantum systems.

\subsection{Quantum machine learning}

Quantum machine learning is an active but still exploratory direction for high-energy and nuclear physics \cite{Guan:2020bdl,Boehnlein:2021eym,Yin:2026,Fang:2025}. Variational quantum circuits, quantum kernel methods, and quantum autoencoders have been studied for classification, anomaly detection, and data encoding in particle-physics settings \cite{Chen:2020qcnn,Bal:2025oneparticle}. For \HENP, potential applications include lattice-field-theory toy models, event classification, generative modeling, and hybrid quantum-classical inference. 

For an example, researchers have proposed 1P1Q, a quantum encoding scheme for high-energy physics \cite{Bal:2025oneparticle}. Unlike classical compression methods, 1P1Q maps each particle's kinematics directly onto a qubit, preserving full information for quantum algorithms. The scheme was validated on two tasks: a Quantum Autoencoder (QAE) for anomaly detection, and a Variational Quantum Classifier (VQC) for jet classification. The QAE outperformed classical autoencoders in distinguishing signal from QCD background jets while using far fewer parameters. The VQC achieved competitive performance with greatly reduced complexity. Real-world tests using 2016 CMS collision data confirmed the method's robustness.
Figure \ref{fig:Bal} illustrates the 1P1Q encoding for the ten hardest constituents of a top-quark jet (red) versus a QCD light-flavor jet (blue), with each particle assigned to a distinct qubit. The encoding intrinsically reflects the jet’s kinematic structure: angular coordinates and transverse momentum are mapped onto the Bloch sphere, thereby preserving geometric and energy correlations among constituents. The scaling function $f$
in Eq.: $f \rightarrow  1 + \frac{2\pi}{1+e^{-w}}$  serves to amplify the angular separation on the Bloch sphere.
These results demonstrate that 1P1Q offers a promising path for integrating quantum machine learning into collider data analysis, potentially reducing computational overhead while maintaining high performance in experimental physics workflows.

However, in general the field remains limited by hardware noise, data-loading costs, circuit depth, and the difficulty of demonstrating a quantum advantage for realistic scientific tasks.

The \QML\ in \HENP\ may become important when quantum simulation and quantum computing mature for real-time dynamics, finite-density systems, or Hamiltonian field theory. In the near term, \QML\ benchmarks should be small, transparent, and compared against strong classical baselines. Claims of advantage should include total resource accounting, not only circuit-level performance.

\subsection{Benchmarking and community infrastructure}
\label{subsec:benchmarks}

The field needs benchmarks designed around physics, not only architecture comparison. We propose four benchmark families. First, \LFT\ benchmarks should include scalar $\phi^4$ theory, two-dimensional $U(1)$ gauge theory, small-volume $SU(2)$ and $SU(3)$ gauge theories, sign-problem toy models, and eventually standardized lattice-QCD ensemble tasks. Metrics should include observable bias, autocorrelation time, acceptance rate, topological tunneling, continuum-scaling relevance, and wall-clock cost.

Second, heavy-ion benchmarks should include event-level datasets from transport, hydrodynamic, and hybrid simulations with standardized observables and known hidden parameters. Data formats should support particle clouds, event images, spectra, jets, and detector-smeared outputs. Metrics should include spectra, flow, correlations, centrality dependence, rare-event coverage, calibration, and performance under generator shift.

Third, neutron-star benchmarks should include synthetic and real-data-inspired ensembles of \EoS\ curves, mass-radius points, tidal deformabilities, X-ray spectra, and gravitational-wave likelihood summaries. Metrics should include posterior coverage, calibration, physical constraint satisfaction, sensitivity to priors, and detection power for phase transitions or exotic degrees of freedom.

Fourth, functional, EFT, and holographic benchmarks should include differential-equation solutions, FRG or DSE toy systems, spectral reconstruction tasks, and bottom-up model-inference datasets. Metrics should include residuals of physical equations, boundary-condition satisfaction, extrapolation to new parameter regimes, and uncertainty under truncation.

Benchmarks should also include negative controls. A classifier for criticality should be tested on datasets with no critical signal but similar multiplicity and flow backgrounds. A generator for hydrodynamic events should be tested on observables deliberately withheld from training. A lattice sampler should be tested on observables sensitive to topology and long-distance modes. A neutron-star inference method should be tested on mock data generated from \EoS\ families absent from the training set. These negative controls are essential because many AI models succeed by learning shortcuts.

\begin{figure}[t]
    \centering
    \includegraphics[width=0.94\linewidth]{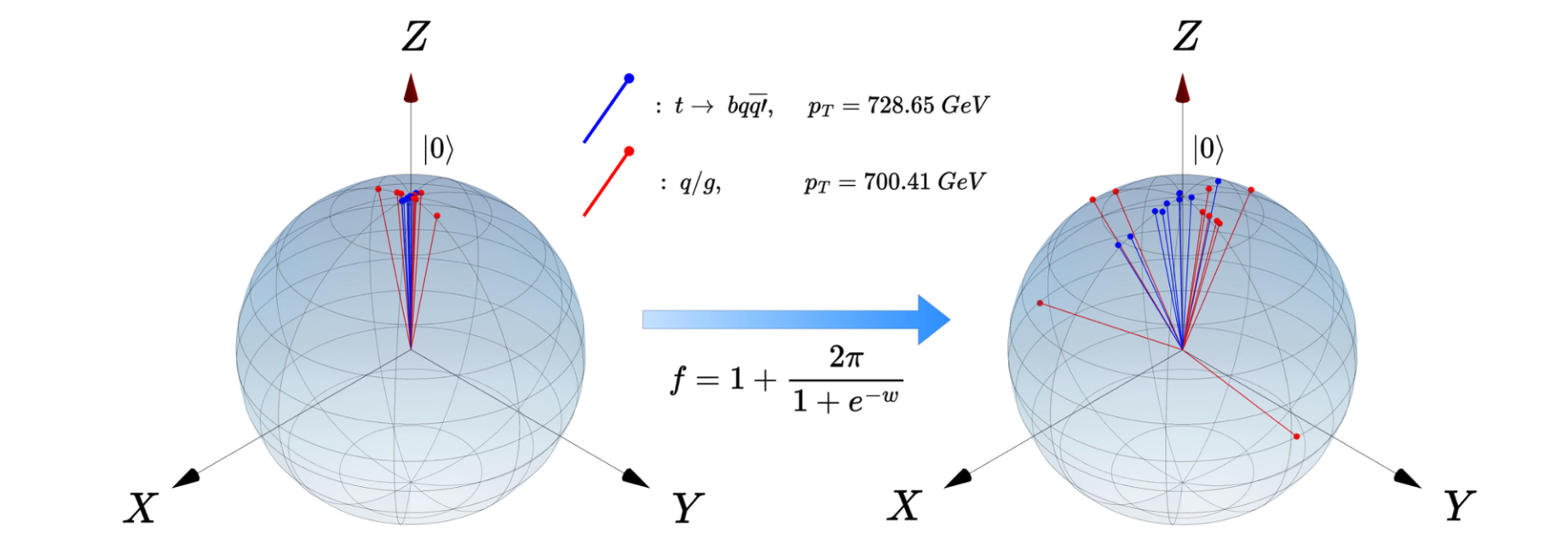}
    \caption{Bloch sphere: Effect of the input scaling described in $f \rightarrow 1 + \frac{2\pi}{1+e^{-w}}$ when applied to the ten hardest particles of a QCD and top jet with
comparable jet $p_T$. The value of $w$ converges to a scaling factor $f$ = 7.268 for the VQC. Adapted from Ref.~\cite{Bal:2025oneparticle}.}
    \label{fig:Bal}
\end{figure}

Open data standards would greatly accelerate progress. Heavy-ion event data should record generator provenance, physics parameters, random seeds where possible, detector or smearing assumptions, and observable definitions. Lattice benchmark datasets should record action, coupling, lattice size, boundary conditions, algorithm history, and autocorrelation estimates. Neutron-star datasets should distinguish prior samples, posterior samples, observational likelihood summaries, and physical filters. Without such metadata, benchmark comparisons are difficult to interpret.

\begin{table*}[t]
\caption{Suggested benchmark families for AI in high-energy nuclear physics. Detailed metrics are discussed in the text.}
\label{tab:benchmarks}
\resizebox{\textwidth}{!}{%
\begin{tabular}{llll}
\hline\hline
Sector & Candidate tasks & Main metrics & Key constraints \\
\hline
Lattice field theory & $\phi^4$, $U(1)$, $SU(2)$, $SU(3)$ sampling & Bias, autocorrelation, topology, cost & Gauge symmetry, exactness \\
Heavy-ion collisions & Events, spectra, flow, jets, CME/criticality & Correlations, calibration, generator shift & Conservation, acceptance \\
Neutron stars & EoS, mass-radius-tidal, GW/X-ray data & Coverage, phase-transition detection & Stability, causality \\
Functional/holographic & FRG/DSE, spectra, inverse holography & Residuals, extrapolation, uncertainty & Ward identities, UV/IR limits \\
\hline\hline
\end{tabular}}
\end{table*}

%% file: summary_outlook.tex
\section{Conclusions and outlook}
\label{sec:outlook}

AI in \HENP\ is moving from pattern recognition toward scientific workflow design. Three developments define this shift. Learned representations extract information from high-dimensional events, configurations, and observations that is difficult to isolate manually. Probabilistic ML and differentiable programming propagate uncertainties through expensive inverse problems. Generative models accelerate simulation and sampling when they respect physical constraints and are validated at the level of observables and inference.

The most important future direction is a closed \QCD\ inference loop. On the forward side, physics-respecting generative models should produce lattice configurations, hydrodynamic events, particle clouds, correlators, or \EoS\ functions faster and with quantified bias. On the inverse side, \SBI\ and Bayesian pipelines should extract posteriors for \EoS, transport coefficients, phase-structure hypotheses, and effective interactions from collider and astrophysical data. Across regimes, these posteriors should be compared and combined with lattice, continuum, EFT, perturbative, and holographic constraints. The long-term value of AI will be measured by its ability to enable such cross-regime consistency tests. The phrase ``closed inference loop'' should not be understood as a self-confirming autonomous system. A loop trained, calibrated, and validated on the same simulator family can amplify shared biases and converge toward an opaque interpolation of the existing model rather than a more faithful description of QCD. Other risks include overfitting to a fixed benchmark suite, repeatedly reusing theory-generated data for both training and validation, losing provenance as data products are transformed, and optimizing predictive scores while discarding mechanistic interpretability. Benchmarks are therefore necessary but not sufficient. Mitigation requires independent experimental and cross-model holdouts, negative controls, explicit model-discrepancy and out-of-distribution tests, interpretable intermediate physics quantities, immutable provenance records, and periodic confrontation with first-principles constraints and genuinely new data. Human review must remain responsible for changing priors, model classes, observables, and acceptance criteria.

Several recommendations follow. Physics constraints should be built into architectures and inference pipelines wherever possible. Gauge equivariance, conservation laws, thermodynamic stability, causality, and differentiable equations are the reason a model can be trusted outside its training set. Uncertainty quantification should be treated as part of the model rather than as a post-processing step. Community benchmarks and open datasets are essential for separating robust progress from benchmark-specific success. Foundation models and agents should be developed with strict provenance, reproducibility, and validation standards. Review articles and collaborative projects should also record negative results and failure modes, because they reveal which physics constraints are missing.

Three scientific questions should guide the next generation of work. Can AI reduce the cost of first-principles or high-fidelity calculations without compromising exactness or controlled uncertainty? This is the central question for lattice samplers, hydrodynamic emulators, and differentiable solvers. Can AI improve the information extracted from existing and future data? This is the central question for heavy-ion event-level learning, neutron-star \SBI, and optimized observables for criticality or CME searches. Can AI connect sectors that were historically analyzed separately? This is the central question for unified \QCD\ inference.

The field should also be explicit about current limits. AI cannot by itself solve the sign problem in full \QCD, determine the QCD critical point, or infer neutron-star core composition without assumptions. It cannot replace careful modeling of detector response, hydrodynamic uncertainty, or nuclear-theory priors. Its value is that it can make previously intractable workflows feasible, expose hidden correlations, and force assumptions to be stated in the language of priors, likelihoods, constraints, and validation tests.

The practical motivation will grow with forthcoming data. Beam-energy-scan and fixed-target programs seek rare or subtle signatures of dense \QCD\ matter, where high-dimensional event information and optimized observables may be decisive. Next-generation heavy-ion experiments will produce large volumes of particle-level data in complex detector environments. The Electron-Ion Collider will add complementary information on nucleon and nuclear structure, where event-level and detector-level ML will be important even though the present review focuses on QCD matter. Astrophysical data streams will also expand: additional gravitational-wave detections, X-ray radius constraints, and multimessenger observations will update dense-matter posteriors. AI methods validated now on controlled simulations should be designed to ingest these future data without changing the scientific standard of inference.

This also motivates infrastructure. A benchmark created for one paper is useful but limited; a benchmark maintained across experiments and theory groups can become a community reference. The long-term goal is an ecosystem in which new AI methods are compared on common physics tasks, with shared metrics and documented systematics.

The opportunity is substantial. New data from RHIC, the LHC, the Electron-Ion Collider, FAIR/CBM, NICA, J-PARC-HI, HIAF, neutron-star observatories, and gravitational-wave detectors will arrive in forms that are increasingly high dimensional and heterogeneous. At the same time, theoretical calculations are becoming more precise and more computationally demanding. AI can help bridge this gap when it is developed as part of the physics workflow, with \QCD\ reasoning setting the constraints and validation standards. The most promising future is one in which AI expands the range of \QCD\ questions that can be asked quantitatively.